\documentclass[manuscript]{aastex}

\usepackage{graphics}
\usepackage{mathptmx}
\usepackage{natbib}

\newcommand{\teff}{$T_{\!\mbox{\scriptsize\em eff}}$}

\newcommand{\msun}{$M_\odot$}

\newcommand{\siii}{Si\,{\sc ii}}
\newcommand{\siiii}{Si\,{\sc iii}}
\newcommand{\siiv}{Si\,{\sc iv}}

\def\kpc{\mbox{${\rm kpc}^{-1}$}}
 \newcommand{\hii}{\ion{H}{2}}

\slugcomment{To appear in Astrophysical Journal}

\shorttitle{Blue Supergiants in M81}
\shortauthors{Kudritzki et al.}

\begin{document}


\title{Quantitative Spectroscopy of Blue Supergiant Stars in the Disk of M81:\\
    Metallicity, Metallicity Gradient and Distance}


\author{Rolf-Peter Kudritzki\altaffilmark{1,2}, Miguel A. Urbaneja, Zachary Gazak, Fabio Bresolin}
\affil{Institute for Astronomy, University of Hawaii, 2680 Woodlawn Drive, Honolulu, HI 96822}
\email{kud@ifa.hawaii.edu, urbaneja@ifa.hawaii.edu, zgazak@ifa.hawaii.edu, bresolin@ifa.hawaii.edu}
\author{Norbert Przybilla}
\affil{Dr. Remeis-Sternwarte Bamberg \& ECAP, D-96049 Bamberg, Germany}
\email{przybilla@sternwarte.uni-erlangen.de}

\and

\author{Wolfgang Gieren, Grzegorz Pietrzy\'nski\altaffilmark{3}}
\affil{Universidad de Concepcion, Departamento de Astronomia, Casilla 160-C, Concepcion, Chile}
\email{wgieren@astro-udec.cl, pietrzyn@astrouw.edu.pl}

\altaffiltext{1}{Max-Planck-Institute for Astrophysics, Karl-Schwarzschild-Str.1, D-85741 Garching, Germany}
\altaffiltext{2}{University Observatory Munich, Scheinerstr. 1, D-81679 Munich, Germany}
\altaffiltext{3}{Warsaw University Observatory, Al. Ujazdowskie 4, 00-478 Warsaw, Poland}


\begin{abstract}

The quantitative spectral analysis of low resolution ($\sim 5$ \AA) Keck LRIS spectra of blue 
supergiants in the disk of the giant spiral galaxy M81 is used to determine stellar effective 
temperatures, gravities, metallicities, luminosites, interstellar reddening and a new distance 
using the Flux-weighted Gravity--Luminosity Relationship (FGLR). Substantial reddening and 
extinction is found with E(B-V) ranging between 0.13 to 0.38 mag and an average value of 0.26 mag.  
The distance modulus obtained after individual reddening corrections is 27.7 $\pm$0.1 mag. The 
result is discussed with regard to recently measured TRGB and Cepheid distances. The 
metallicities (based on elements such as iron, titanium, magnesium) are supersolar 
($\approx$ 0.2 dex) in the inner disk (R $\la$ 5 kpc) and slightly subsolar ($\approx$ -0.05 dex) 
in the outer disk (R $\ga$ 10 kpc) 
with a shallow metallicity gradient of 0.034 dex \kpc. The comparison with published oxygen abundances of
planetary nebulae and metallicities determined through fits of HST color-magnitude diagrams indicates
a late metal enrichment and a flattening of the abundance gradient over the last 5 Gyrs. This might be 
the result of gas infall from metal rich satellite galaxies. Combining these M81 metallicities with published 
blue supergiant abundance studies in the Local Group and the Sculptor Group a 
galaxy mass metallicity-relationship based solely on stellar spectroscopic studies is presented and compared 
with recent studies of SDSS star forming galaxies.

\end{abstract}





\section{Introduction}

The determination of the chemical composition and distances of galaxies is crucial for 
constraining the theory of galaxy formation and evolution in a dark energy 
and cold dark matter dominated universe. Ultimately, these measurements lead to ever stronger 
constraints on the cosmological
parameters and the history of cosmic chemical enrichment, from the primordial metal-free universe to the
present-day chemically diversified structure. For instance, the relationship between central metallicity and galactic
mass appears to be a Rosetta stone to understand chemical evolution and galaxy formation 
\citep{lequeux79, tremonti04, maiolino08}. In a similar way, the observed metallicity gradients
in spiral galaxies, apparently large for spirals of lower mass and shallow for high mass galaxies 
\citep{garnett97, skillman98, garnett04}, provide crucial insight into galaxy formation and evolution.
Both the observed mass-metallicity relationship and the abundance gradients are used to test the
theoretical predictions of hierarchical clustering, galaxy formation, merging, infall, galactic winds and
variability of star formation activity and IMF obtained in the framework of a $\Lambda$CDM dominated universe
\citep{prantzos00, naab06, colavitti08, yin09, sanchez09, delucia04, derossi07, finlator08, brooks07, koeppen07, 
wiersma09, dave11a, dave11b}. Note that this is only a small selection of papers relevant to the subject, 
others are found in the references therein.

However, as intriguing the observations of the mass-metallicity relationship and the metallicity gradients
of galaxies are, the published results are highly uncertain. They rely on observations of \hii~region
emission lines, mostly restricted to oxygen, and the analysis method applied is the so-called ``strong-line
method'', which uses the fluxes of the strongest forbidden lines of (most commonly) [\ion{O}{2}] and [\ion{O}{3}]
relative to  H$_{\beta}$. Unfortunately, abundances obtained with the strong-line method depend heavily on the
calibration used. As a striking example, \citet{kewley08} have demonstrated that the quantitative
shape of the mass-metallicity relationship of galaxies can change from very steep to almost flat depending
on the calibration used. In the same way, as shown by \citet{kud08} and
\citet{bresolin09} in their study of the Sculptor spiral galaxy NGC 300, 
metallicity gradients of spiral galaxies can change from steep to
flat and absolute values of metallicity can shift by as much as 0.6 dex, again as the result of different
calibrations of the strong line method. In consequence, galaxy metallicities are uncertain by 0.6 to 0.8 dex because of the
systematic uncertainties inherent in the strong line methods used. This major problem requires a fresh
approach and is begging for the development of a new and independent method less affected by
systematic uncertainties.

An obvious alternative method to constrain metallicity is the detailed quantitative
spectroscopic analysis of individual blue supergiant stars (BSGs) in galaxies. BSGs of spectral type A and B are massive
stars in the mass range between 12 to 40 \msun in the short-lived evolutionary phase ($10^{3}$ to $10^{5}$ years)
when they leave the hydrogen main sequence and cross the HR-diagram at constant luminosity and
almost constant mass to become red supergiants. Because of Wien's law massive stars increase their brightness in visual light
dramatically when evolving towards lower temperatures and reach absolute visual magnitudes up to M$_{V}
\approx$ -9.5 mag in the BSG phase \citep{bresolin03}, rivaling with the integrated light of globular clusters and dwarf galaxies.
Because of their extreme brightness they are ideal tools to accurately determine the chemical composition
of young stellar populations in galaxies.

BSG spectra are rich in metal absorption lines from several elements (C, N, O, Mg, Al, S, Si, Ti, Fe,
among others). As young objects with ages of 10 Myrs they provide important probes of the current
 composition of the interstellar medium. Based on detailed high resolution, very high signal-to-noise (S/N) studies of blue 
supergiants, which yield abundances as accurate as 0.05 dex \citep{przybilla06, schiller08, przybilla08},
\citet{kud08} developed an efficient new spectral diagnostic technique for
low resolution spectra (FWHM  $\sim 5$ \AA)  with good S/N ratio (50 or better), which allows for an
accurate determination of effective temperature, gravity, metallicity, interstellar reddening and
extinction. Metallicities accurate to 0.1 to 0.2 dex for each individual target can be obtained at this
lower resolution and S/N. The method has  been applied to irregular and spiral galaxies in the Local Group 
(WLM -- \citealt{bresolin06}; \citealt{urbaneja08}; NGC
3109 -- \citealt{evans07}; IC 1613 -- \citealt{bresolin07}; M33 -- \citealt{u09}) and beyond (NGC 300 -- \citealt{kud08}).

In this paper we present the spectral analyis of low resolution Keck LRIS spectra of 26 
BSGs in the disk of the giant spiral galaxy M81. M81 is one of the most massive spirals in the 
Local Volume \citep{mccommas09}. It has low foreground extinction with a galactic luminosity of 2.5 L$_{*}$ 
(corresponding to M$_{K}$ = - 24 mag and M$_{K*}$ = -23 mag)
and is characteristic of disk galaxies seen at redshift surveys out to z $\sim$ 1 \citep{williams09}.
The star formation history and chemical evolution of this galaxy have been subject to extensive recent 
photometric studies \citep{dalcanton09, davidge09, williams09, barker09, durell10}. 
\hii~regions and Planetary Nebulae have been studied by \citet{stanghellini10} extending the classical 
work by \citet{garnett87} and \citet{stauffer84}. With our work we provide for the first time direct 
quantitative spectroscopic information about stellar metallicity of the young disk population.

An important additional aspect of the quantitative spectroscopy of BSGs is their use as accurate 
distance indicators through the Flux-weighted Gravity--Luminosity Relationship (FGLR). This new distance 
determination method has been introduced by \citet{kud03} and \citet{kud08}. 
It uses stellar gravity 
and effective temperature as a measure of absolute bolometric magnitude and provides a distance estimate
which is free of the uncertainties caused by interstellar reddening, since the determination of reddening 
is a by-product of the quantitative spectral analysis. First distance determinations using this method 
have been carried out by \citet[][ WLM]{urbaneja08} and \citet[][ M33]{u09}.

There has been a long history of attempts to measure the distance to M81 from Hubble (1929) to the present 
(see \citealt{mccommas09} for references and a plot of distance modulus as a function of time). The work 
published over the last decade gives a range between 27.60 to 28.03 mag in distance modulus indicating an 
uncertainty of 20\%. Our BSG spectroscopy and the effective temperatures and gravities determined will 
give us a FGLR distance which we can then compare with most recent HST work on Cepheids
and the tip of the red giant branch, TRGB. M81 has been used as one of the 
calibration galaxies for the Tully-Fisher and the surface fluctuation methods in the HST Key project 
\citep{freedman01} and by \citet{mould08, mould09}. 

In section 2 of this paper we describe the observations and data reduction. Section 3 discusses the 
quantitative spectroscopic analysis and the determination of extinction, effective temperature, gravity 
and metallicity. Section 4 and 5 discuss interstellar reddening and compare the spectroscopically 
determined stellar parameters with evolutionary tracks in order to constrain the evolutionary status of 
the objects observed. Section 6 compares metallicity and metallicity gradient of the BSGs with published 
metallicity constraints for the older disk population of M81 and discusses chemical evolution over the last 
Gyrs. In section 7 we provide a galaxy mass-metallicity relationship based on BSG spectroscopic studies
and compare with published work using \hii~region emission lines. In section 8 we determine a new distance 
to M81 using the FGLR-method and discuss recent Cepheid and TRGB work. Section 9 sumarizes the results 
and discusses aspects of future work.

\section{Observations and Data Reduction}

\begin{figure}[!]
 \begin{center}
  \includegraphics[width=0.90\textwidth]{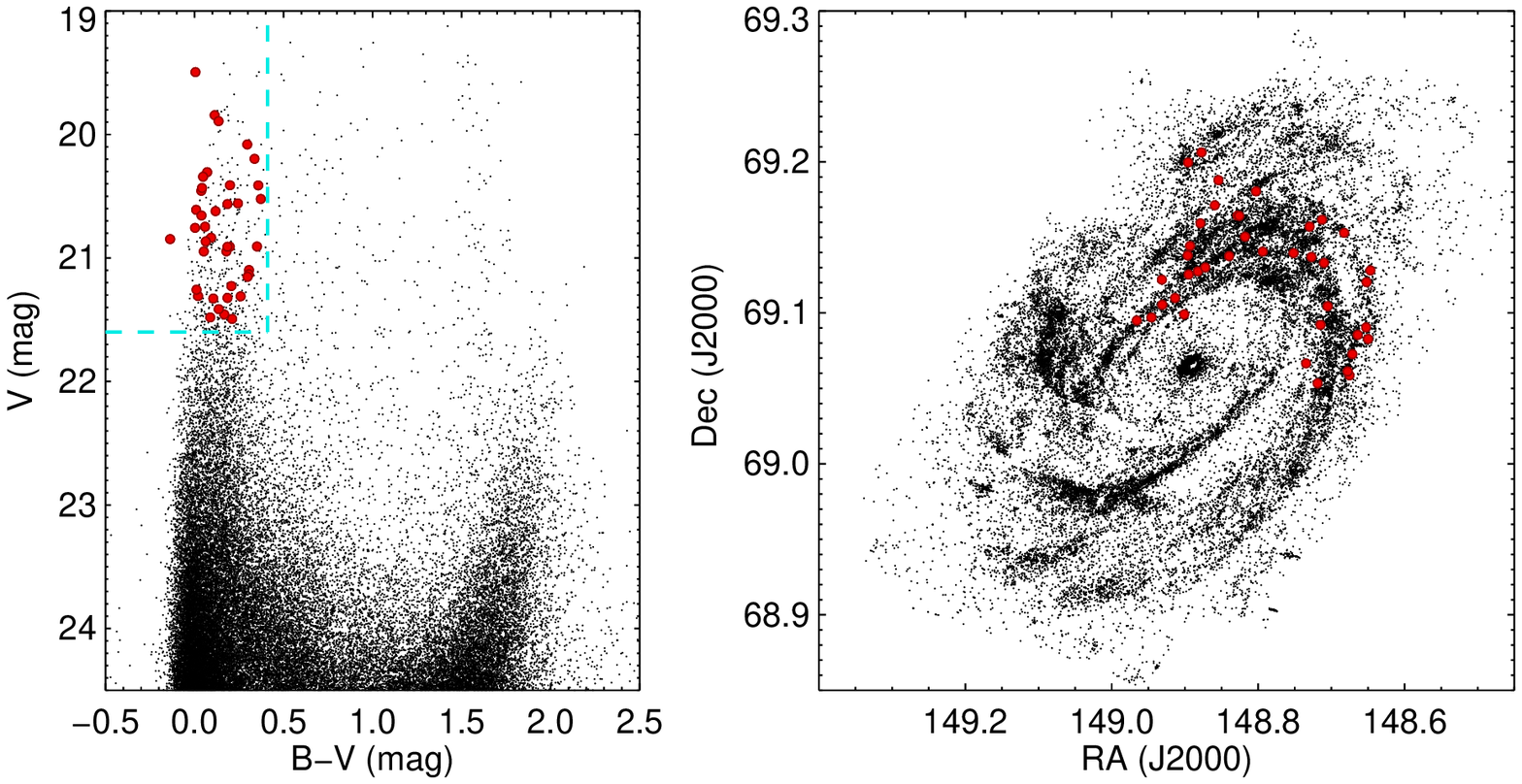}
  \caption[]{Selection of M81 BSG targets. Left: Color magnitude diagram (photometry from \citealt{dalcanton09}) 
with selection box (blue dashed) and selected targets (red). Right: Location of selected targets within M81.\label{m81cmd} }
 \end{center}
\end{figure}

The observations were carried out with the Keck 1 telescope on Mauna Kea and the Low Resolution Imaging Spectrograph
\citep[LRIS,][]{oke95} using the atmospheric dispersion corrector, a slit width of 1.2 arcseconds, the D560 dichroic and  
the 600/4000 grism (0.63~\AA\,pix$^{-1}$) and the  900/5500 grating (0.53~\AA\,pix$^{-1}$) in the 
blue and red channel, respectively. In this paper, we 
will discuss and analyze the blue channel (LRIS-B) spectra only, which have a resolution of 5 \AA~FHWM. Because of the 
UV sensitivity of the LRIS-B configuration the spectra extend to shortward of the Balmer discontinuity at 
3640~\AA, which is crucial for the determination of T$_{\rm eff}$ from the Balmer jump (see section 3).
Three MOS fields were prepared with 20 to 25 targets each. The
BSG candidate targets were selected from HST ACS B,V images obtained within the ANGST project 
\citep{dalcanton09}, which covers the whole galaxy. Published B,V photometry of the 
M81 ANGST fields was used to preselect targets with point source PSF characteristic and with -0.2 mag $\la$ B-V $\la$ 0.4 mag and V $\la$ 21.5 mag.
Each target was carefully inspected with regard to multiplicity. Fig.~\ref{m81cmd} shows the selection 
from the color-magnitude diagram (CMD) and the location of our targets within the galaxy. None of our targets 
is related to one of the stellar clusters investigated by \citet{chandar01} or \citet{santiago10}.

The observations were scheduled for three dark nights in 2010 (February 14 to 16). The first night had 
perfect conditions with 0.75 arcsec seeing yielding reasonably exposed 
spectra with a total exposure time of 6.75 hours (observed in exposure segments of 45 minutes each) of 
the first field (field Z). The observing conditions degraded significantly during the second and third nights 
with poor seeing (1.3 arcsec) and occasional clouds. As a result, almost one half of these two nights was lost 
and only one additional field (field C) could be observed with a total of 11.3 hours exposure time under 
mediocre conditions.

Data reduction was performed using a custom pipeline written in IDL
designed to efficiently extract faint objects observed over a full night.
LRIS science and calibration frames were flat fielded and bias subtracted.  For each reduced frame, 
object spectra were traced along the dispersion axis and extracted using the optimal extraction method 
\citep{horne86} meant to maximize the S/N of faint spectra.  For this technique we utilized 
a Moffat function which was determined to best fit the 2-D spectral profile at each pixel (wavelength) 
perpendicular to the dispersion.  The Moffat fit was modified to include a measure of the background 
level for subtraction.  The spectra were then wavelength calibrated using techniques in the idlspec2d 
IDL package developed for SDSS.  

Each science object spectrum was flux calibrated by performing corrections for wavelength dependent 
extinction at varying airmass over Mauna Kea \citep{beland88} and then multiplying by a sensitivity 
function to convert extracted data numbers into units of ergs/s/cm$^2$/\AA.  The sensitivity function 
was calculated by scaling airmass-corrected observed flux standard stars (GD 50, Feige 34, HZ 44, 
and BD+33d2642) to the published spectral energy distributions of \cite{oke90}.  A final 
spectrum for each target was produced by taking the median of all wavelength and flux calibrated spectral 
frames.  Those spectra were normalized by manually selecting continuum regions and dividing by a high 
order polynomial fit to the continuum flux levels. The S/N values of our spectra vary between 40 to 80.

Table 1 provides the information about the objects used for this spectroscopic study. While we selected 
25 targets in each field, we could not use all of them. A few turned out to be blue foreground objects 
in the Milky Way halo, some had composite spectra indicating the presence of several objects in the 
slit and for some the S/N was not sufficient. For the remaining objects we list coordinates, galactocentric 
distance, spectral type, V magnitude, B-V color and the measured Balmer jump D$_{B}$ in Table 1. The way, how
D$_{B}$ is defined and measured, is described in \citet{kud08}. 

\section{Spectroscopic Analysis}

The analysis method has been described in detail in \citet{kud08}. A comprehensive grid of line-blanketed 
model atmospheres and very detailed NLTE line formation calculations is used to calculate  
spectral energy distributions (SEDs), including the Balmer jump, and normalized synthetic spectra. Relative 
to the work presented in \citet{kud08} the grid has been extended to cover temperatures from 16000K down to 
7900K at gravities between log g =3.0 to 0.8 (cgs). The lower limit of log g is a function of T$_{\rm eff}$ parallel 
to the Eddington-limit. Models are calculated for 14 metallicities [Z] = $\log\left(\mathrm{Z}/\mathrm{Z}_{\odot}\right)$ : -1.30, 
-1.15, -1.00, -0.85, -0.70, -0.60, -0.50, -0.40, -0.30, -0.15, 0.00, 0.15, 0.30, 0.50 dex. Z/Z$_{\odot}$ is the 
metallicity relative to the sun in the sense that the abundance for each element is scaled by the same 
factor relative to its solar abundance. Solar abundances were taken from \cite{grevesse98}, except for oxygen
where we adopt the value from \citet{allende01}. For all further details of the model grid we refer the 
reader to \citet{kud08}. The physics of the model atmospheres and the NLTE line formation calculations 
are described in detail by \citet{przybilla06} and references therein.

\begin{figure}
\begin{center}
  \includegraphics[scale=0.3,angle=90]{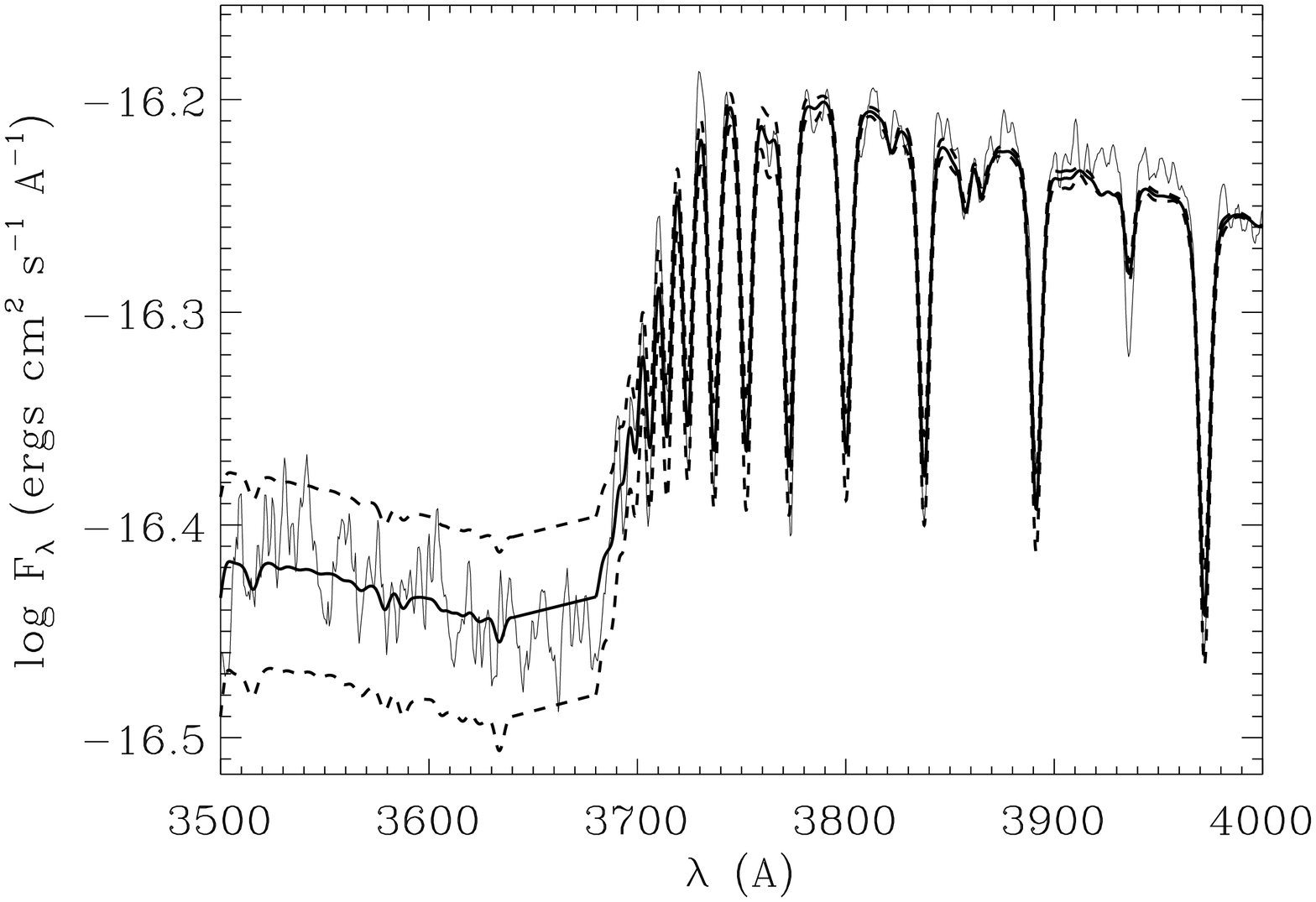}
 \includegraphics[scale=0.3,angle=90]{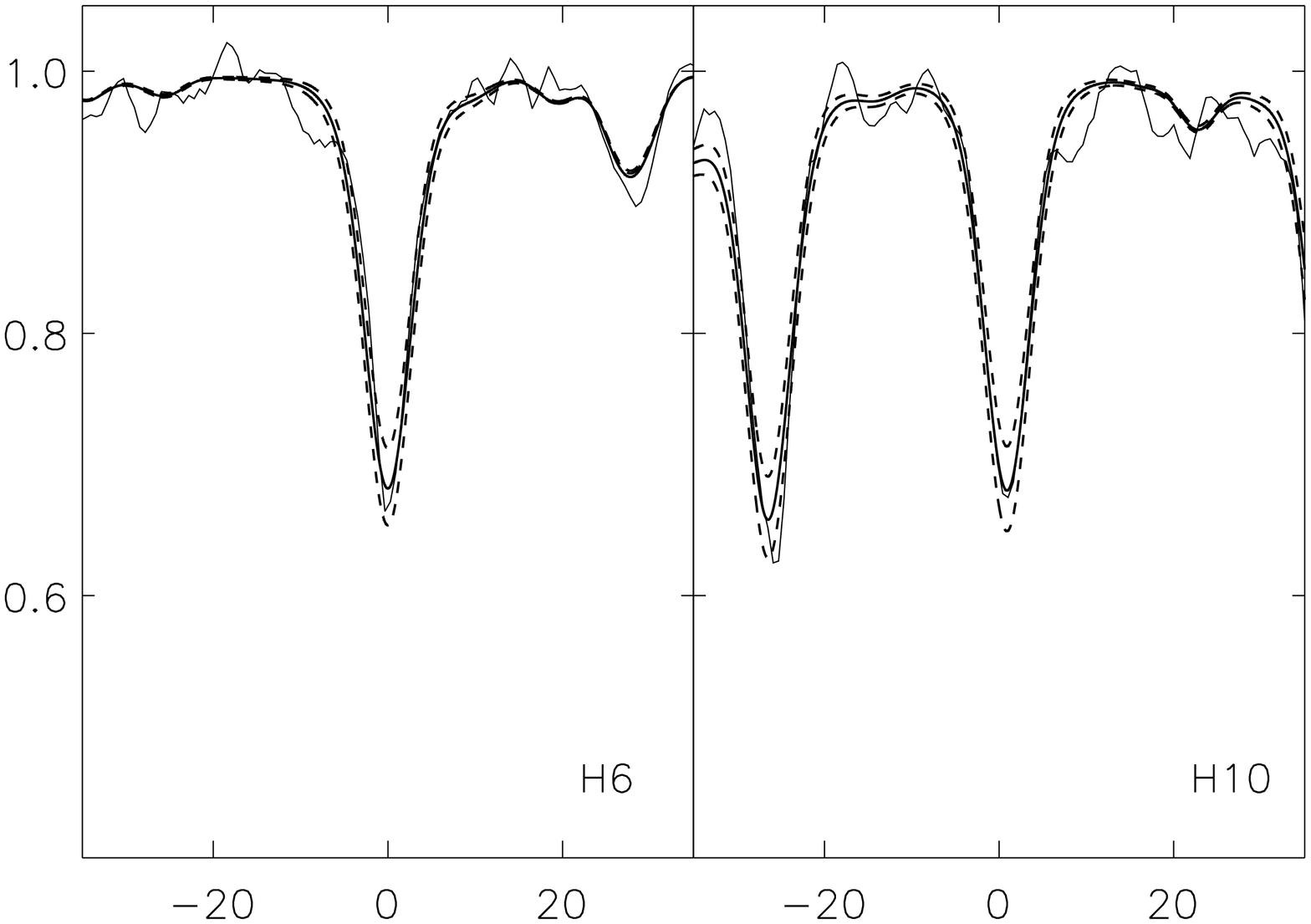} 
\caption{Analysis of object Z10. Left: Fit of the observed Balmer jump. The final model with the 
parameters given in Table 2 (thick solid curve) fits the Balmer discontinuity well. Two models 
with T$_{\rm eff}$ higher/lower by 500K are also shown (dashed) to demonstrate the temperature 
sensitivity of the D$_{B}$ fit. Right: Fit of two Balmer lines with the final model. Two models 
with log g higher/lower by 0.1 dex are shown (dashed) to demonstrate the gravity sensitivity of 
the Balmer line fits. Note that the strong spectral line at the left edge of the panel for H$_{10}$ 
is H$_{11}$, which is not used for the fits,
because it is at the edge of the normalized spectrum, where continuum rectification 
becomes difficult.  \label{z10}}
 \end{center}
\end{figure}

\begin{figure}
\begin{center}
  \includegraphics[scale=0.3,angle=90]{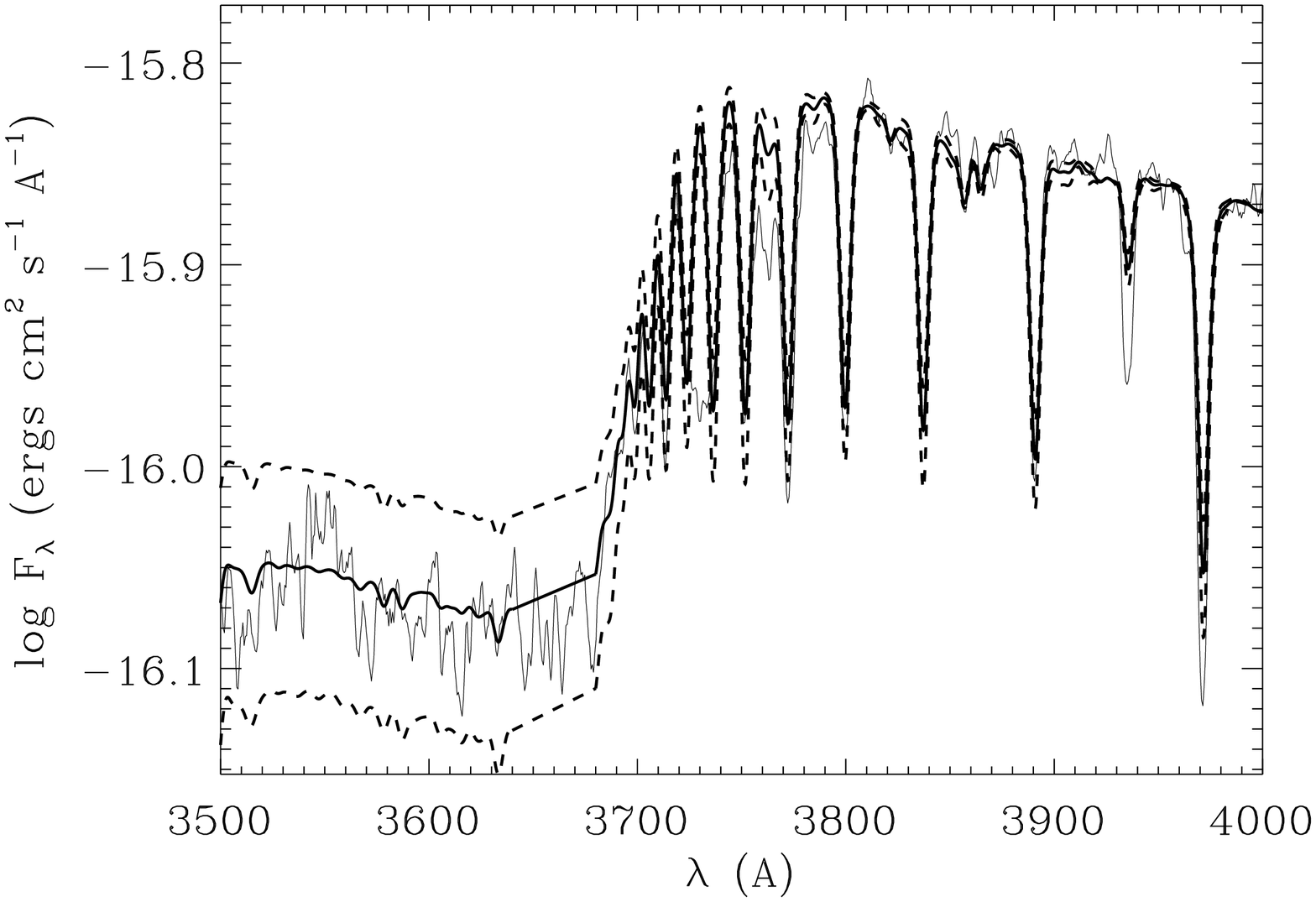}
 \includegraphics[scale=0.3,angle=90]{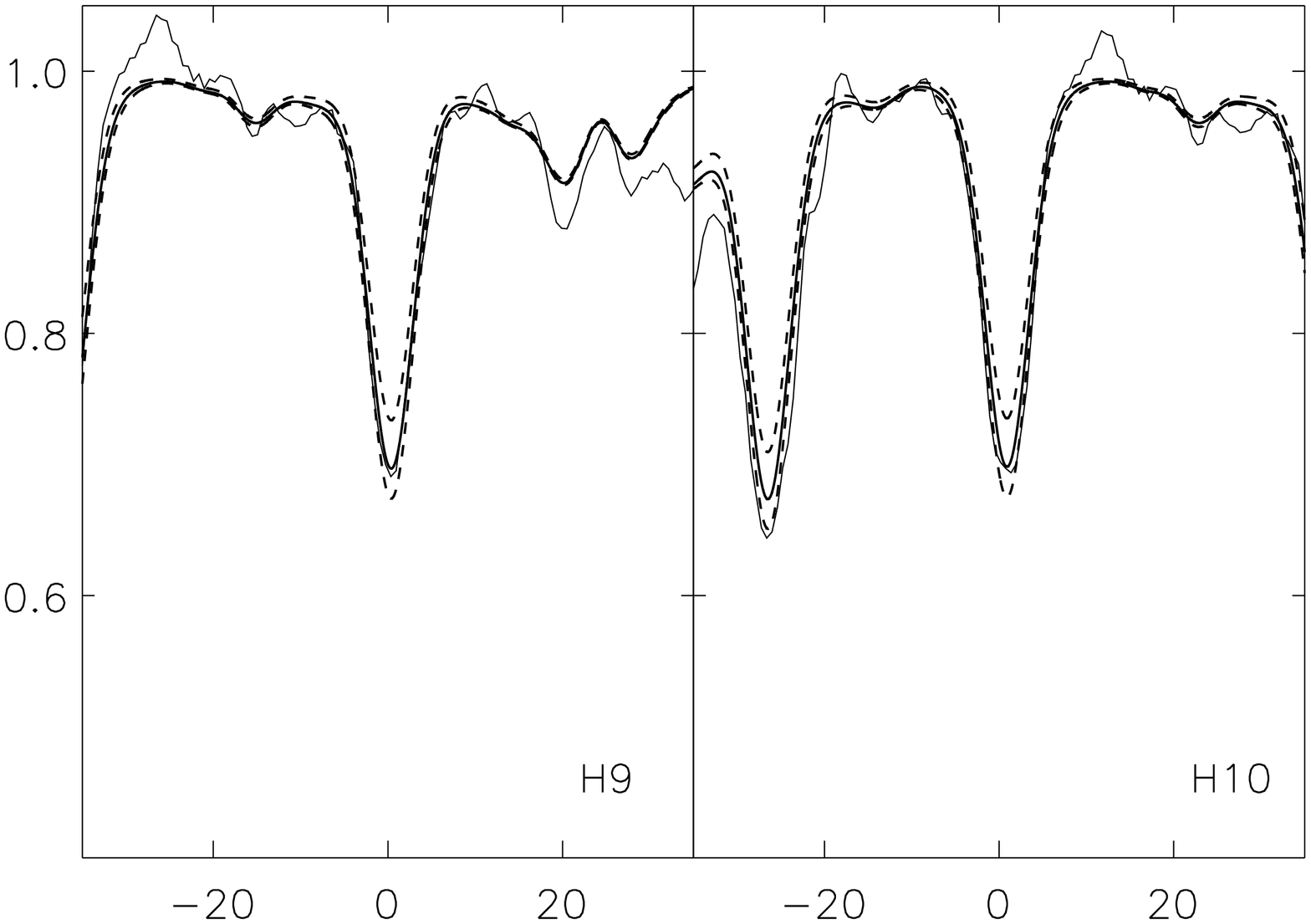} 
\caption{Same as Fig. 2 but for object C20. \label{c20}}
 \end{center}
\end{figure}

\begin{figure}
\begin{center}
  \includegraphics[scale=0.3,angle=90]{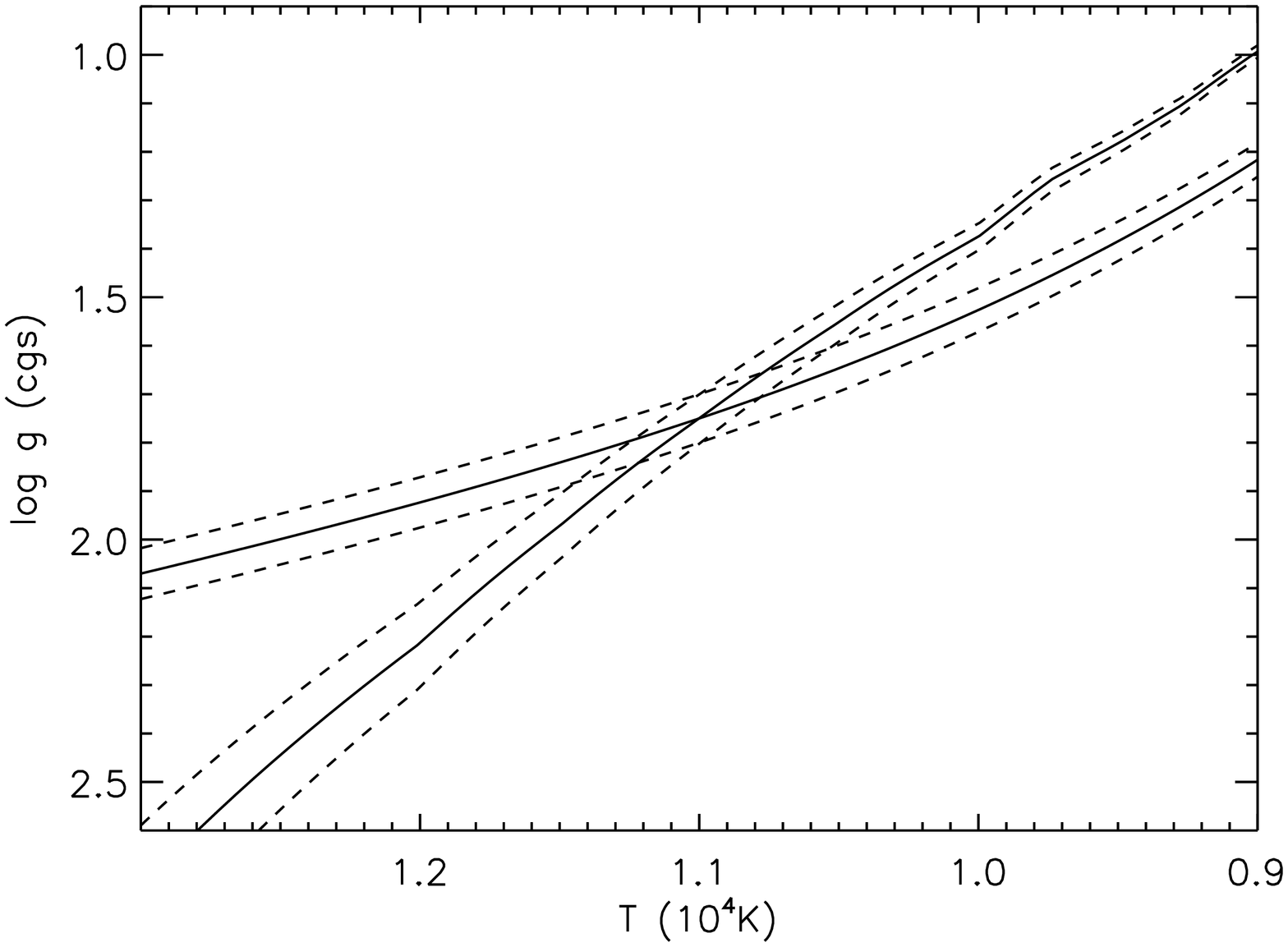}
 \includegraphics[scale=0.3,angle=90]{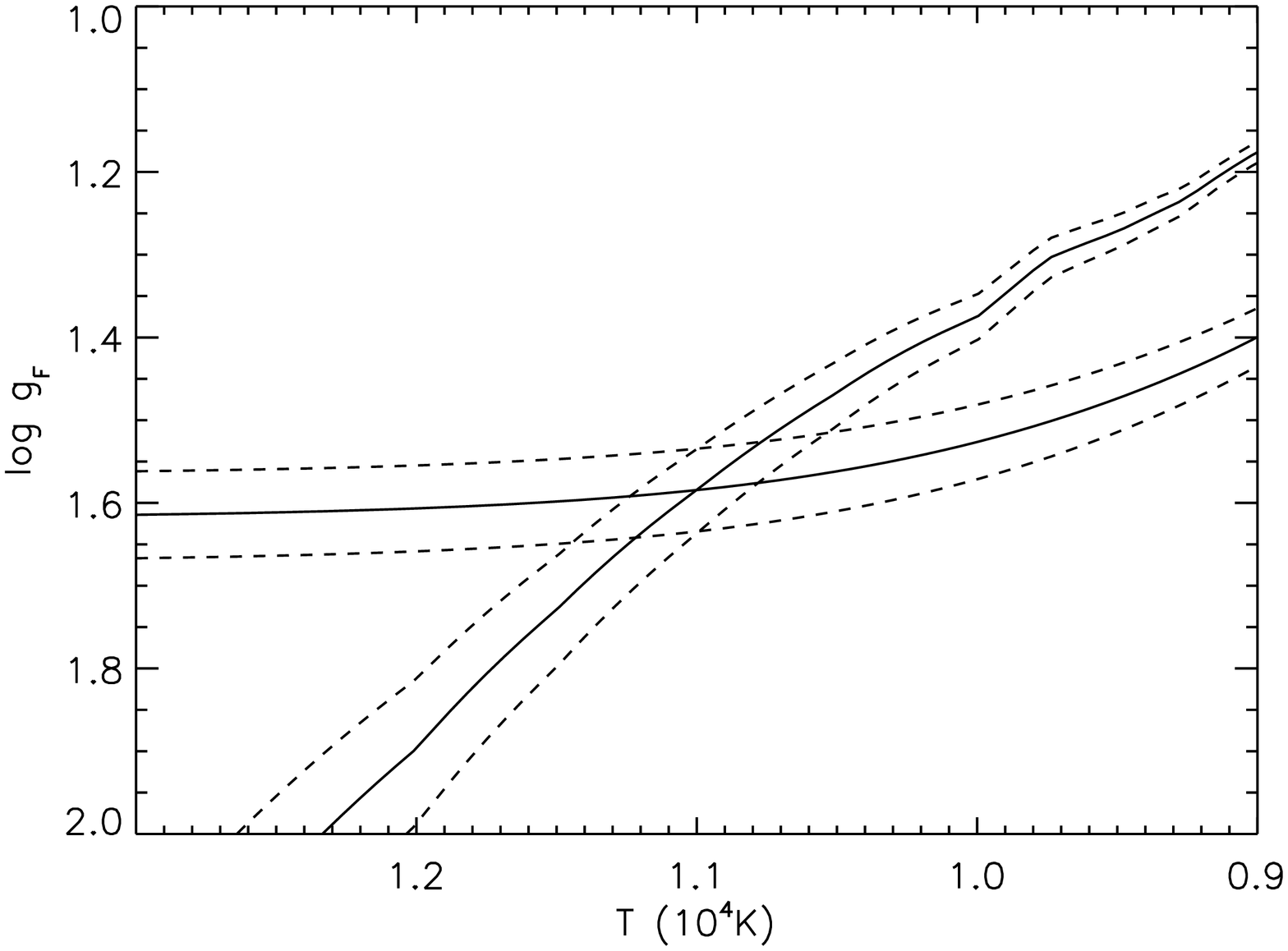} 
\caption{Fit diagrams for the fit of the Balmer jump (steeper curves) and the Balmer lines. Left: the (log g, T$_{\rm eff}$) diagram, right: (log g$_{F}$, T$_{\rm eff}$). Teff is given in 10$^4$ K. The dashed curves indicate maximum errors of the fits. For discussion, see text.\label{z10fit}}
 \end{center}
\end{figure}

\begin{figure}
\begin{center}
  \includegraphics[scale=0.8]{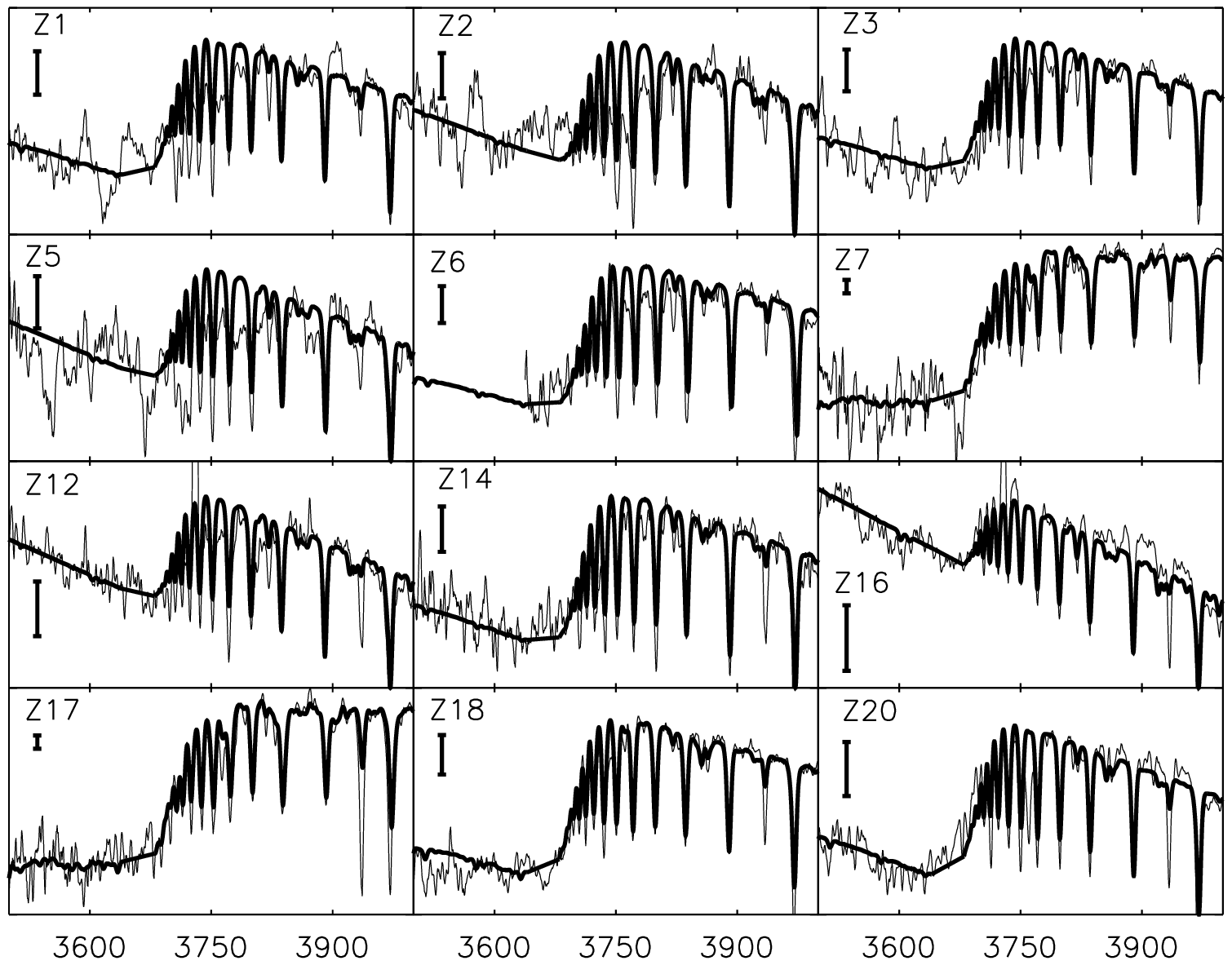}
\caption{Balmer jump fit for 12 objects in field Z. Logarithm of flux is plotted vs. wavelength in \AA. 
The bar in each panel indicates 0.05 dex changes in flux level. \label{zdb}}
 \end{center}
\end{figure}

\begin{figure}
\begin{center}
  \includegraphics[scale=0.7, angle=90]{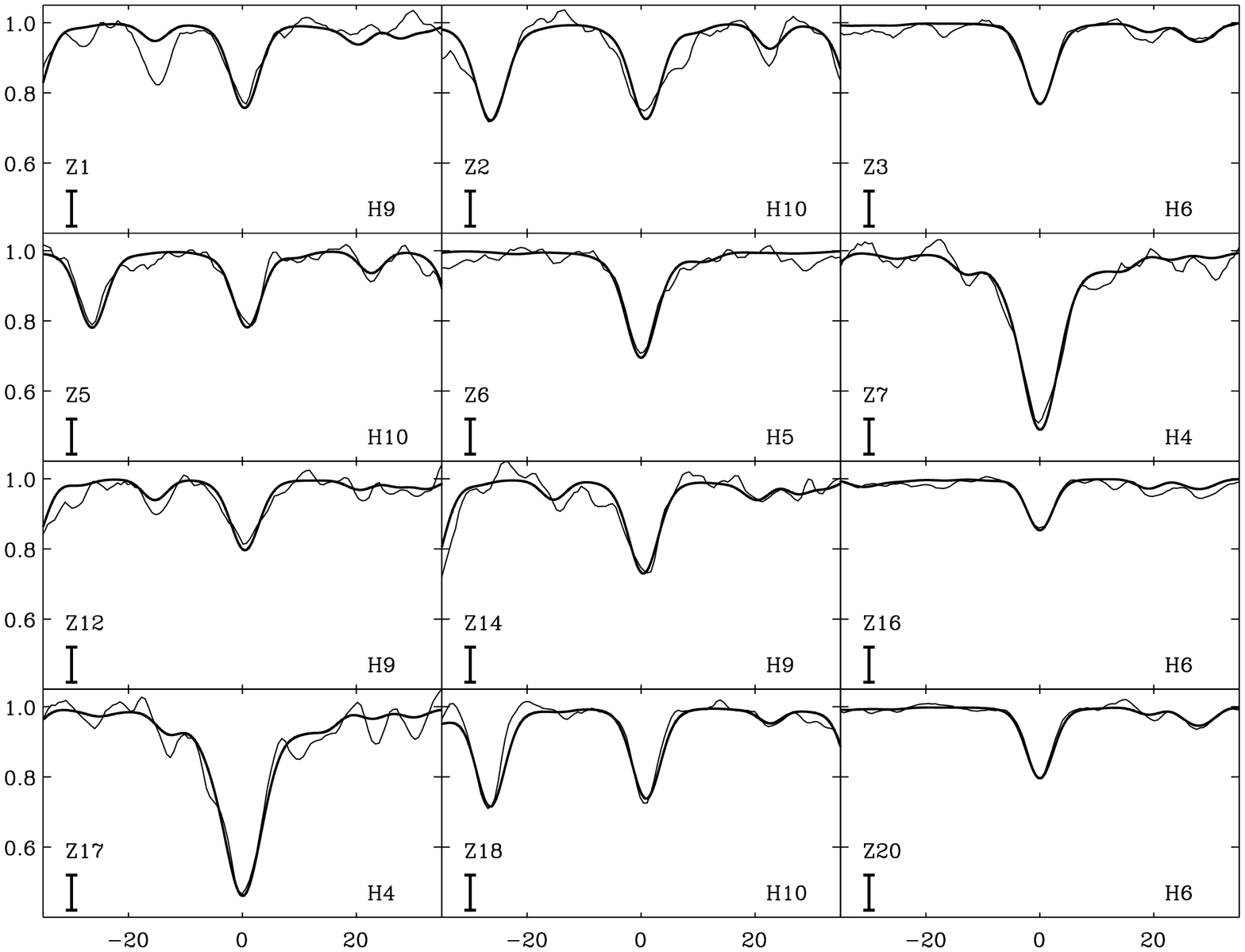}
\caption{Balmer line fit for 12 objects in field Z. Normalized flux is plotted vs. wavelength displacement from the line center in \AA. \label{zbalm}}
 \end{center}
\end{figure}

\begin{figure}
\begin{center}
  \includegraphics[scale=0.7, angle=90]{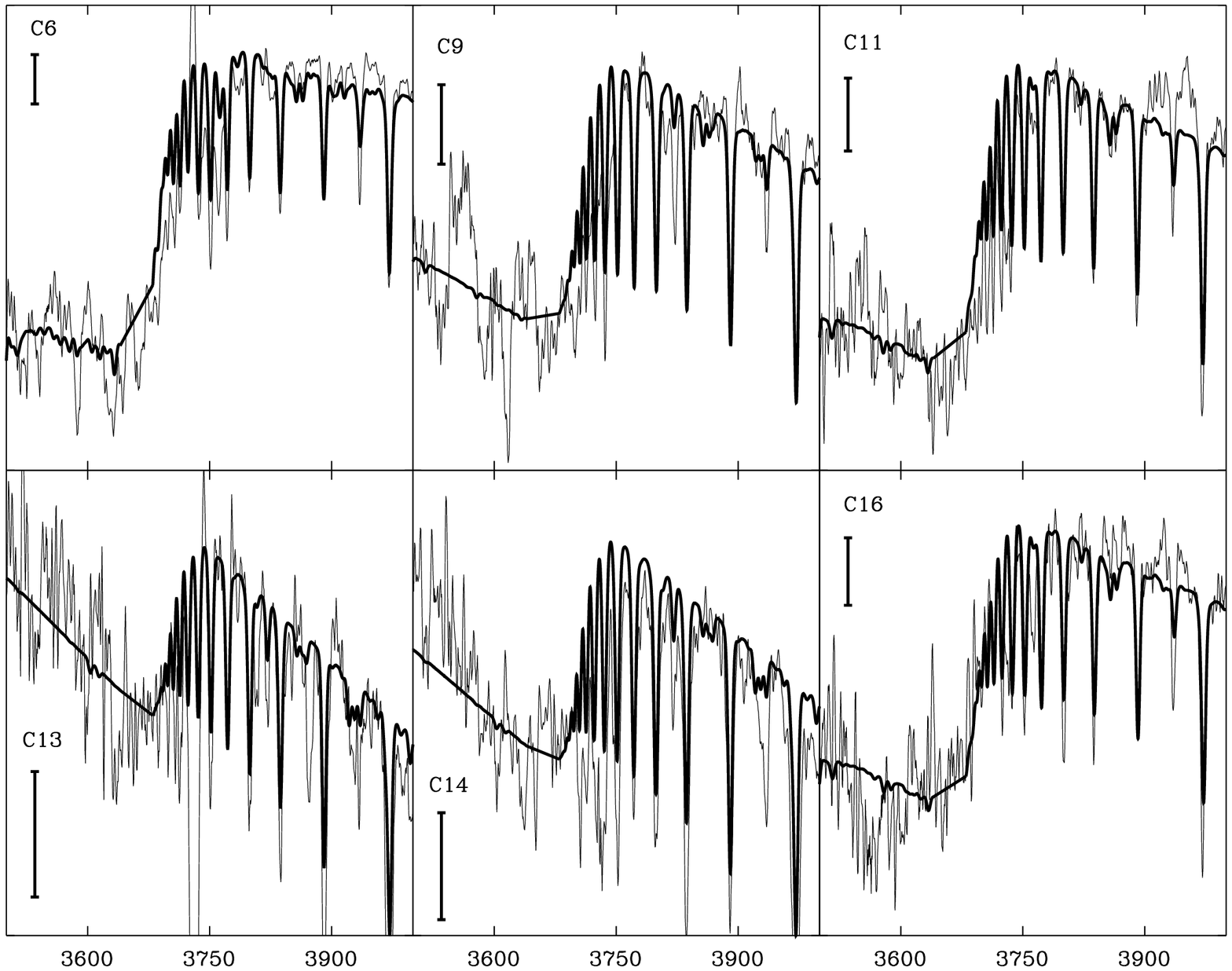}
\caption{Balmer jump fit for 6 objects in field C. Logarithm of flux is plotted vs. wavelength in \AA. 
The bar in each panel indicates 0.05 dex changes in flux level. \label{cdb}}
\end{center}
\end{figure}

\begin{figure}
\begin{center}
  \includegraphics[scale=0.8]{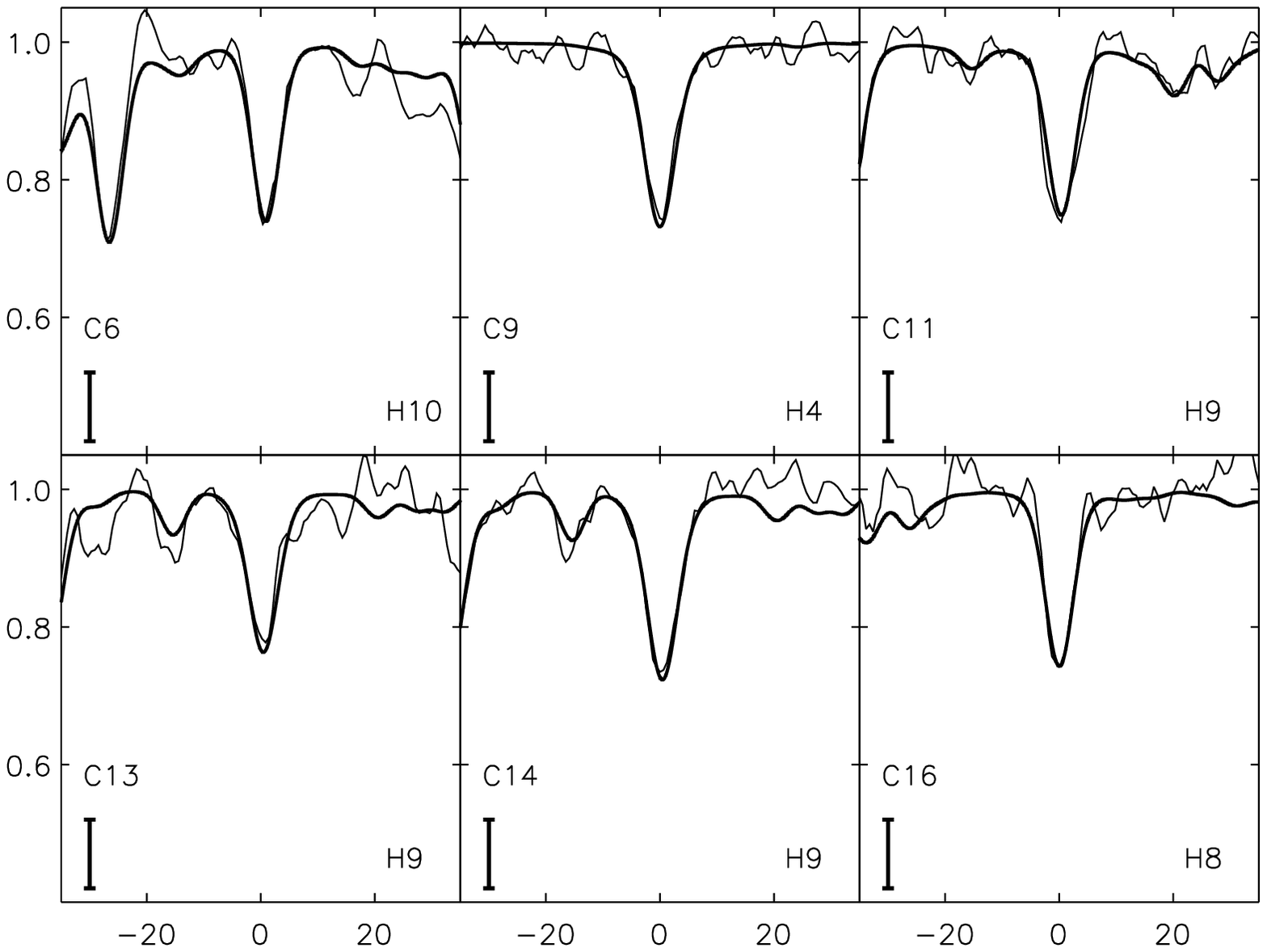}
\caption{Balmer line fit for 6 objects in field C. Normalized flux is plotted vs. wavelength displacement from the line center in \AA. \label{cbalm}}
\end{center}
\end{figure}

The spectral analysis proceeds in several steps. First, fit curves in the (log g, T$_{\rm eff}$)-plane are 
constructed, along which the models reproduce the observed Balmer jump and the Balmer lines. The Balmer jump 
is mostly a function of temperature, but also depends weakly on gravity, whereas the Balmer lines depend 
mostly on gravity and weakly on temperature. Fig.~\ref{z10} and~\ref{c20} demonstrate the dependence of 
the Balmer jump on temperature and of the Balmer lines on gravity. The intersection area of these fit 
curves determines the stellar effective temperatures and gravities and the corresponding uncertainties 
(see Fig.~\ref{z10fit}). The fact that the fit curves for the Balmer jump and the Balmer lines are not 
orthogonal leads to relatively large error boxes, in particular with regard to gravity log g. On the other hand,
the flux weighted gravity

\begin{equation}
        \log~g_{F} = \log~g - 4\log (T_{\rm eff} \times 10^{-4})
\end{equation}
\noindent
is determined much more accurately, since the Balmer lines depend solely on $\log\,g_{F}$ for temperatures 
higher than 9000K (for an explanation of the physics behind this behaviour, see \citealt*{kud08}). This is important 
for the use of flux weighted gravity as an indicator of absolute magnitude and distance (see section 8).
Fig~\ref{zdb},~\ref{zbalm},~\ref{cdb},~\ref{cbalm} show fits of D$_{B}$ and one Balmer line for the remaining objects in 
Table 1 (with spectral types later than or equal to B3) to give an impression of the quality of the data.
We note that we usually try to use all Balmer lines from H$_{4}$ to H$_{10}$ to
constrain gravity. However, varying from star to star we may encounter 
difficulties with individual Balmer lines. H$_{7}$, for instance, is many 
times corrupted by interstellar CaII absorption. H$_{4}$, H$_{5}$ and even H$_{6}$ are
sometimes affected by HII emission. Another problem are strong stellar winds,
which can fill H$_{4}$ and H$_{5}$ with broad emission. Spectral flaws by improper
corrections of comic ray hits may also affect line profiles. However,
in general, we have more than one Balmer line per star to constrain
gravity, usually three to four. For Fig. 2, 6, 8 and 9 we have selected 
the best fitting cases.

\begin{figure}
\begin{center}
  \includegraphics[scale=0.8]{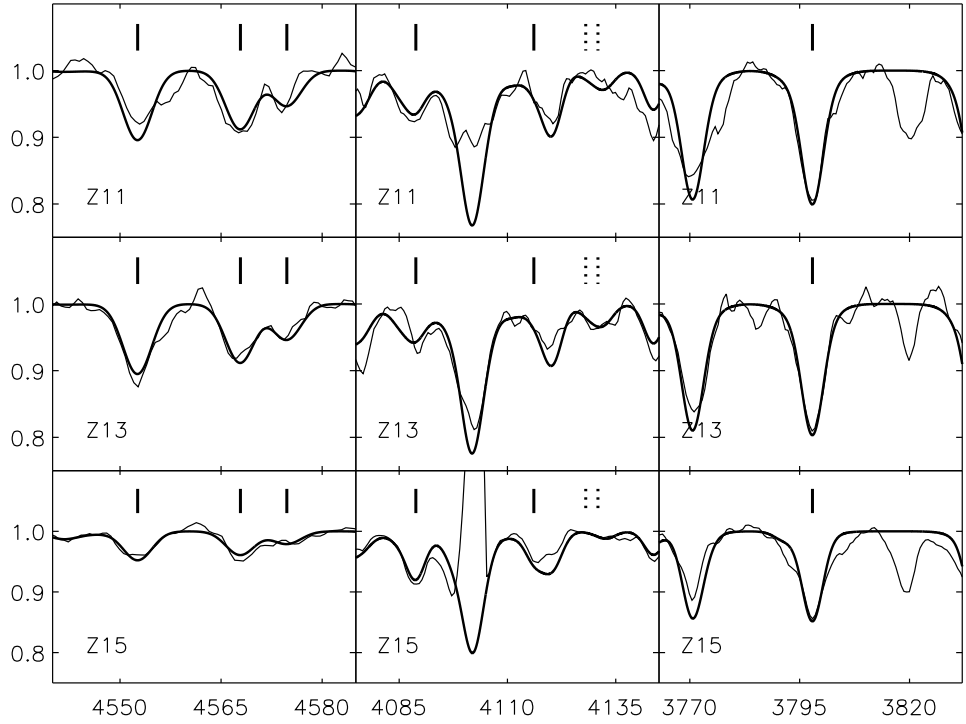}
\caption{\ion{Si}{3} line fits (left), \ion{Si}{4} (solid bars) and \ion{Si}{2} (dotted bars) line fits (middle), and hydrogen H10 line fits (right) for the
three early B supergiants of our sample. Note that \ion{Si}{4} 4116\AA~is blended by \ion{He}{1}. Normalized flux is plotted vs. wavelength in \AA. \label{bfit}}
\end{center}
\end{figure}

Three objects of our sample are of earlier spectral type (B0.5 to B1.5). For those, the Balmer jump is not a 
good temperature indicator. We use the ionization equilibrium of \ion{Si}{2}, \ion{Si}{3}, and \ion{Si}{4}~lines instead 
and apply the analysis method developed by \citet{urbaneja05}, which relies on the use of 
line-blanketed NLTE model atmospheres including the effects of stellar winds. Fig~\ref{bfit} shows the 
spectral fits for the key lines of these objects.

For three objects of later spectral type (Z4, Z9, C21) the wavelength range of the observed spectra 
does not cover the region of the Balmer jump. Thus, the only way to estimate their temperature is the 
relationship between effective temperature and spectral type (see \citealt{kud03}). As shown by 
\citet{kud08} this method works only, as long as the metallicity is about solar. From the galactocentric 
distance of these objects and our study of metallicity and metallicity gradient (see section 6) for the other 
objects in our sample this seems to be a reasonable assumption and, thus, temperature, gravity and luminosity of 
these objects are very likely well determined. Nevertheless, we will not make use of these objects for the 
determination of the distance to M81 from the flux weighted gravity. 

We note that with the fit of either the Balmer jump D$_{B}$, or the silicon equilibrium, or  the spectral 
type in the (log g, T$_{eff}$)-plane we can always calculate a reddening correction E(B-V) along the fit curve at each effective 
temperature by comparing the observed value of B-V with the one calculated by the atmospheric model. For fitting 
the Balmer jump, we then correct for reddening using 
the \cite{cardelli89} reddening law with R$_{V}$ = 3.1. Once we have found the intersection with the fit curve for the Balmer lines, 
we can then also determine the final reddening value E(B-V) for the final values of T$_{eff}$ and log g. It is a 
big advantage of this spectroscopic determination of stellar parameters that it yields interstellar reddening for free.
For an estimate of distances, this is a fundamental advantage of the method.

\begin{figure}
\begin{center}
  \includegraphics[scale=0.4]{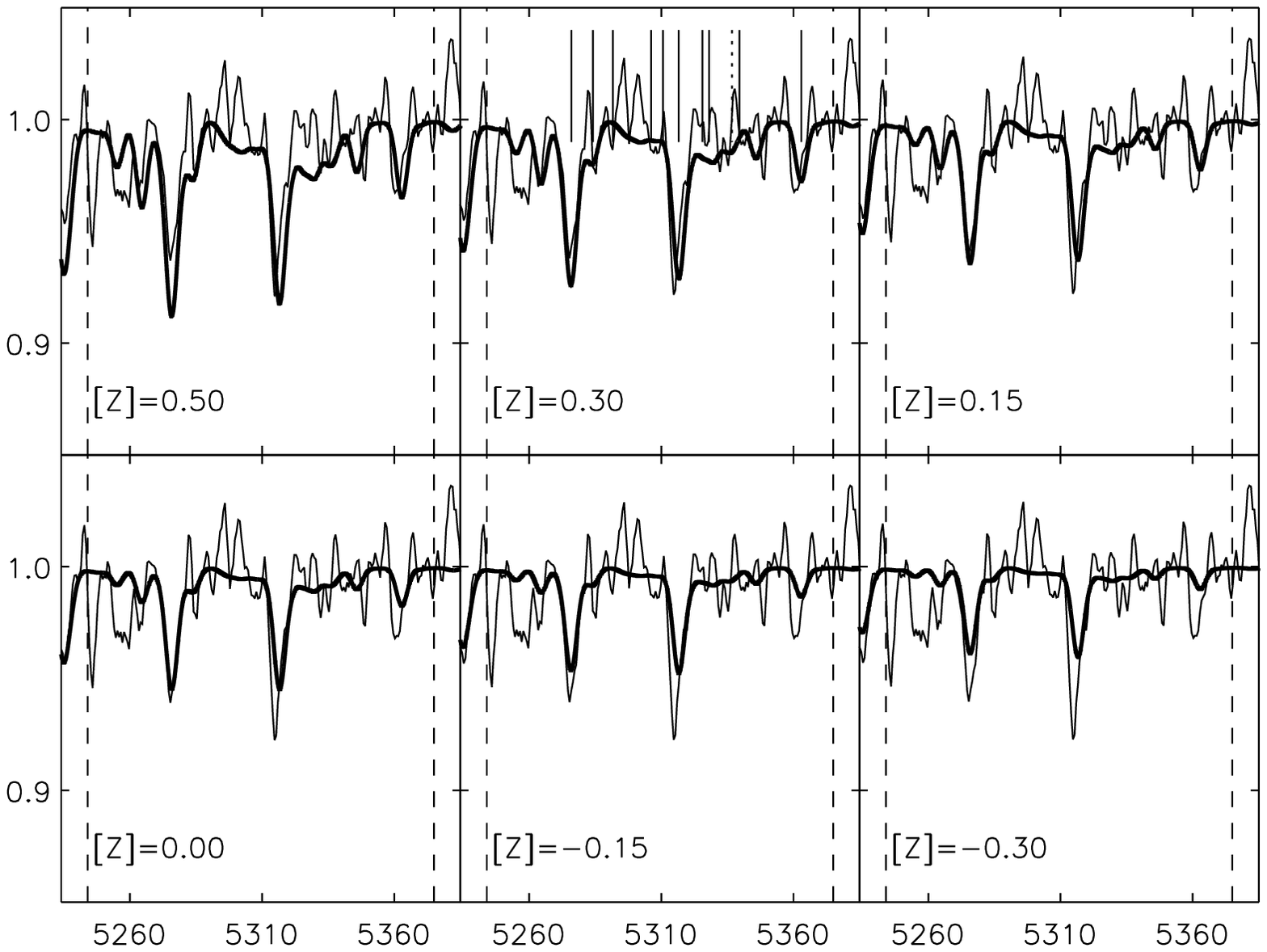}
   \includegraphics[scale=0.4]{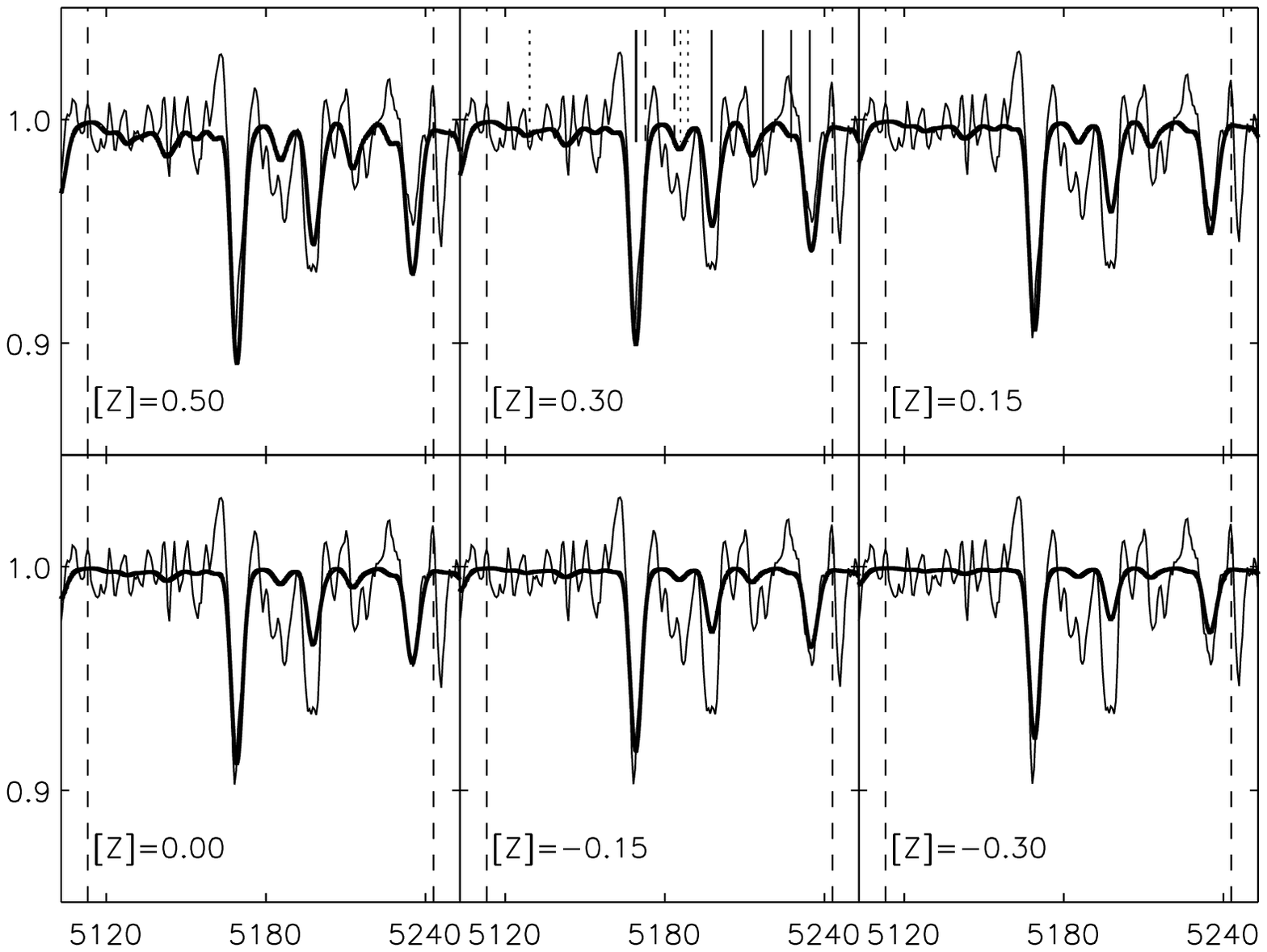}
 \includegraphics[scale=0.4]{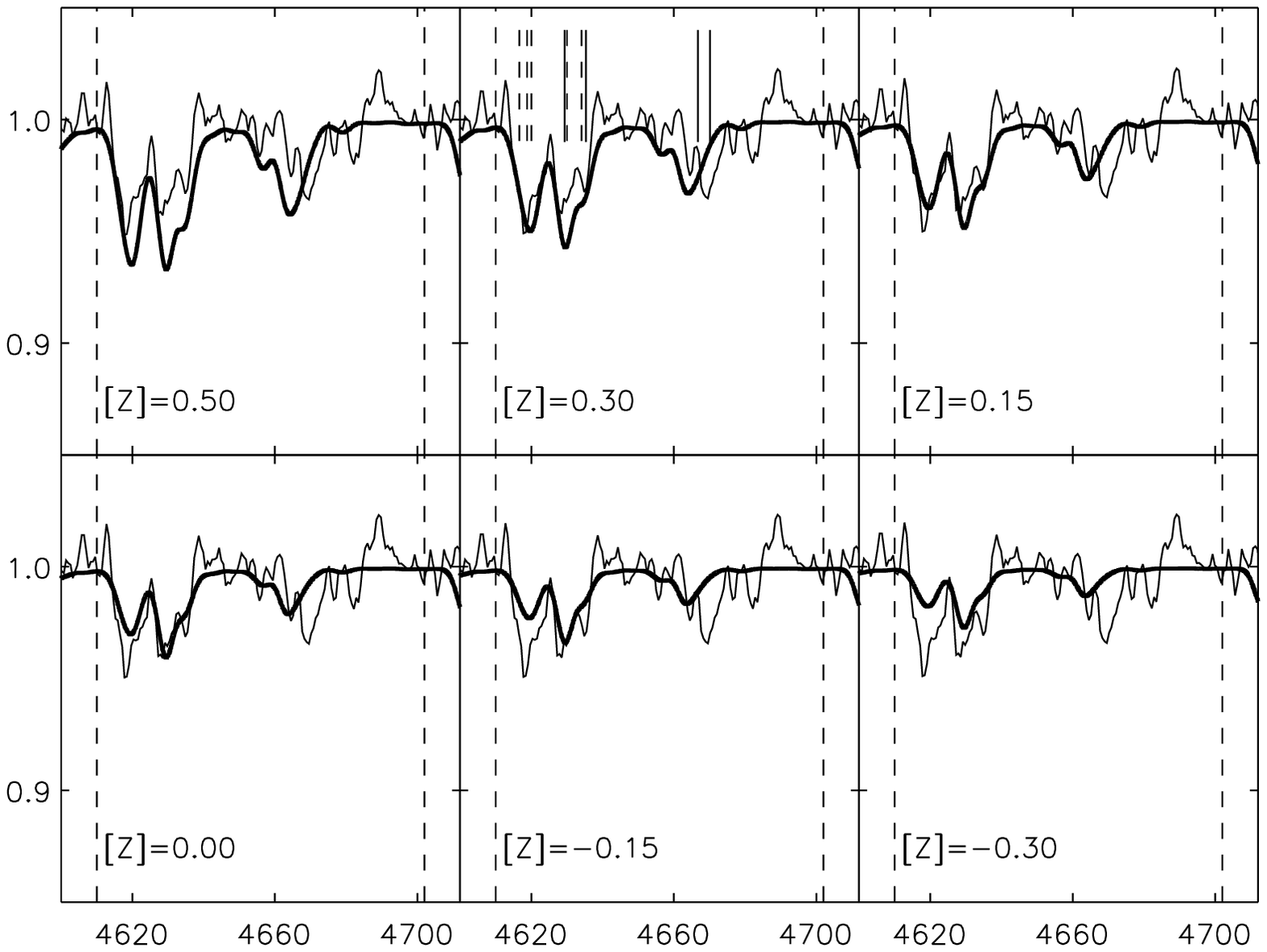} 
\includegraphics[scale=0.4]{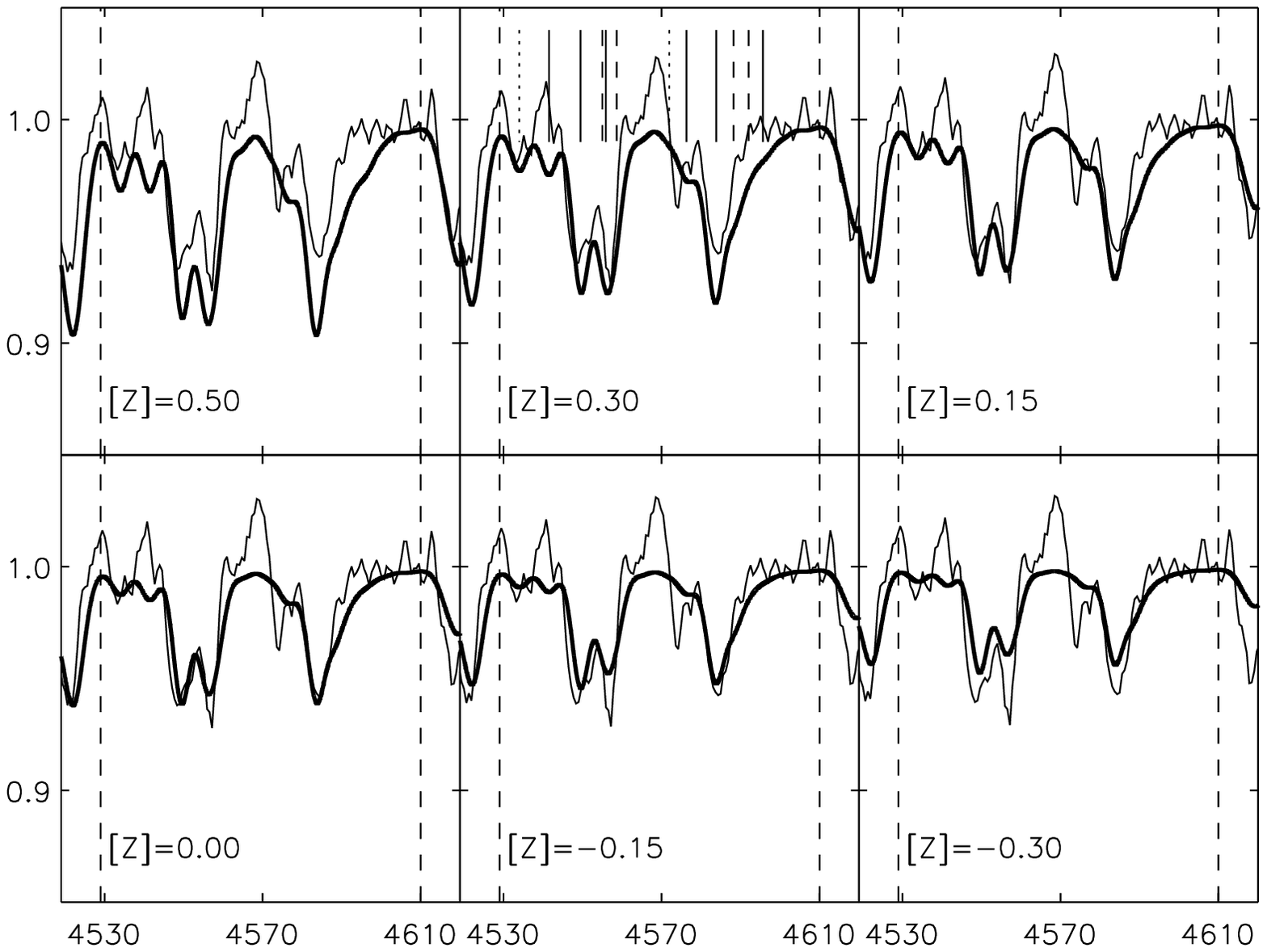} 
 \includegraphics[scale=0.4]{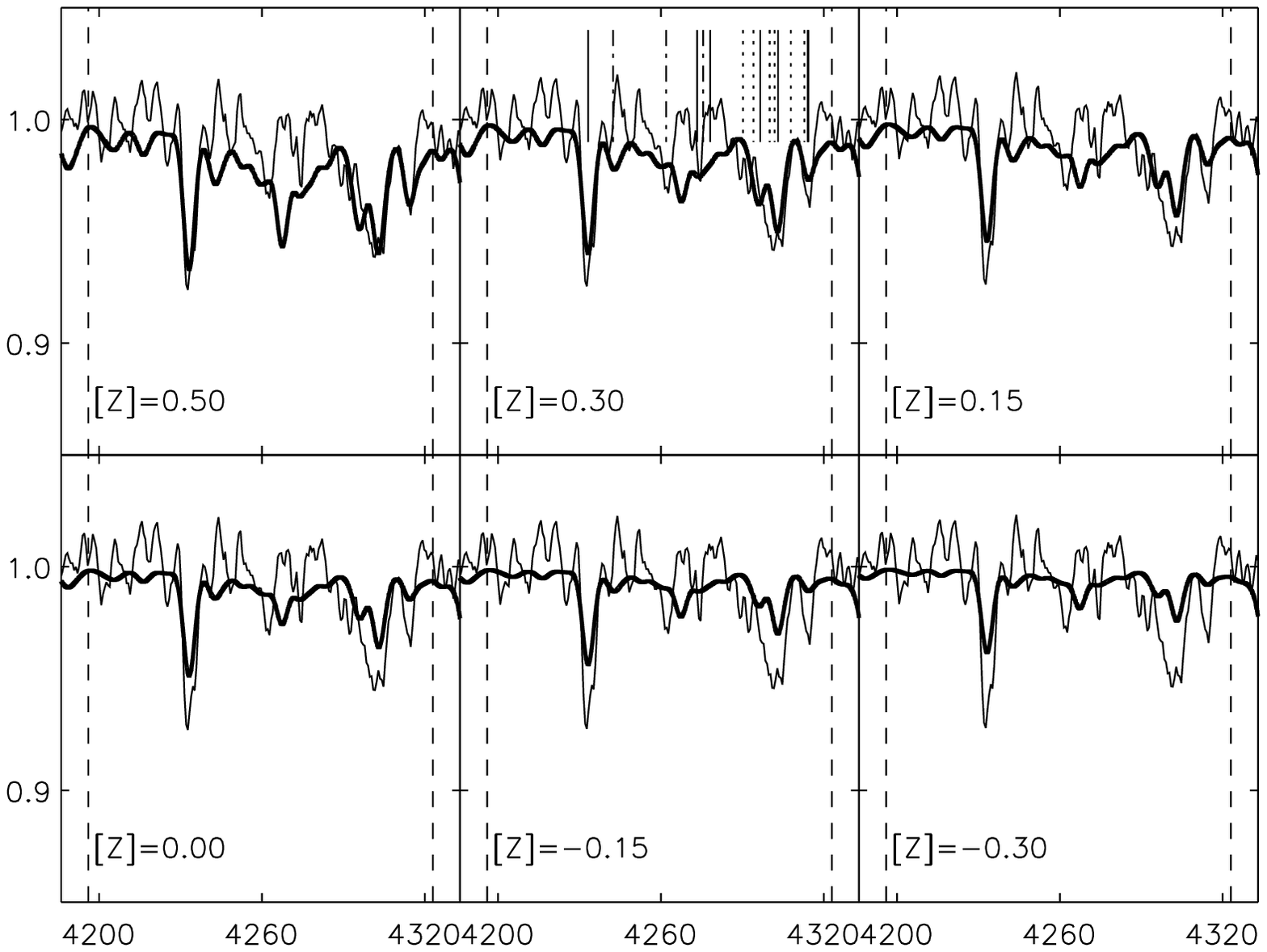} 
\includegraphics[scale=0.4]{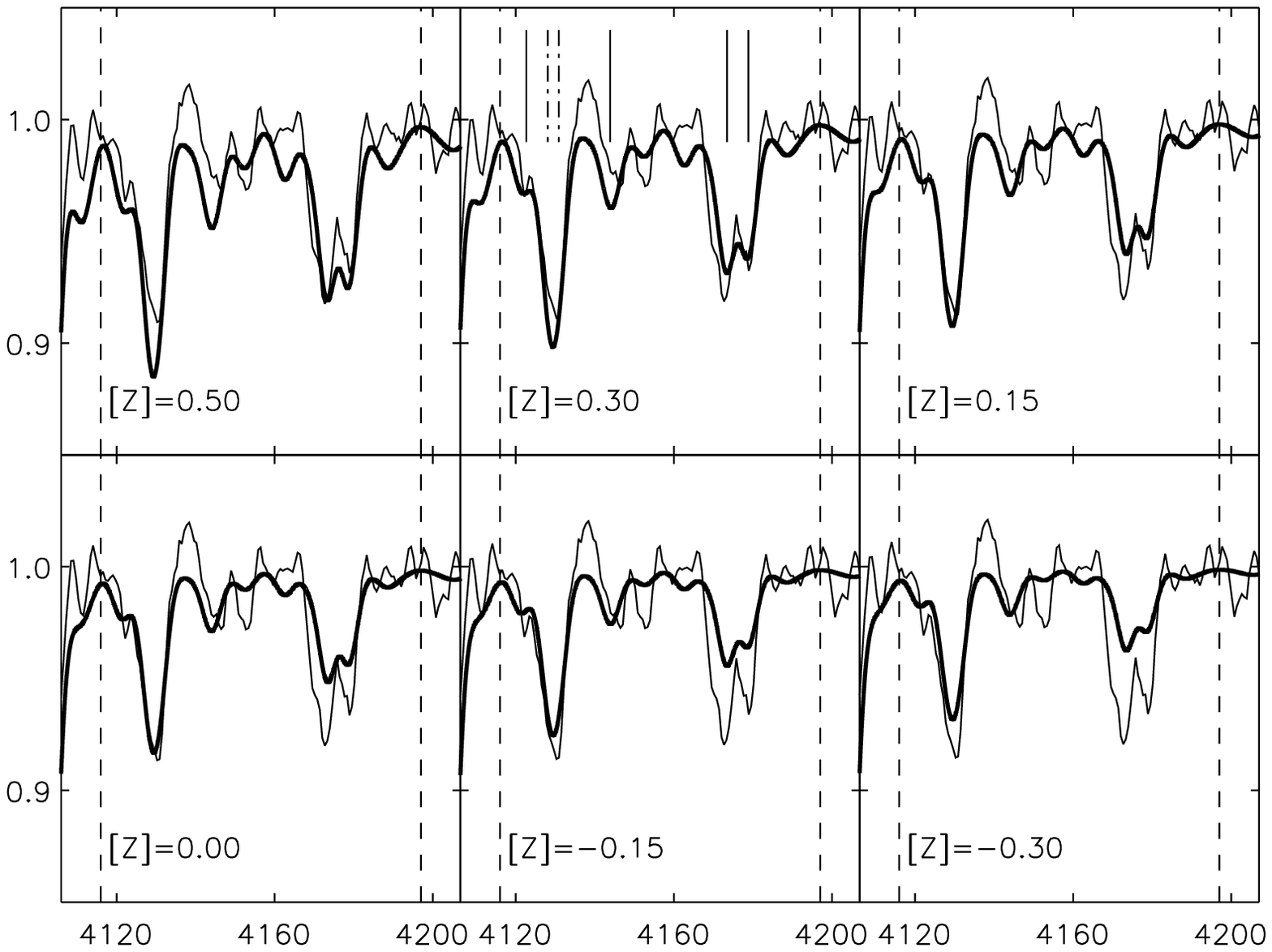}

\caption{Metallicity fits in 6 spectral windows of object C20. The synthetic spectra are 
plotted in bold and the metallicity is indicated at the left bottom of each plot. Normalized flux is plotted vs. wavelength in \AA.
Fe lines are indicated by solid bars, Cr: dashed, Ti: dotted, Si: dash-dotted, Mg: dash-triple dotted. Only 6 of the 14 available
metallicities are shown.
\label{c20metfit1}}
 \end{center}
\end{figure}

\begin{figure}
\begin{center}
  \includegraphics[scale=0.7,angle=90]{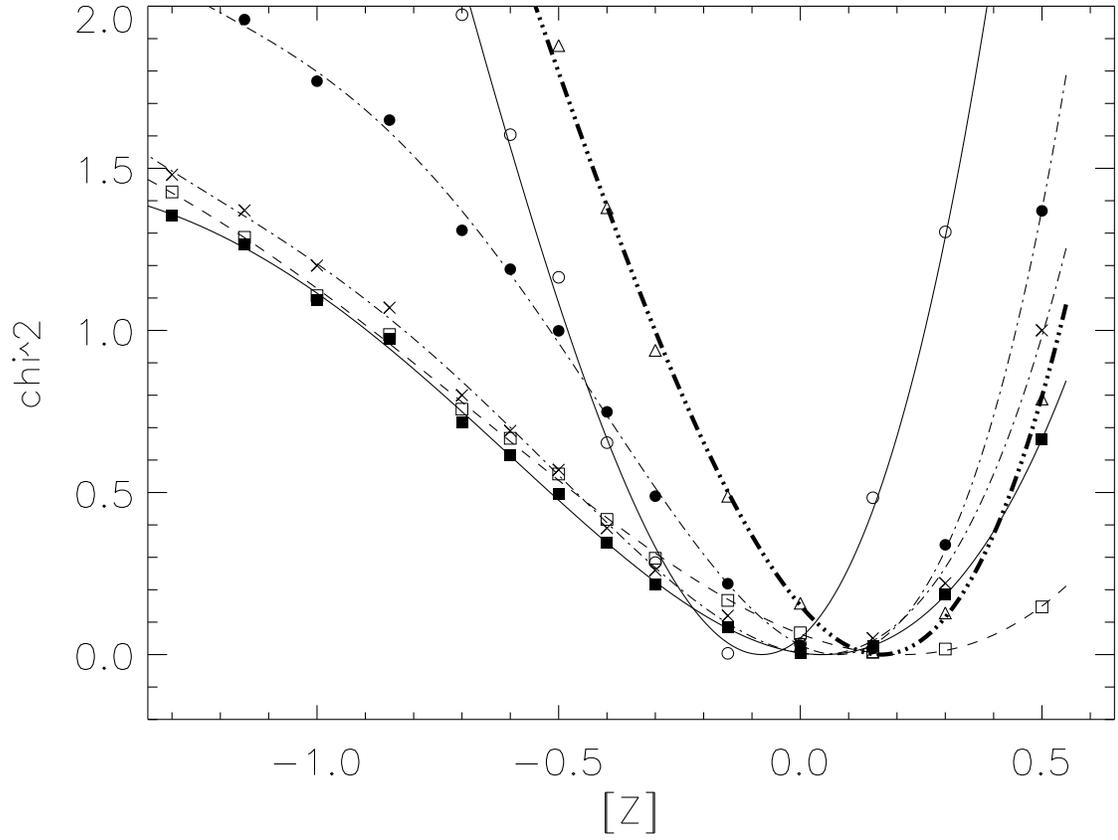}
\caption{$\chi^{2} ([Z])$ for each spectral window of object C20 as a function of metallicity [Z]. 
The curve for each window has a well defined minimum abcissa [Z]$_{i}$. The average of
all [Z]$_{i}$ is adopted as the stellar metallicity value.
\label{c20metfit2}}
 \end{center}
\end{figure}

In the next step, with effective temperature and gravity measured we can use our synthetic spectra to determine metallicity. 
For this purpose, we concentrate on the objects cooler than 17000K, since the S/N is not high enough for the hotter 
objects (see \citealt*{urbaneja05}). Three of the cooler objects cannot be used for this purpose, because their effective 
temperature is not constrained by a Balmer jump measurement but by the use of the spectral type already assuming solar abundance. 
In addition, two more objects (Z11, C16) have spectra too noisy for a metallicity fit. 
Object Z20 shows a metal line spectrum at longer wavelengths, which indicates a spectral type somewhat 
cooler (A0) than the temperature we obtain from the Balmer jump. There is a slight chance that this is a composite spectrum, thus, 
this target is also not used for the determination of metallicity (and also not for distance determination, see below).
While this reduces the number of targets suitable for a metallicity determination, 
it still leaves us with a sub-sample of 15 objects
large enough to constrain average metallicity and metallicity gradient of M81, as we will show below.
For the measurement of metallicity we apply the technique developed by \citet{kud08}. For each star we 
identify spectral windows in the observed spectrum, which are free of strong Balmer lines, 
nebular emission lines or spectral flaws caused by improper correction for cosmic ray hits and for which the continuum of the
normalized spectrum can be easily matched with the one of the synthetic spectra. A pixel-by-pixel comparison 
of observed and calculated normalized fluxes as a function of metallicity then allows for a 
calculation of $\chi ^{2}([Z])$ in each spectral window i and the determination of [Z]$_{i}$ at which $\chi ^{2}$ 
is minimal. For this comparison, the observed spectra are renormalized for each metallicity
so that the synthetic spectrum always intersects the observations at the same 
value at the edges of the spectral window (see also \citealt{kud08}).
An average of all [Z]$_{i}$ is then used as the measure of metallicity (for details, see again \citealt*{kud08}).
An example is given in Fig~\ref{c20metfit1} and ~\ref{c20metfit2} for target C20.

While the analysis method is straighforward and has been tested carefully in 
previous work, two obvious issues, unresolved binarity and blending with
fainter sources in the galaxy studied, need to be discussed as possible
sources of systematic uncertainties. Unresolved binarity can affect the 
analysis in two ways, first, through the contribution of a secondary to the 
photometric fluxes and the spectrum and, second, through the effects of close 
binary evolution with mass transfer or mass loss. In the first case, it is 
very unlikely that both components have a very similar spectral type and 
luminosity because of the very short lifetime in the supergiant stage. The 
most likely case is a secondary of lower mass still on the main sequence. 
However, such an object would be much fainter by several magnitudes and not 
affect the spectroscopic analysis or the photometry. The second case is more 
serious, but would affect only the FGLR-distance determination. Binary induced 
mass transfer or mass-loss would change the stellar mass at a given luminosity 
and create outliers from the FGLR-relationship. Such outliers have been found 
by \citet{kud08}, and \citet{u09}. They are usually also outliers, 
when the mass-luminosity relationships of the targets are 
plotted. We will investigate this latter relationship in section 5 (Fig.14).

Blending does not appear to be a problem because of the enormous optical 
brightness of the supergiants as already discussed in \citet{kud08}. The 
study by \citet{bresolin05} shows that at the distance of NGC 300 at 2 Mpc 
even ground-based photometry of blue supergiants is accurate and not 
affected by blending. Thus, at 3.5 Mpc for M81 with HST imaging
and with our careful selection of targets 
(see section 2) we do not expect blending effects influencing the photometry 
and, therefore, also not the spectroscopy. Of course, in individual cases there
is always the very small chance of an unresolved coincidence of a target with
another bright source. In such cases, the likelyhood that the unresolved blends have
the same spectral type is extremely small, again because of the short 
lifetime of blue supergiants. Thus, significant blends should be recognized in 
the spectrum. Target Z20 might be such a case. We also refer the reader to the 
careful modeling of blending effects in the HST imaging of Cepheids out to galaxies with 30 Mpc
distances \citep{riess09b, riess11} resulting in magnitude corrections of the order of only
0.1 mag. Cepheids are 3 to 6 mag fainter than blue supergiants. Thus, since 
Cepheids are only very weakly affected by blending, we do not expect significant effects for supergiants.

The results of the spectroscopic analysis are summarized in Table 2. Generally, the stellar parameters and their 
uncertainties are comparable to those obtained in our previous work for galaxies less distant (see \citealt*{kud08,urbaneja08}). 
We conclude that for this type of low resolution quantitative spectroscopy the step from one Mpc (WLM), over 2 Mpc 
(NGC 300) to now 4 Mpc is entirely feasible. In the following, we discuss the results in detail.

\section{Reddening and Extinction}

\begin{figure}
\begin{center}
  \includegraphics[scale=0.3,angle=90]{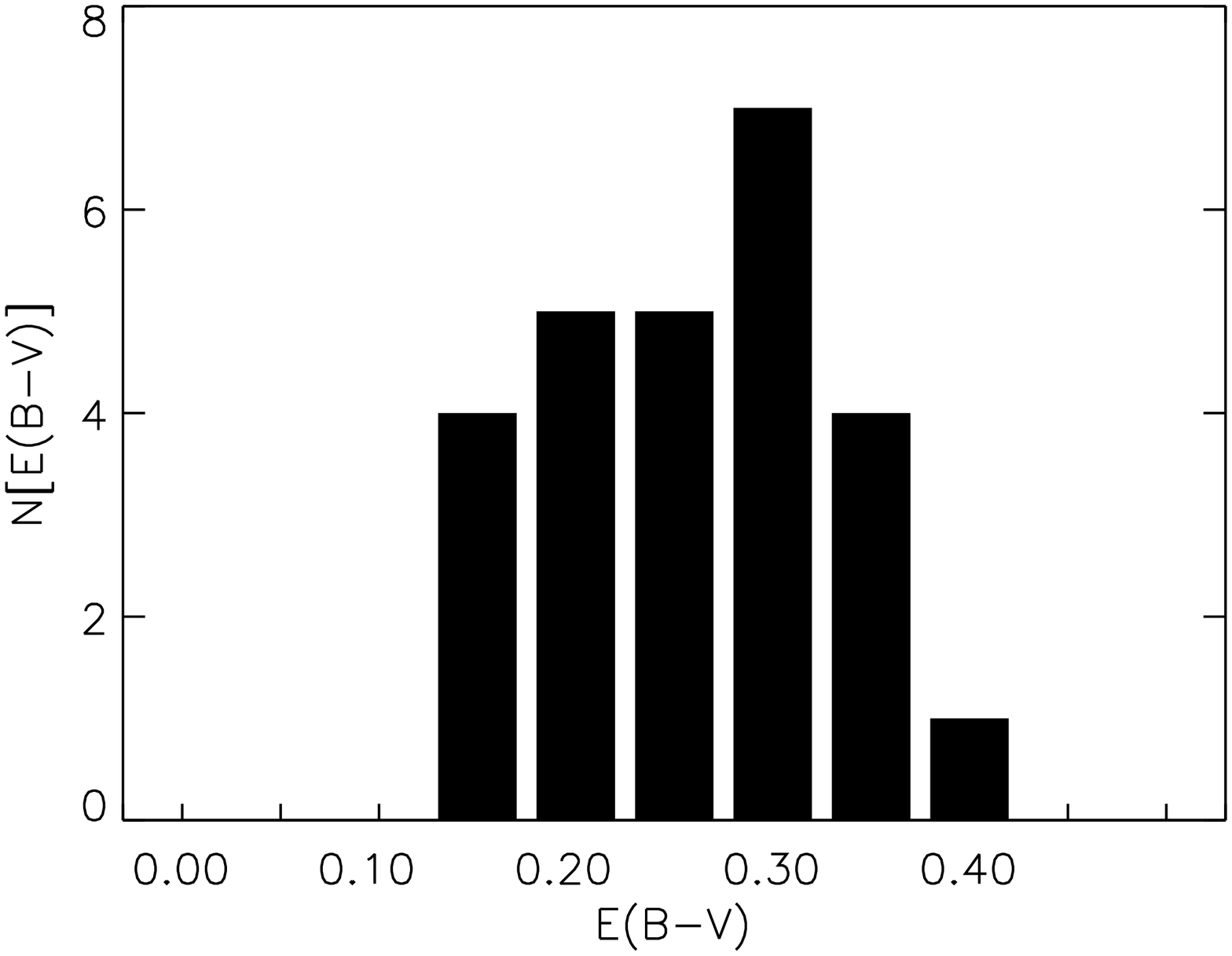}
 \includegraphics[scale=0.3,angle=90]{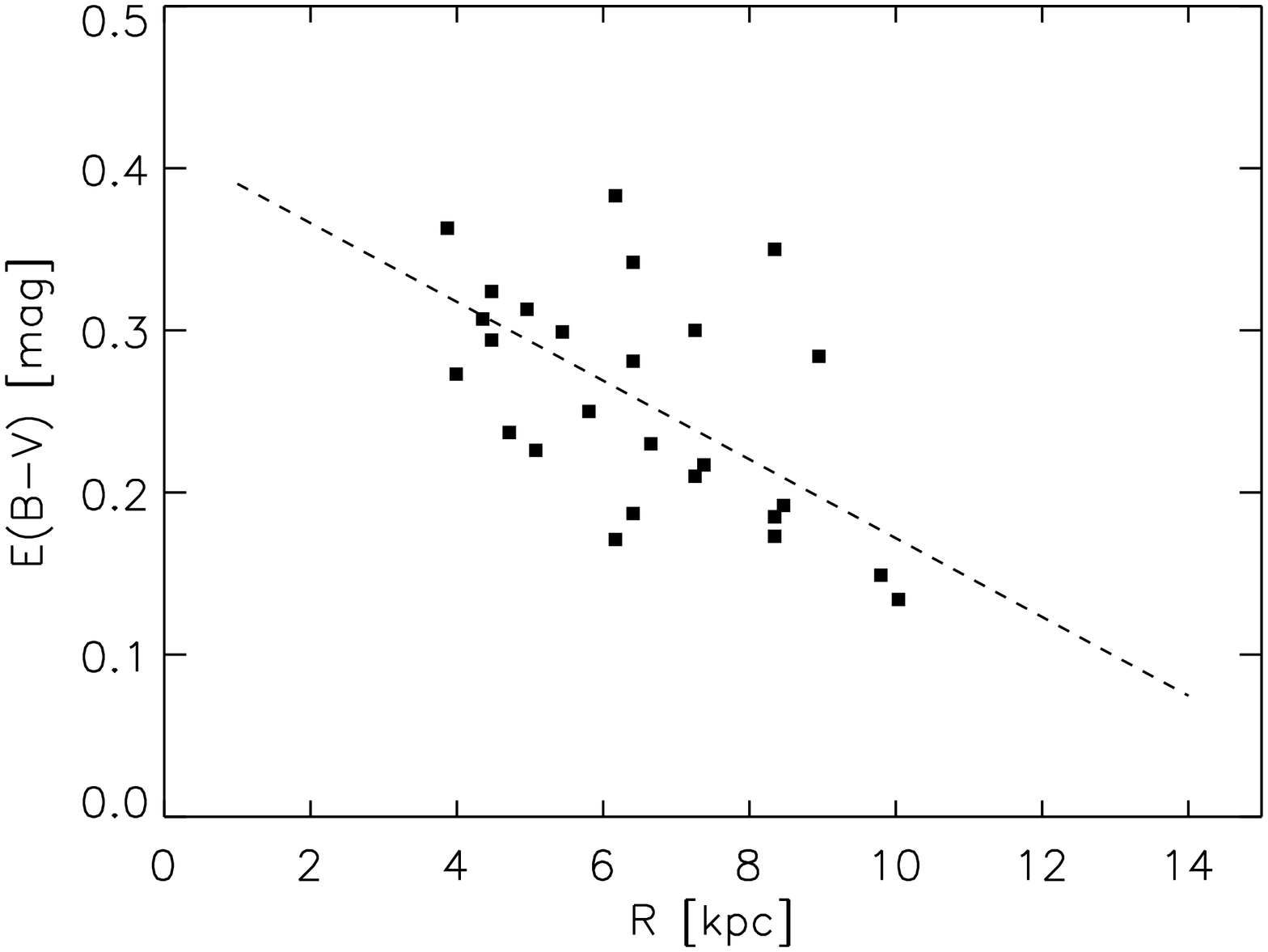}
\caption{Interstellar reddening in M81. Left: Histogram of the E(B-V) distribution. Right: E(B-V) as a function of 
galactocentric distance with the regression curve (dashed) discussed in the text..
\label{ebv}}
 \end{center}
\end{figure}

As described above, one of the advantages of the spectroscopic analysis is that it provides information 
about interstellar reddening. For massive stars imbedded into the dusty disk of a star-forming spiral galaxy 
we expect a wide range of interstellar reddening. Indeed, we find a range from E(B-V) = 0.13 to 0.38 mag. 
Fig~\ref{ebv} shows the distribution of interstellar reddening among our targets. The average value is 
E(B-V)$_{av}$ = 0.26 mag. The foreground reddening is 0.08 mag \citep{schlegel98}.
Our reddening values include both, intrinisc and foreground reddening. 
We stress that our average value of E(B-V) may underestimate the average 
reddening in M81, as our target selection (see Fig. 1) is biased towards 
lower reddening.

Fig~\ref{ebv} shows reddening as a function of galactocentric distance. While the scatter is large, 
it is still tempting to fit a regession to the data. We find

\begin{equation}
        E(B-V) = \left(0.415\pm{0.025}\right) - \left(0.0243\pm{0.0037}\right) \times d/kpc
\end{equation}
 
The lower reddening beyond 10 kpc indicated by this regression is in agreement with the results found 
by \citet{williams09}, who investigated star formation history and metallicity with HST color-magnitude 
diagrams in the outer fields of M81 and found E(B-V) = 0.14 mag at 14 kpc galactocentric distance.

We note that the reddening values found in our study are much larger than the value of 0.03 mag originally 
assumed in the HST distance scale key project \citep{freedman94} for Cepheids at inner fields between 
3 to 6 kpc galactocentric distance. The final key project study \citep{freedman01} obtained an average 
value of E(B-V) = 0.15 mag, still significantly smaller than our value, in particular in view of the fact 
that a difference of 0.1 mag in reddening results in a difference of 0.3 mag in distance modulus if 
the ratio of total to selective extinction is R$_{V}$ = 3.1.  

\section{Stellar Properties and Evolution}

\begin{figure}
\begin{center}
  \includegraphics[scale=0.3,angle=90]{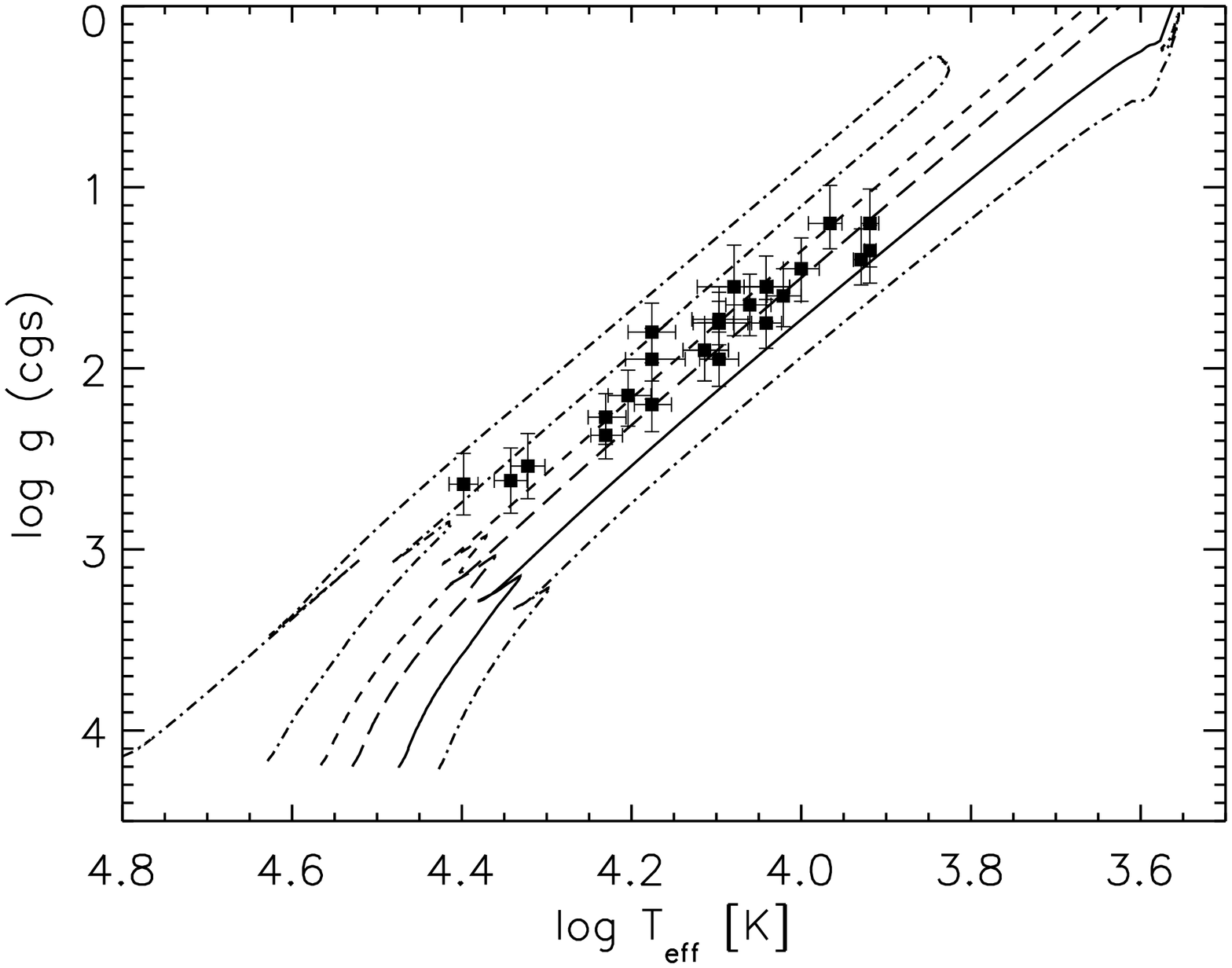}
 \includegraphics[scale=0.3,angle=90]{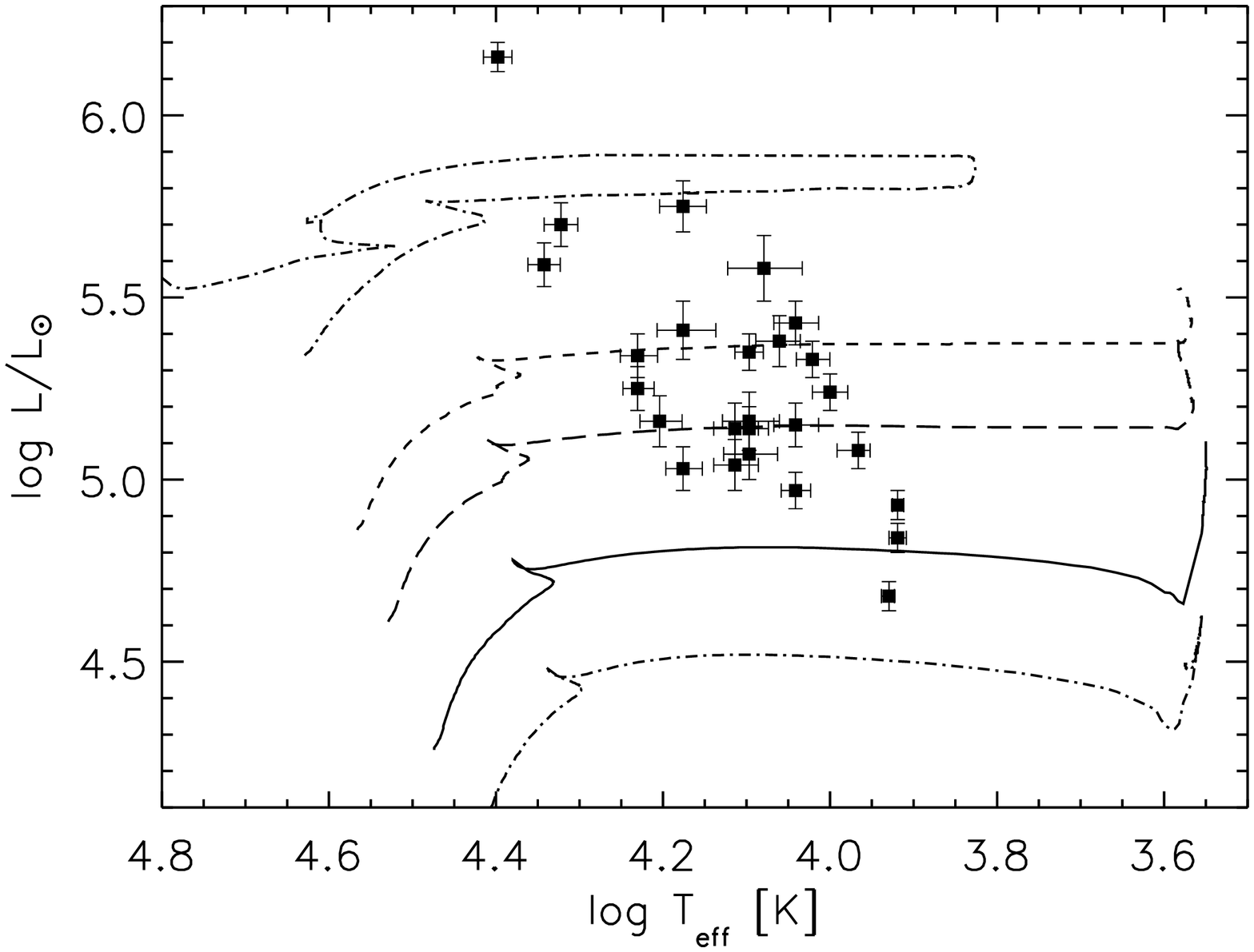}
\caption{Stellar parameters of the observed sample of M81 supergiants compared with evolutionary tracks 
for the Milky Way metallicity including the effects of rotational mixing \citep{meynet03}. Left: (log g, log T$_{\rm eff}$) - diagram.
Right: Hertzsprung-Russel diagram. The zero-age main sequence masses are (in increasing luminosity/decreasing gravity) 12, 15, 20, 25, 40 solar masses, respectively.
\label{hrd}}
 \end{center}
\end{figure}

Fig.~\ref{hrd} (left panel) shows the location of all targets in the (log g, log T$_{eff}$)-plane compared 
with evolutionary tracks \citep{meynet03}, which were calculated for solar metallicity and which include 
the effects of rotational mixing and anisotropic mass-loss. The advantage of a diagram of this type is that it
is independent of any assumption on distance and relies completely on the results of the spectroscopic 
analysis (on the other hand, systematic effects in the evolutionary tracks might affect
the comparison). The targets form an evolutionary sequence crossing from the main sequence towards the red supergiant 
stage with initial zero age main sequence (ZAMS) masses between 15 to 50 M$_{\odot}$ and the majority of 
objects with ZAMS masses about 20 to 25 M$_{\odot}$.

A complementary way to discuss stellar evolution and stellar properties is the Hertzprung-Russell diagram (HRD). This requires 
information about the distance. In section 8 we will use the FGLR to determine a distance modulus of 
$\mu$ = 27.7$\pm{0.1}$. With this distance and using the spectroscopically determined reddening and 
extinction and the bolometric corrections provided by the model atmospheres for the final parameters 
of temperature, gravity and metallicity we can determine absolute bolometric magnitudes, luminosities 
and stellar radii. In the calculation of stellar radii from luminosities we take into account
that the errors in luminosity are dominated by the errors in effective 
temperature and are, thus, correlated (maximum luminosity corresponds to 
maximum temperature and, thus, minimum radius, wheras minimum luminosity at minimum
temperature yields maximum radius). The results are given in Table 3 and the resulting HRD is 
shown in Fig.~\ref{hrd} (right panel).

The HRD confirms that the majority of targets is in the ZAMS mass-range of about 20 to 25 M$_{\odot}$ and is 
generally consistent with the (log g, log T$_{eff}$)-diagram. However, one object (Z15) sticks out as 
very luminous. We recall that the spectroscopic analysis of this object was difficult because of extremely 
strong contamination with nebular \hii~emission, which might affect the determination of gravity in a 
systematic way which is difficult to assess. In consequence, we have not included this object in the 
FGLR determination of the distance.

\begin{figure}
\begin{center}
  \includegraphics[scale=0.3,angle=90]{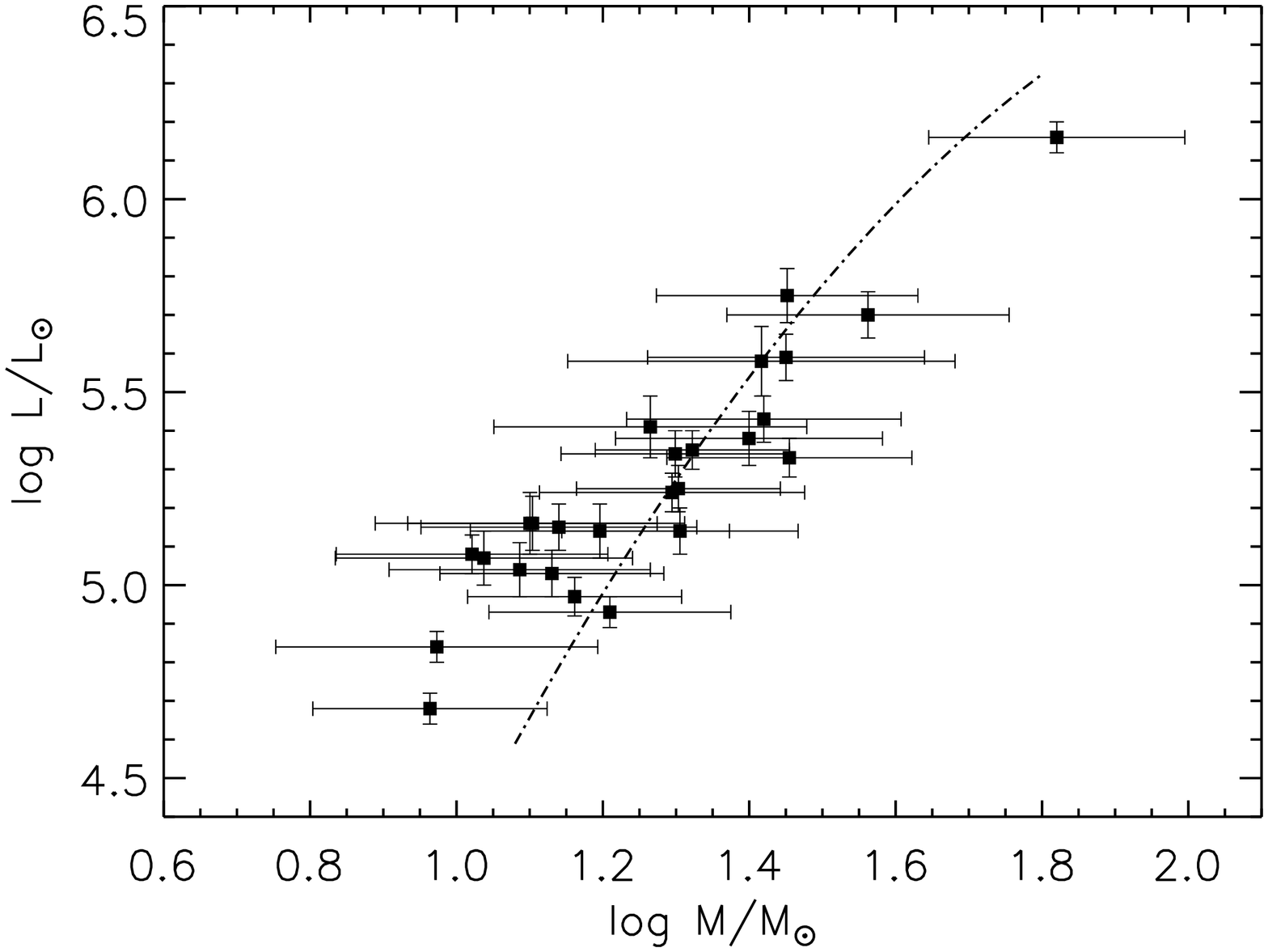}
 \includegraphics[scale=0.3,angle=90]{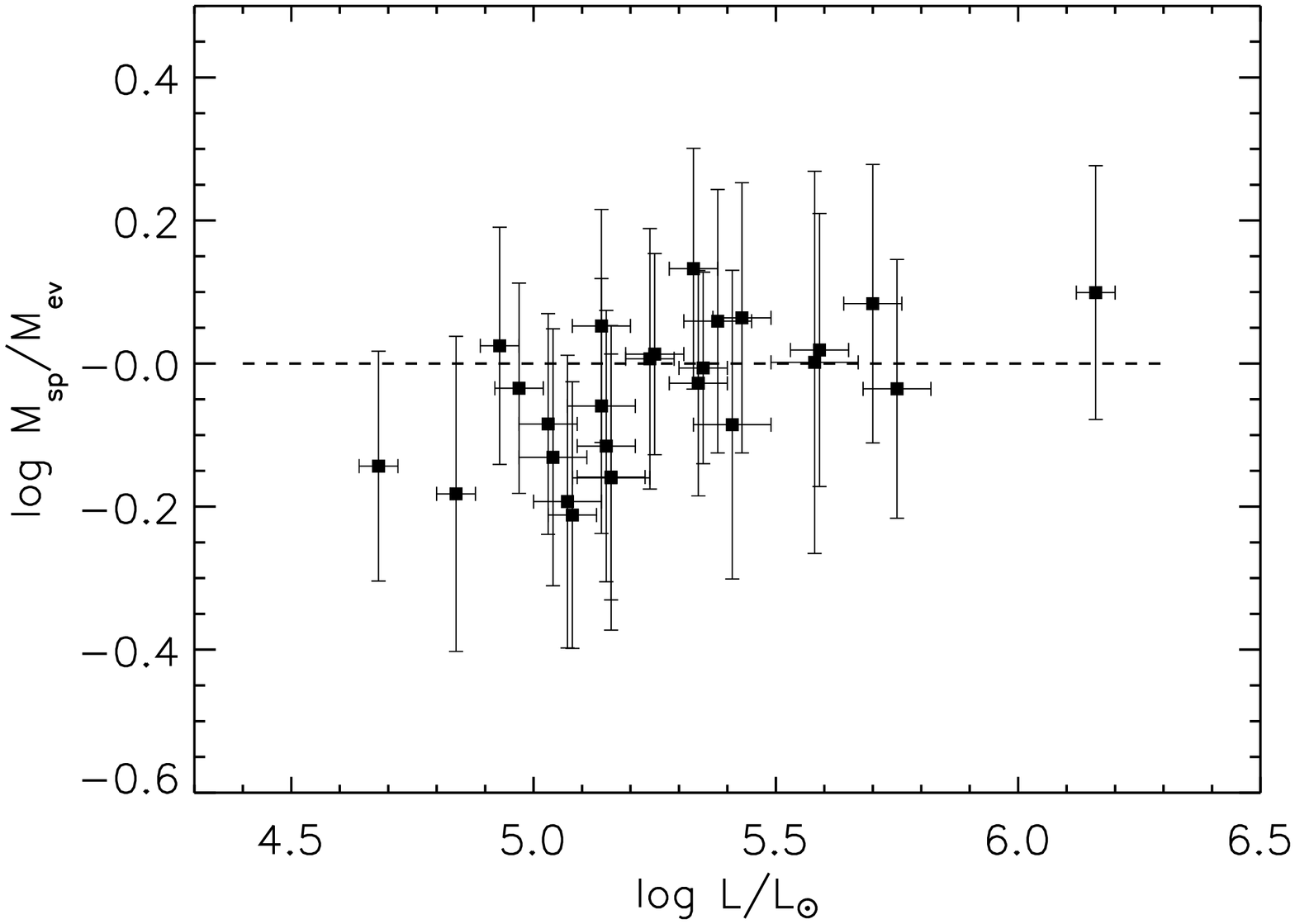}
\caption{Left: Observed mass-luminosity relationship compared with stellar evolution theory using the tracks from Fig.~\ref{hrd} at an effective temperature of 10$^{4}$K. 
Right: Logarithmic ratio of spectroscopic to evolutionary masses as a function of luminosity.
\label{mass}}
 \end{center}
\end{figure}

With the stellar radii determined from the luminosities we can use the gravities to estimate spectroscopic 
stellar masses. Those are also given in Table 3. An alternative way to estimate masses is to use stellar 
luminosities and to compare with the luminosities and actual masses at the BSG temperatures 
of evolutionary tracks. Evolutionary masses are also given in Table 3. They are determined from the BSG 
mass-metallicity relationship given by \citet{kud08} (for Milky Way metallicity and
including the effects of rotational mixing).
We emphasize that both spectroscopic and evolutionary masses are present-day masses and are generally expected to be lower than
the initial ZAMS masses through the effects of mass-loss.
Since the early work by \citet{herrero92} it has been found that spectroscopic masses are often 
significantly smaller than evolutionary masses, although with the development of fully line-blanketed 
model atmospheres and improved NLTE line formation the effect has become much smaller (see \citealt*{kud09} 
for a review, and references therein). In Fig.~\ref{mass} we check our sample for this effect by 
comparing the observed spectroscopic mass-luminosity relationship with the prediction of stellar evolution 
and by directly plotting the ratio of spectroscopic to evolutionary mass as a function of luminosity. We 
find a small effect only at the lower mass end, were spectroscopic masses appear to be somewhat smaller 
than evolutionary masses. However, we conclude that our sample is not significantly different 
from the one studied by \citet{kud08} in NGC 300 and \citet{u09} in M33.

\section{Metallicity, Metallicity Gradient and Chemical Evolution}

\begin{figure}
\begin{center}
  \includegraphics[scale=0.3,angle=90]{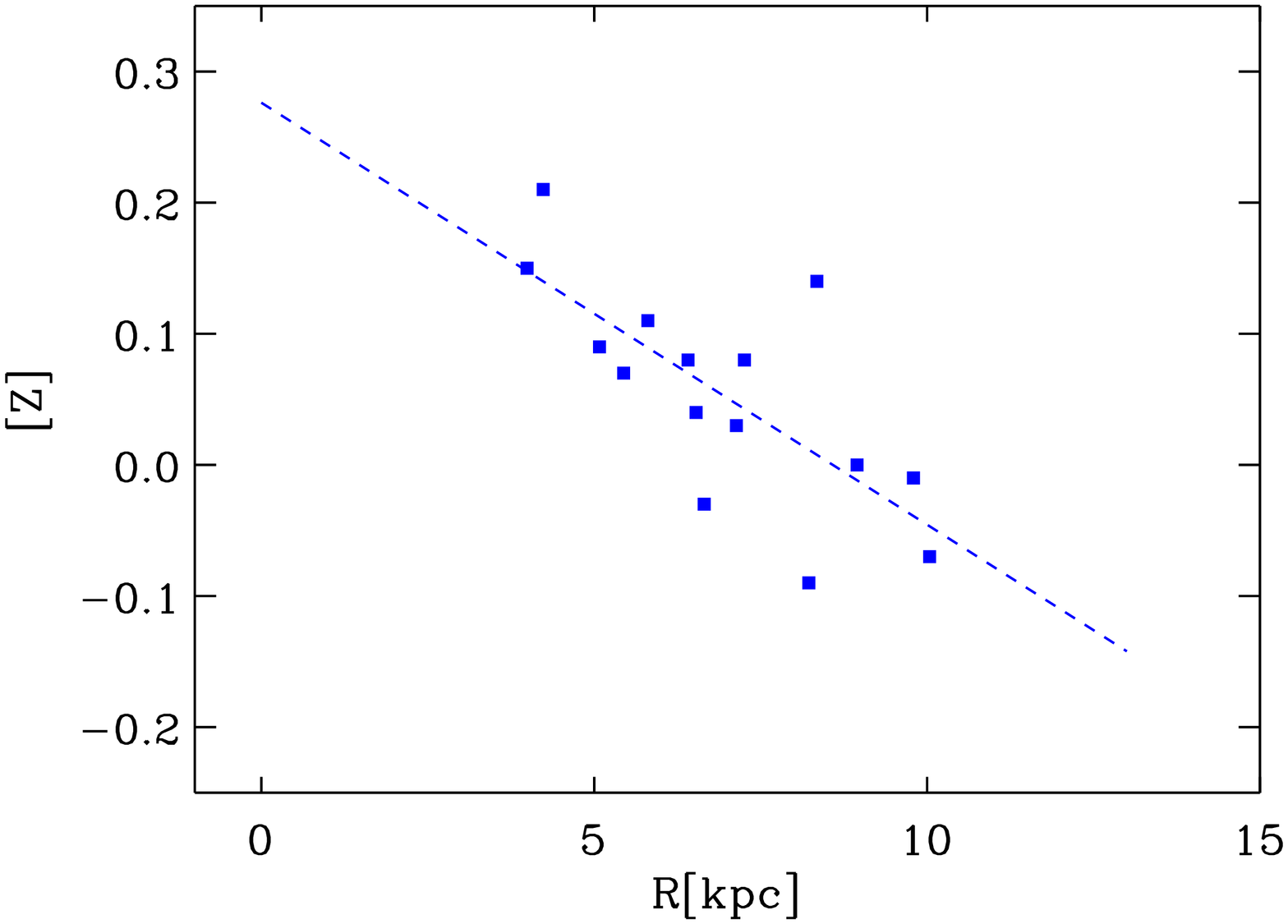}
 \includegraphics[scale=0.3,angle=90]{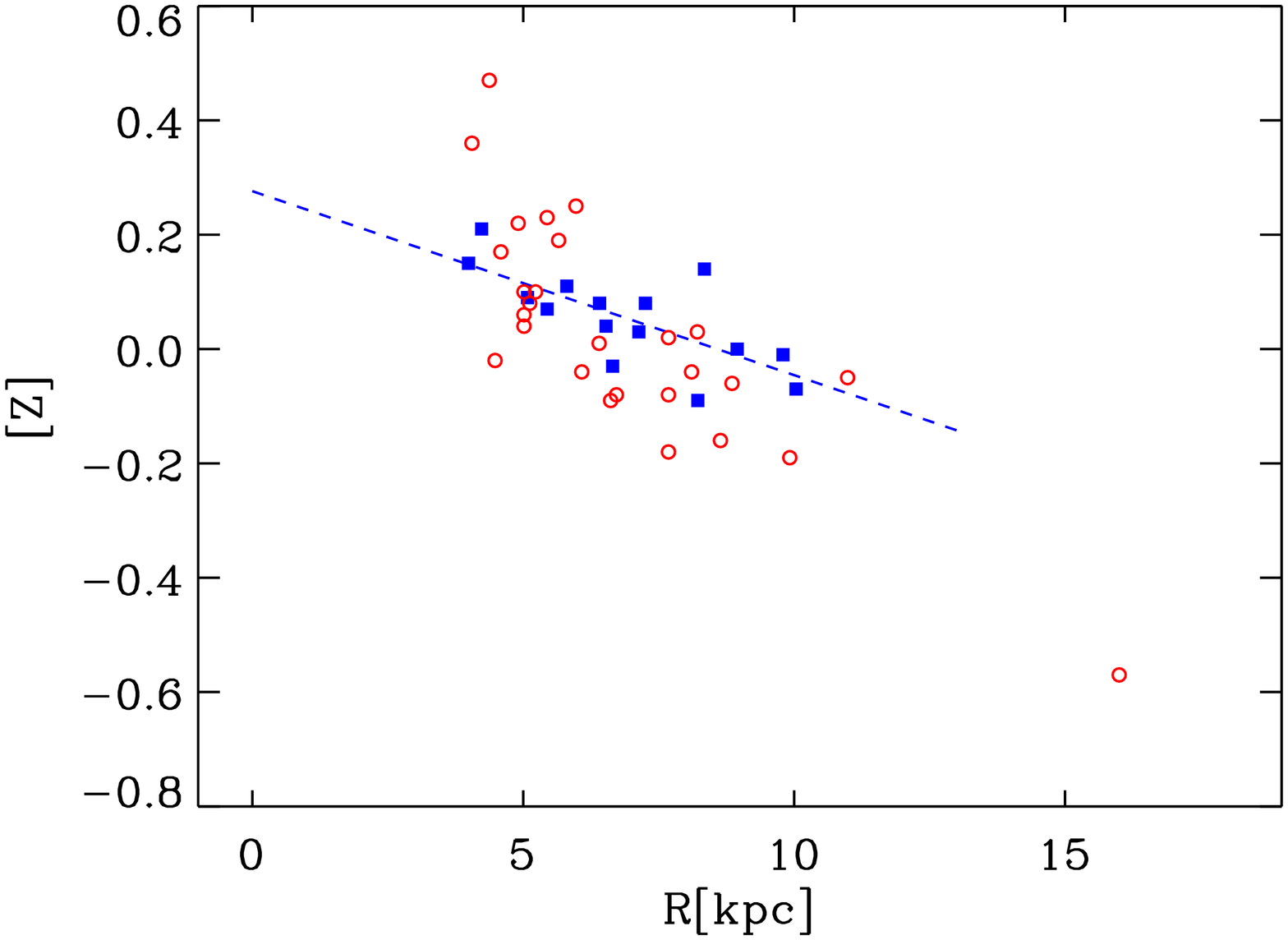}
\includegraphics[scale=0.3,angle=90]{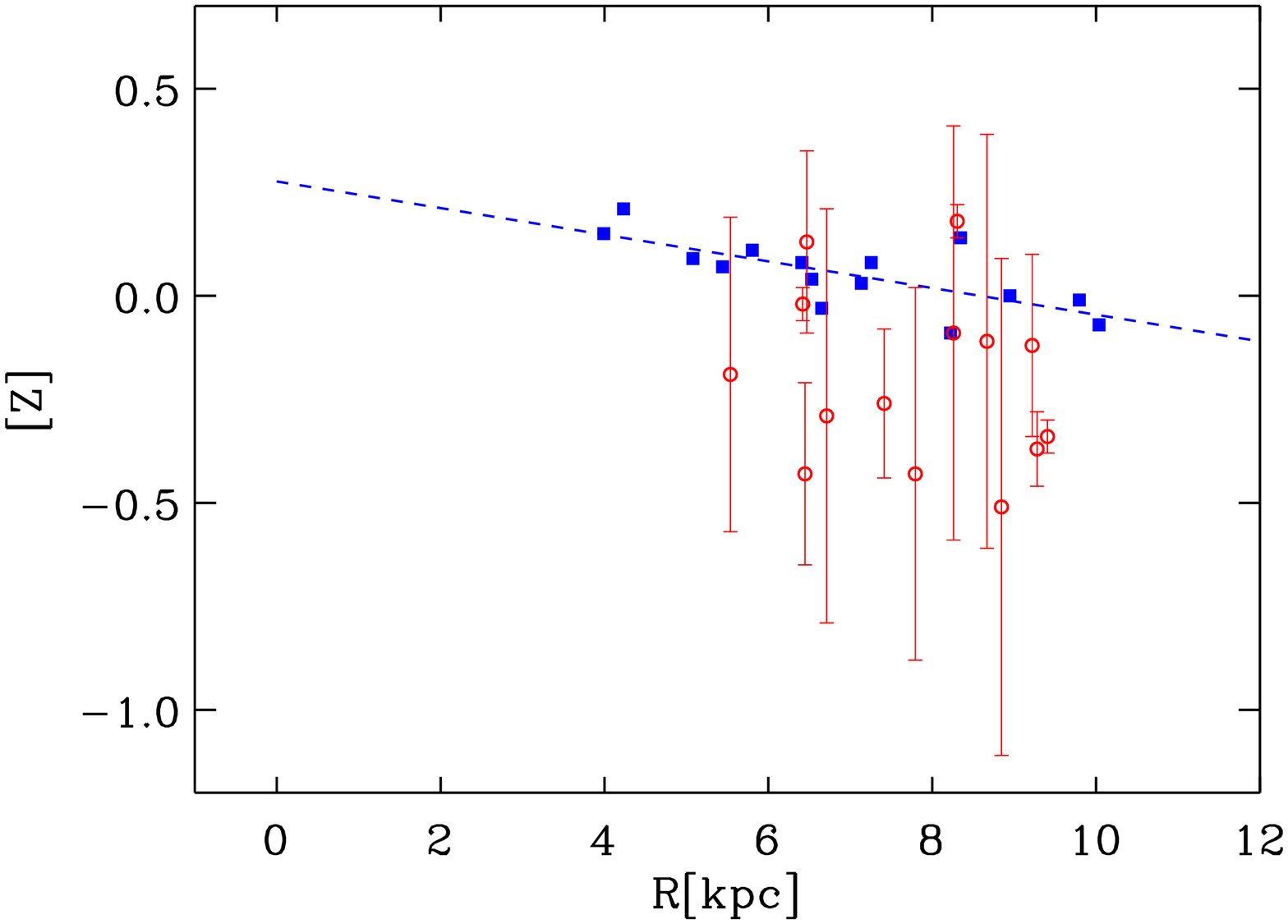}
 \includegraphics[scale=0.3,angle=90]{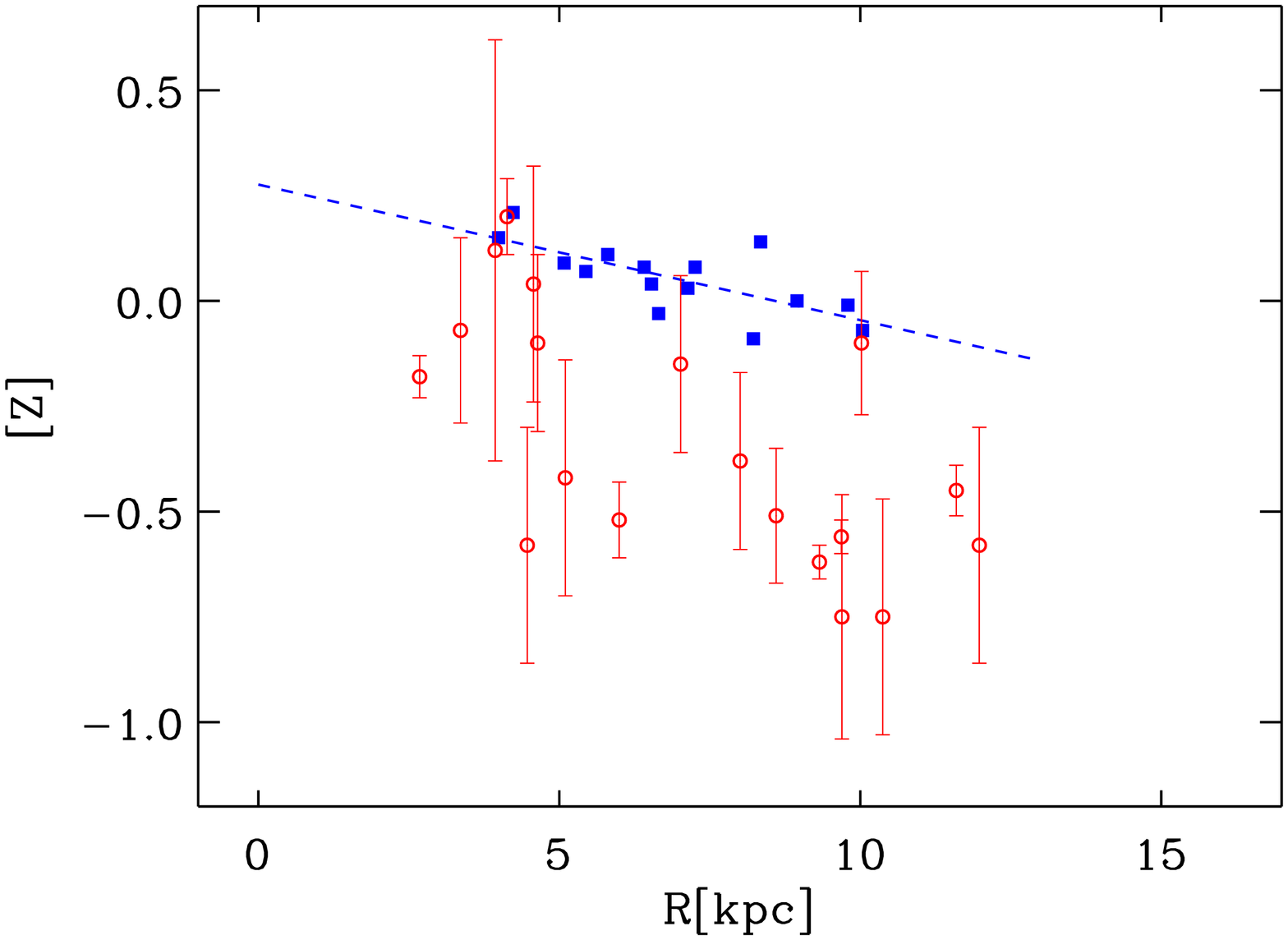}
\caption{Upper left: Metallicity of BSGs in M81 as function of galactocentric distance in kpc. The dashed curve 
is the regression discussed in the text. Uncertainties are given in Table 2 and not plotted. 
Upper right: Same as upper left with the the oxygen abundances of \hii~regions from the strong line 
studies by \citet{garnett87} and \citet{stauffer84} overplotted. Random uncertainties of the 
\hii~region data are between 0,1 to 0.2 dex, systematic uncertainties are discussed in the text. 
Lower left: Same as upper left but with the \hii~region oxygen abundances by \citet{stanghellini10} 
overplotted. Lower right: Same as upper left, but oxygen 
abundances of PNe obtained by \citet{stanghellini10} overploted with error bars. The dashed line in all four panels is the BSG regression obtained in this work.
For a detailed discussion, see text.
\label{zgrad}}
 \end{center}
\end{figure}

Metallicities of 15 targets together with their galactocentric distance are given in Table 2. This allows 
us to discuss stellar metallicity and the metallicity gradient in M81. Fig.~\ref{zgrad} (upper left panel) shows a plot of 
logarithmic metallicity relative to the sun [Z] as a function
of galactocentric distance (at the distance of 3.47 Mpc - see section 8 - R$_{25}$ = 11.99 arcmin 
corresponds to 12.09 kpc). A metallicity gradient of the young disk population in M81 is clearly visible. 
A linear regression (using the routine fitexy, Numerical Recipes, \citealt*{press92}) yields

\begin{equation}
\left[Z\right] = \left(0.286\pm{0.061}\right) - \left(0.033\pm{0.009}\right) R/kpc
\end{equation}

With respect to the distance independent normalized angular galactocentric distance R/R$_{25}$ we obtain

\begin{equation}
\left[Z\right] = \left(0.286\pm{0.061}\right) - \left(0.411\pm{0.109}\right) R/R_{25}
\end{equation}

As is evident from the plot and the regression, young massive stars in the disk of M81 have slightly super-solar metallicities 
at the inner regions and slightly sub-solar metallicity in outer parts. The gradient is very 
shallow, though, compared to the less massive galaxies studied in our BSG project. For NGC 300 and M33 metallicity  
gradients were determined of 0.08  and 0.07 dex \kpc, respectively by \citet{kud08} and \citet{u09}. 
On the other hand, for the Milky Way, which has a mass comparable to M81, \citet{daflon04} in their spectroscopy 
of massive stars obtain a gradient of -0.031$\pm{0.012}$ dex \kpc~very similar to our result. (We note, however, 
the results by \citet{rolleston00} for B-stars and \citet{luck06,luck11} for Cepheids, who obtained 0.07 
dex \kpc~and 0.055 dex \kpc, respectively).

\citet{garnett87} and \citet{stauffer84} have analyzed \hii~region emission line spectra of M81 to derive 
oxygen abundances as a function of galactocentric radius. They used a strong line method following 
the calibration by \citet{pagel79} (\citealt*{garnett87} also used photo-ionization models for an independent 
check of the abundances obtained). In Fig.~\ref{zgrad} (right upper panel) we overplot these results with 
the galactocentric distances corrected to the distance used in our work. In the range of 5 kpc to 11 kpc there 
is a large number of objects in a similar abundance range as the BSGs with a slight off-set of -0.1 dex. However, 
at 5 kpc and below there are several objects with very high oxygen abundance. This result might be an artefact 
of the strong-line calibration used. These inner data points together with the \hii~region Muench 1 at 16 kpc 
(carefully discussed in \citealt*{garnett87}) lead to an oxygen abundance gradient of -0.064$\pm{0.020}$  dex \kpc~with 
a significantly higher value of [O] = 0.46$\pm{0.14}$ dex at the center, where [O] is defined in the same way
as [Z], namely [O] $\equiv\,$ log (O/O$_{\odot}$) = [O/H] - [O/H]$_{\odot}$ with [O/H] = 12 + log (O/H)  
and [O/H]$_\odot$ = 8.69 dex \citep{allende01}.

The \hii~regions of M81 have also been included in the work by \citet{zaritsky94} who developed a different 
strong-line calibration method. Their central metallicity is even higher, [O] = 0.51$\pm$0.11 dex, and the gradient is
0.042$\pm{0.015}$ dex \kpc~somewhat higher than the result of our BSG work. As has already been shown by 
\citet{bresolin09} and \citet{bresolin11}, this calibration leads to metallicities, which are too high when 
compared with \hii~region oxygen abundances based on electron temperature determinations with auroral lines 
(see also \citealt*{kud08} for a comparison with BSG metallicities). Our results support this conclusion.

\citet{henry95} used published emission line fluxes of M81 and a series of
of photoionization models for a study of the oxygen abundance gradient. Their
results yield a central value of [O] = 0.26 and a gradient of 
-0.074 dex kpc$^{-1}$ (Henry, private communication). The central value agrees 
with our BSG work, but the gradient is steeper. 

\citet{stanghellini10} have recently studied planetary nebulae (PNe) and \hii~regions in M81 and used the detection 
of auroral lines to determine nebular electron temperatures and abundances. Since according to \citet{bresolin09} 
this approach leads to more reliable results, a comparison with the \citet{stanghellini10} \hii~region oxygen 
abundances is important. This is done in Fig.~\ref{zgrad} (lower left panel). At first glance, there seem to 
be two groups of \hii~regions, one group with abundances comparable to the BSGs and another with abundances 0.4 dex 
smaller. However, for many of the objects the abundances are too uncertain with individual errors as large as up 
to 0.6 dex estimated by \citet{stanghellini10} and, thus, no clear conclusions are possible with regard to 
abundance and abundance gradient from this sample. \citet{stanghellini10} combine their sample 
with the one by \citet{garnett87} to discuss 
metallicity and metallicity gradient. However, while the random errors of the \citet{garnett87} sample are 
small (0.1 to 0.2 dex), the abundances are  affected by the systematic uncertainties of the strong-line method. 
On the other hand, for the \citet{stanghellini10} abundances the situation is opposite, the random errors are 
large and the systematic errors are strongly reduced. Thus, we think the combination of the two samples is subject 
to uncertainties which are difficult to estimate.

Contrary to their \hii~region observations, the PNe analyzed by \citet{stanghellini10} have 
abundances generally more accurate. In Fig.~\ref{zgrad} (lower right panel) they are also compared with the 
BSG metallicities. The average difference in metallicity between the PNe and the BSGs 
is about -0.4 dex and seems to be significant. The 
metallicity gradient is -0.057$\pm{0.007}$ dex \kpc~and steeper than for the BSGs. This is a very interesting result, 
since the PNe of this sample do not contain type I PNe objects and consist only of type II and III, which means 
that they are significantly older than the BSGs with average ages of 3 and 6 Gyrs, respectively 
\citep{maciel09,stanghellini10b}. This means that over the last 5 Gyrs the metallicity must have increased 
substantially and the metallicity gradient of the disk has become shallower.

Photometric investigations of the disk of M81 confirm this conclusion. \citet{williams09} in their comprehensive 
study of star formation and metallicity analyzing HST color-magnitude diagrams of an outer disk field at R/R$_{25}$ = 1.17
find metallicities in the range between [Z] = -0.6 to -0.3 dex, for the population with ages between 10 Gyrs to 50 Myrs age. 
They also find solar metallicity for the younger population. This result is in agreement with the \citet{tikhonov05}, 
who investigated HST CMDs of a different disk field, and \cite{davidge09}, who used the red giant branch from CFHT 
MegaCam CMDs over the whole disk of M81 to also estimate a metallicity of [Z] =-0.4 dex. While metallicities obtained 
in this way might suffer from uncertainties in the extinction adopted and the systematics of the isochrones used, 
the picture emerging from the combination of our BSG results, the PNe observed and CMDs studied indicates that 
for a long period the metallicity of the M81 disk remained roughly constant and subsolar, but obviously, before 
the birth of the young population of masssive stars, there must have been a phase of enrichment. 

This situation is different from the Milky Way. Young massive stars have a metallicity very similar to the sun 
\citep{przybilla08}. PNe metallicities are also very close to the one of the sun and to massive stars 
\citep{henry10, stanghellini10b}. The metallicity enrichment of the thin disk has been very slow with an 
estimated increase of metallicity $\Delta$[Z] = 0.017 dex Gyr$^{-1}$ and the metal poor ([Z] -0.58 dex) thick disk may have formed
12 to 13 Gyrs ago in a single starburst \citep{fuhrmann11}. We also note that the case of M33 is similar to the Milky Way \citep{bresolin10, urbaneja05b}.
At this point, one can only speculate what caused the 
late enrichment of the very young population in M81. An interesting thought has been formulated by \citet{williams09}. 
M81 has satellite galaxies such as NGC 3077 and M82, which are gas and metal rich \citep{martin97} and 
are involved in tidal interaction with M81 \citep{appleton81, heckman90}. Recent inflow from such satellites 
or the tidal interaction induced by them and leading to recent bursts of star formation
could then have influenced the chemical evolution.

Chemical evolution models of galaxies also predict changes of the metallicity gradients as a function of time, 
however, many times with qualitatively different results. For instance, \citet{chiappini01} predict gradients to become steeper with time, 
whereas \citet{hou00} predict the opposite. Simulations of disk evolution including the effects of stellar migration by
\citet{roskar08} also predict a flattening of the gradient through the
homogenization of the population in the disk as a function of time.

The comparison of planetary nebulae with a younger stellar generation 
such as massive stars or \hii~regions offers, in principle, an opportunity to provide observational constraints. In the case of 
the Milky Way \citet{stanghellini10b} conclude that the gradient is steepening with time. 
However, \citet{maciel08} find the opposite, whereas \citet{henry10} do not find any hints of evolution at all. 
Thus, the situation of the temporary evolution of the Milky Way abundance gradient remains controversial. 
In M81 comparing our BSG results with the PNe abundances determined by \citet{stanghellini10} we find a weak 
indication that the abundance gradient became shallower with time.

\section{Mass -- metallicity relationship of galaxies from BSG spectroscopy}

Since the early work by \citet{lequeux79} the mass-metallity relationship of star forming galaxies has 
been regarded as an important observational constraint for understanding galaxy formation and evolution 
(see references introduced in the discussion). While these pioneering investigations were restricted 
to a relatively small sample of galaxies, the recent spectroscopic surveys such as SDSS opened the 
opportunity to study a large number of such objects. \citet{tremonti04} have analyzed more than 50,000 
galaxies observed within SDSS and obtained a well defined relationship between oxygen abundance and 
total stellar mass. However, the oxygen abundances are again based on the use of strong \hii~region 
emission lines only. While \citet{tremonti04} took special care of this problem and developed their 
own calibration of their strong line method, the systematic uncertainties are important to be investigated. 
\citet{bresolin09} found that this calibration very likely overestimates oxygen abundances. 
In a more general approach, \citet{kewley08} demonstrated very clearly that the mass-metallicity relationship 
obtained from the standard strong lines of \hii~regions depends very strongly on the calibration of the 
strong line method used. Applying ten different calibrations, which are frequently used in \hii~region 
abundance studies, on the same data set of emission lines of about 20,000 SDSS galaxies \citet{kewley08} 
obtained the shocking result that the mass-metallicity relationship can change from steep to almost flat 
just dependent on the calibration used. Since all the work published with regard to this relationship 
seems to rely on strong line \hii~region data and given these systematic uncertainties, it seems 
appropriate to start an investigation based on stellar spectroscopy only. With the results obtained here 
and compiling the metallicities of the BSG quantitative spectroscopy work for other galaxies published so 
far we have made a first attempt.

\begin{figure}
\begin{center}
  \includegraphics[scale=0.3,angle=90]{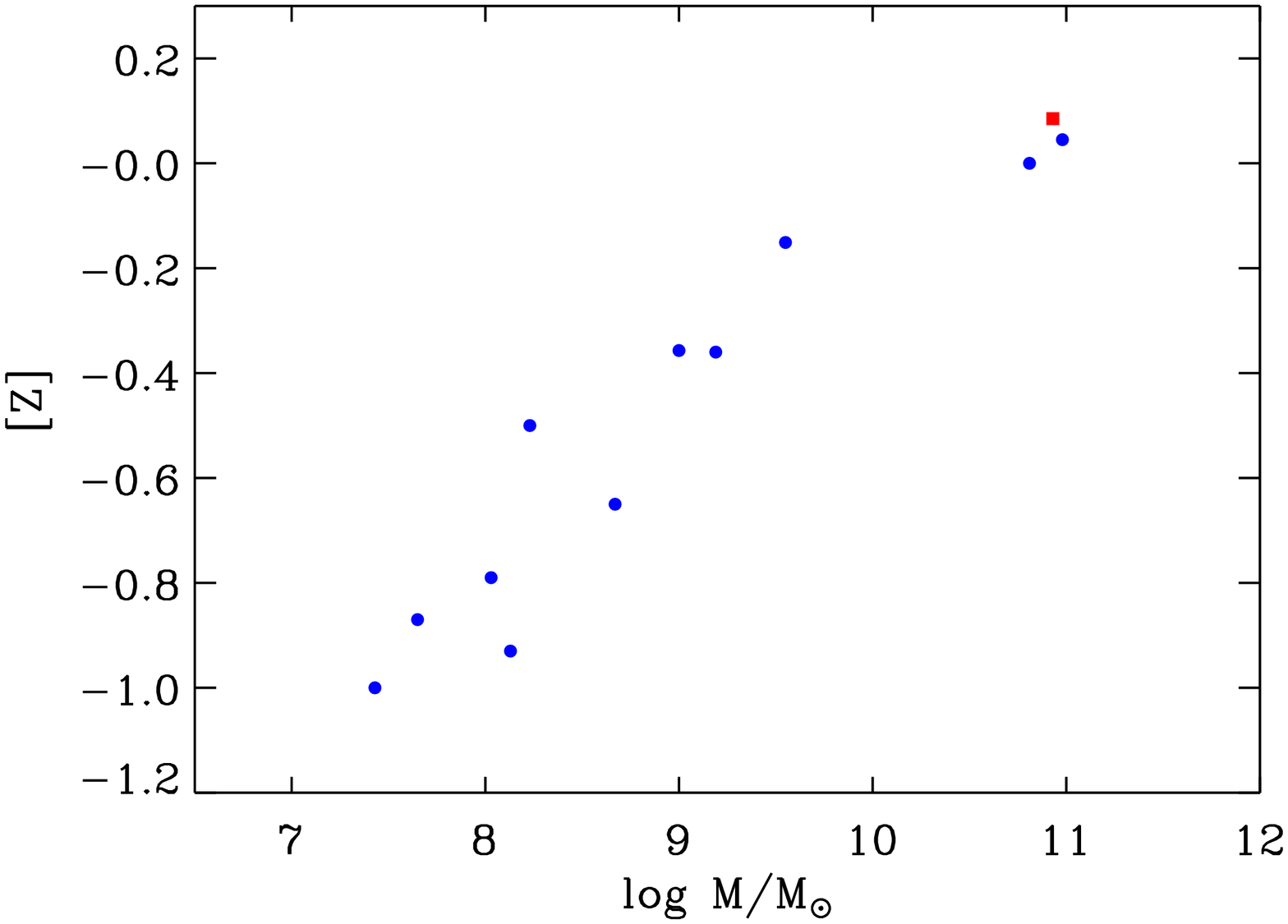}
 \includegraphics[scale=0.3,angle=90]{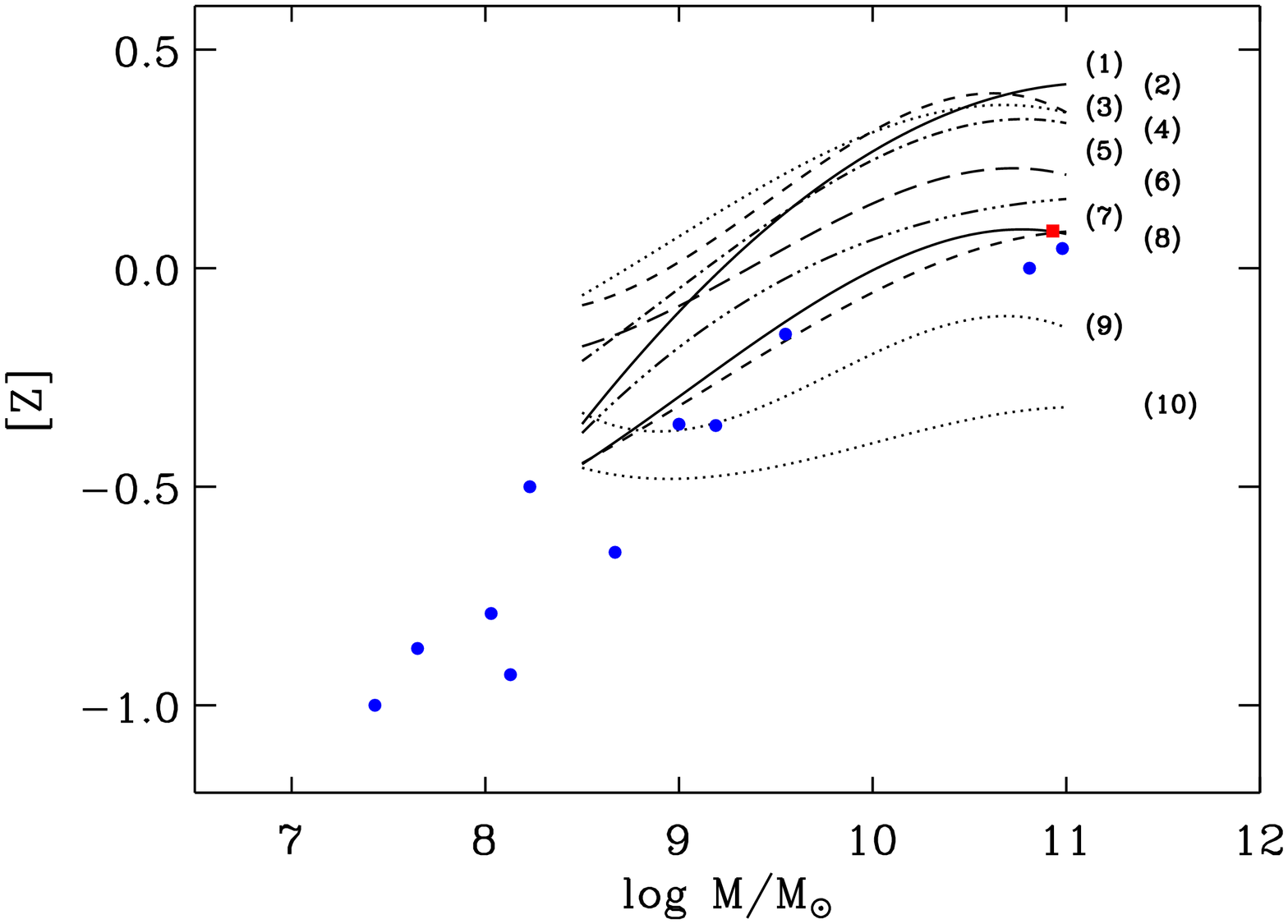}
\caption{Left: Observed mass-metallicity relationship of galaxies obtained from spectroscopic 
studies of blue supergiants. The red square is the M81 result from this paper. Right: Same as left, but now with
the average relationships obtained by \citet{kewley08} for the ten different \hii~region strong line calibrations used in 
their study of 20,000 SDSS galaxies. The ten calibrations are: [1] (solid) \citet{tremonti04}, [2] (dashed) \citet{zaritsky94},
[3] (dottet) \citet{kobulnicky04}, [4] (dash-dotted)\citet{kewley02}, [5] (long-dashed) \citet{mcgaugh91}, [6] (dash-triple-dotted) \citet{denicolo02}, 
[7] (solid)\citet{pettini04} (using [\ion{O}{3}]/H$_{\beta}$ and [\ion{N}{2}]/H$_{\alpha}$), [8] (dashed) \citet{pettini04} 
(using [\ion{N}{2}]/H$_{\alpha}$), [9] (dotted) \citet{pilyugin01}, [10] (dotted) \citet{pilyugin05}.
\label{massmet}}
 \end{center}
\end{figure}

The compilation of galaxy masses and metallicities is given in Table 4. For the spiral galaxies with a clear 
metallicity gradient (NGC300, M33, MW, M31, M81) metallicity values were taken at galactocentric distances 
of two disk scale lengths. For the irregular Local Group galaxies average 
values were used. The data are plotted in Fig.~\ref{massmet} (left panel). A very clear correlation of metallicity 
with stellar mass is obtained.

While the weakness of our approach at this stage is the small size of our sample, it is tempting to compare 
with the SDSS \hii~region based results discussed. For this purpose we have overplotted the average 
mass-metallicity relationships obtained by \citet{kewley08} for the ten different calibrations used in 
their work. It seems that a few of these calibrations \citep{tremonti04, zaritsky94} lead to a much steeper 
relationship than our work, whereas others \citep{pettini04} are in much better agreement.
We note that our sample is probing a larger galaxy mass range than the SDSS studies, going from low-mass 
dwarf irregulars to giant spirals. As pointed out in the study by \citet{lee06} this is important for 
constraining the scenarios for galaxy formation and evolution. (We realize that in \citealt*{lee06} the 
stellar masses of some of the dwarf irregulars overlapping with our sample are significantly smaller than
the masses given by \citealt*{woo08}, which we use for Fig.~\ref{massmet}. This will require further investigation).
In future work we plan to enlarge the sample of galaxies with quantitative studies of BSGs to make 
this comparison more significant.

\section{Distance}

The FGLR is a tight correlation between the flux-weighted gravity 
($g_F\,\equiv\,g/{T^4}_{\rm eff}, T_{\rm eff}$ in units of 10$^{4}$K) 
and the absolute bolometric magnitude M$_{bol}$ of BA supergiants. As described in detail in 
\citet{kud03, kud08} the physical background for this relationship is the fact that massive 
stars evolve at constant luminosity and mass accross the HRD from the hot main sequence to
the red supergiant stage. During this evolution, $g_F$ remains constant, because of the 
constant luminosity and mass. On the other hand, stellar luminosity is a strong function of 
stellar mass (see Fig.~\ref{mass} as an example) and, therefore, also a strong function of 
flux-weighted gravity, which establishes the FGLR. (For all details, we refer the reader to the two papers just cited). 
\citet{urbaneja08} and \citet{u09} were the first to use
the FGLR for distance determination of the metal poor dwarf galaxy WLM and M33, respectively. 
Here, we follow the same procedure as was detailed in these papers.

The FGLR has the form
\begin{equation}
    M_{\rm bol}\,=\,a (\log\,g_F\,-\,1.5)\,+\,b
  \end{equation}
with the recent calibration provided by \citet{kud08}, $a$ = 3.41 and $b$ = -8.02.

\begin{figure}
\begin{center}
  \includegraphics[scale=0.8,angle=90]{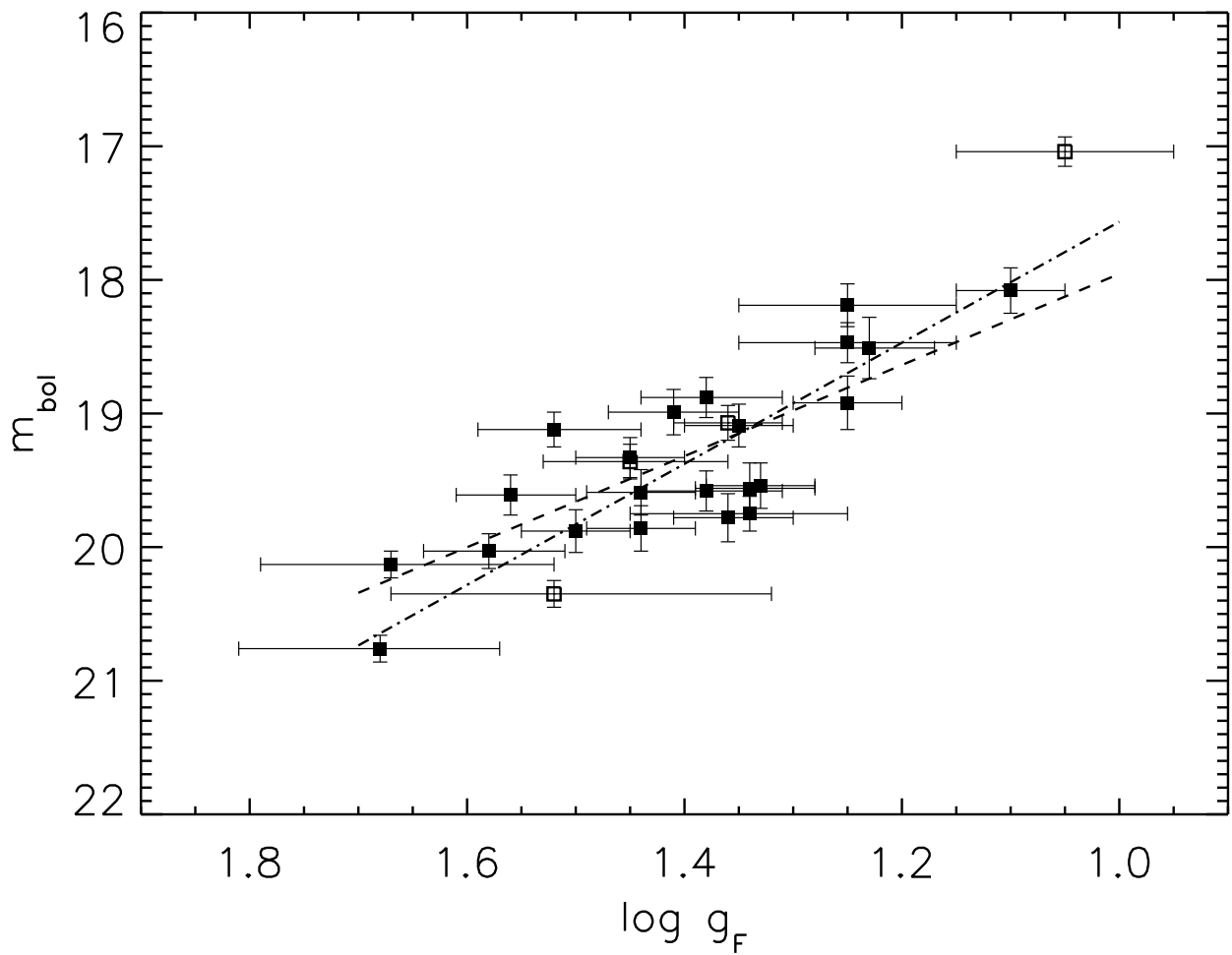}
 \caption{The observed FGLR in M81. Solid squares are targets used for the distance determination fit. Targets plotted as open squares were not included in the fit for reasons explained in the text. The dashed line corresponds to the FGLR calibration by \citet{kud08}. The dashed-dotted line is the new (still preliminary) LMC calibration
(Urbaneja et al, 2011, to be submitted to ApJ) discussed in the text. Both calibrations yield a very similar distance modulus.
\label{fglr}}
 \end{center}
\end{figure}

For each of our targets the spectroscopic analysis yields de-reddened apparent bolometric magnitude $m_{\rm bol}$
and flux-weighted gravity, which are given in Table 2. 
These data are plotted in Fig.~\ref{fglr}. Very obviously, there is a clear relationship between 
flux-weigthed gravity and apparent bolometric magnitude. We can use these data to fit a 
regression of the form

 \begin{equation}
    m_{\rm bol}\,=\,a (\log\,g_F\,-\,1.5)\,+\,b_{\rm M81}~.
  \end{equation}

The fit result is also shown in Fig.~\ref{fglr}. Since our targets span only 
a limited range in $g_{F}$ compared to the \citet{kud08} calibration sample, we adopt
the slope value provided by this calibration and fit only the intercept $b_{\rm M81}$. 
The difference between $b$ and  $b_{\rm M81}$ yields the distance modulus, which we determine to be
$\mu$ = 27.71$\,\pm\,$0.08 mag (the error is calculated similarly as in \citealt*{urbaneja08}).

The \citet{kud08} calibration of the FGLR is based on data from eight galaxies with distances 
mostly determined from using Cepheids. Recently, we have started the study of a large sample of BA 
supergiants in the LMC using high resolution, high S/N spectra with the goal to provide a new 
calibration of the FGLR based on the LMC only. This work is almost completed and will be published 
soon (Urbaneja et al. 2011, to be submitted to ApJ). With an adopted distance modulus 
to the LMC of m-M = 18.50 mag we obtain the calibration 
values a$_{LMC}$ = 4.53 and b$_{LMC}$ = -7.88. While this is a significantly steeper FGLR at the 
low luminosity/high g$_{F}$ end, this change in calibration does barely affect our 
distance determination, because most of our targets are at lower g$_{F}$/higher 
luminosity. A regression fit with these (still preliminary) calibration values yields a distance modulus of 
$\mu$ = 27.68$\,\pm\,$0.09 mag. We, thus, adopt 
a distance modulus of  $\mu$ = 27.7$\,\pm\,$0.1 mag.

We compare this value with previous distance determinations based on Cepheids. In addition to the HST Key 
Project work on M81 \citep{freedman94, freedman01} there are two recent studies by \citet{mccommas09} 
and by \citet{gerke11}. Cepheid distance studies typically apply  the Wesenheit method \citep{madore82} with a combination of V and I 
band magnitudes  which is assumed to be reddening free and then compare with the corresponding period luminosity 
relationship of LMC Cepheids. Following \citet{kennicutt98}, distances are corrected
for the difference in abundance between the target Cepheids and those in the
LMC. This so-called "metallicity correction" has the form $\Delta \mu$ = $\gamma$([O/H] - 8.5) 
where [O/H] = 12 + log (O/H) is the logarithmic oxygen abundance of the
young stellar population in the target galaxy at the galactocentric distance
of the observed Cepheid field relative to hydrogen. $\gamma$ is a fit parameter and has been determined
by \citet{kennicutt98} from the fact that Cepheids in inner fields of the 
spiral galaxy M101 are brighter and yield a shorter apparent distance modulus than those in outer fields.
Attributing this difference to a metallicity dependence of the period luminosity relationship and adopting
stellar metallicities and metallicity gradients from the oxygen \hii~region strong line studies by 
\citet{zaritsky94}, \citet{kennicutt98} obtained $\gamma$ = -0.29 mag dex$^{-1}$. [O/H] = 8.5 dex in this metallicity correction is the
adopted value of this abundance for the LMC. It refers to the "old" oxygen abundance scale where where [O/H]$_\odot$=8.9 dex.
(We will show below that this value is too high independent of the actual value of the oxygen abundance for the sun).
\citet{macri06} found a similar value of $\gamma$ for the maser galaxy NGC 4258 again from the different distance moduli obtained 
from inner and outer field Cepheids.
 
\citet{mccommas09} in their Cepheid distance investigation of M81 use HST light curves 
of 11 fundamental and two first overtone short period Cepheids in the outer disk of M81 at 
R = 1.23 R$_{25}$ ($\sim$ = 13.5 kpc) and obtain a distance modulus of M81 relative to the LMC of
$\Delta \mu$ = 9.34$\,\pm\,$0.05 mag. Checking the consistency with the 25 long period Cepheids in two 
inner HST WFPC fields observed by the Key Project located at R = 0.36 R$_{25}$ ($\sim$ = 4.3 kpc)  
\citet{mccommas09} use the same Wesenheit formalism and
obtain a distance modulus 0.23 mag shorter. Following the work by \citet{kennicutt98} and \citet{macri06}
they also apply a metallicity correction with $\gamma$ = -0.29 mag dex$^{-1}$ . This correction introduces 
a small increase of the distance to $\Delta \mu$ = 9.37$\,\pm\,$0.05 mag and reduces the difference in 
distance modulus between outer and inner field Cepheids to 0.09 mag. It is based on the metallicity study
by \citet{zaritsky94} who obtained [O/H] = 9.196 - 0.49~R/R$_{25}$ for 
the oxygen abundance as a function of galactocentric distance as a result of their strong-line analysis of 
\hii~region emission lines. Explaining the full difference in distance modulus between inner and outer field 
Cepheids in terms of metallicity with the \citet{zaritsky94} metallicity gradient requires 
$\gamma$ = -0.55 mag dex$^{-1}$.  

\citet{gerke11} investigate 107 long period Cepheids observed with the LBT in a galactocentric range of 
0.29 $\le$ R/R$_{25} \le$ 0.88 and with ground-based B, V, I photometry. Without applying a metallicity correction they obtain
$\Delta \mu$ = 9.19$\,\pm\,$0.05 mag. They also realize a trend in Cepheid distance modulus as a function of galactocentric distance
and obtain a metallicity correction, which leads to $\gamma$ = -0.56$\,\pm\,$0.36 mag dex$^{-1}$ and a 
distance modulus of $\Delta \mu$ = 9.39$\,\pm\,$0.14 mag. This agrees with with \citet{mccommas09} and also with the original value
of the Key Project of $\Delta \mu$ = 9.30$\,\pm\,$0.15 mag

Our FGLR distance to M81 is based on a LMC distance modulus of 18.5 mag and, thus, a difference of $\Delta \mu$ = 9.2$\,\pm\,$0.1 mag.
This is 0.10 to 0.19 mag or 5 to 8\% shorter than the ones obtained with the Cepheid work. However, we note that there is good agreement with the 
inner field long period Cepheids, when no metallicity corrections are applied. In the following we discuss some 
aspects of this metallicity correction. 

With the solar oxygen abundance [O/H]$_\odot$ = 8.69 dex \citep{allende01} the \citet{zaritsky94} 
logarithmic oxygen abundances relative to the sun are [O] $\equiv\,$ [O/H] - [O/H]$_{\odot}$ = 0.506 - 0.49~R/R$_{25}$. 
If oxygen is taken as proxy for metallicity, this is a significantly higher metallicity than found in our 
BSG spectroscopy in equation (4), while our gradient is shallower. Applying our metallicity gradient to correct for distance modulus 
difference between the inner and outer field Cepheids in M81 would require an even more negative value of 
$\gamma$, namely $\gamma$ = -0.65 mag dex$^{-1}$. Moreover, the LMC oxygen abundance 
[O/H]$_{LMC}$ = 8.50 dex or [O]$_{LMC}$ = -0.19 dex adopted in these corrections is too large compared with the 
LMC oxygen abundance of B-stars found by \citet{hunter07} ([O/H]$_{LMC}$ = 8.33 dex or [O]$_{LMC}$ = -0.36 dex), the iron abundances of LMC 
Cepheids determined by \citet{romaniello08} and \citet{luck98} ([Fe]$_{LMC}$ = -0.33 dex), and the LMC \hii~region oxygen abundances 
obtained by \citet{bresolin11} ([O/H]$_{LMC}$ = 8.36 dex or [O]$_{LMC}$ = -0.33 dex). This means that with our BSG metallicity values 
in M81 the Cepheids in the outer field have a metallicity 0.11 dex higher than the LMC. If one would apply the metallicity correction
with $\gamma$ = -0.65 mag dex$^{-1}$ accordingly, this would enlarge the distance modulus by another 0.07 mag.

However, with such a large negative value of $\gamma$ it is important to note that this empirical correction for the 
metallicity dependence of the period-luminosity relationship, which claims that Cepheids become brighter with 
increasing metallicity, is in striking disagreement with pulsation theory, which predicts exactly the opposite, 
namely that the Cepheid brightness decreases with increasing metallicity \citep{fiorentino02, marconi05, fiorentino07, bono08}.
It also disagrees with the recent high S/N, high spectral resolution quantitative spectroscopy in the Milky Way and the LMC 
carried out by \citet{romaniello08}, which confirms the prediction by pulsation theory. According to this work, 
the value of $\gamma$ should be positive and not negative. In other words, as careful spectroscopic metallicity studies 
compared with observed differences of distance moduli between inner and outer field Cepheids push $\gamma$ to increasingly
negative values, an explanation of that distance modulus differences in terms of metallicity seems unlikely. It must be 
something else and it is an additional systematic effect not understood. 

We also note that \citet{u09} have demonstrated from their quantitative spectroscopy of blue supergiants in M33  
that the difference of distance moduli between inner field and outer field Cepheids found by \citet{scowcroft09} 
would require a $\gamma$-value of -0.55 mag/dex. Even worse, \cite{bresolin10} re-determined \hii~region abundances in M33
using auroral lines and applying their abundance gradient to the Cepheid fields in M33 yields  $\gamma$ = - 1.2 mag dex$^{-1}$ 
(see discussion in \citealt*{bresolin11}).

Another galaxy where the comparison of Cepheids in the inner and outer fields leads to a significantly different 
distance modulus is the maser galaxy NGC 4258. This galaxy is of particular importance, since 
it has been used as the new anchor point for the extragalactic distance scale by \citet{riess09a, riess09b, riess11}  because of its 
accurately known distance from the Keplerian motion of water masers orbiting the central black hole \citep{humphreys08}. 
However, \citet{macri06}, who carried out the HST obervations of Cepheids in NGC 4258 again found the distance modulus 
of the inner field Cepheids to be shorter than in the outer fields and based on the \hii~region strong line method 
oxygen abundances by \citet{zaritsky94} derived a $\gamma$-value of -0.29 mag dex$^{-1}$. Most recently, \cite{bresolin11} 
re-determined the \hii~region metallicities in this galaxy including the observation of auroral lines in a few cases. 
This led to a downward substantial revision of the metallicity, which seems to be close to the LMC and not strongly 
super-solar, and a very shallow abundance gradient. Based on these results, \cite{bresolin11} show that 
$\gamma$ = - 0.69 mag dex$^{-1}$ would be needed to explain the distance modulus difference between inner and outer fields, 
again a value much too negative, when compared with pulsation theory and observational work on Milky Way and LMC 
Cepheids. While the improved \hii~region work on this important galaxy still awaits an independent confirmation through 
a study of BSGs, it is an additional clear indication of a systematic effect on Cepheid distance moduli not understood at this point.
\citet{majaess11} discuss the large metallicity corrections suggested by \citet{gerke11} and by the recent HST/ACS Cepheid 
study of M101 by \citet{shappee11} and demonstrate that such corrections lead to very improbable distances of the LMC and SMC. 
The work by \citet{storm11} indicates that a lower limit for $\gamma$ is -0.2 mag dex$^{-1}$. \citet{majaess11} 
argue that crowding is very likely responsible for the distance modulus differences obtained between inner and outer 
field Cepheids and not metallicity. We think that a careful spectroscopic investigation of galactic metallicities and 
their gradients and distance determinations using the FGLR as an independent method will help to clarify the situation.

Independent of the Cepheid work there have been numerous studies of HST color-magnitude diagrams of M81 to determine 
a distance from the tip of the red giants branch. The distance moduli found were 28.03 mag \citep{sakai04}, 
27.93 mag \citep{tikhonov05}, 27.70 mag \citep{rizzi07}, 27.72 to 27.78 mag (\citealt{dalcanton09}, different fields in 
the halo and the outer disk), 27.81 mag (Extragalactic Distance Database catalogue, \citealt{tully09}) and 
27.86 mag \citep{durell10}. The more recent work since 2007 has converged on an improved 
methodology and seems to agree, within the uncertainties, with the distance 
modulus found in our study.

\section{Conclusions and Future Work} 

In this paper we have demonstrated that the quantitative spectroscopy of BSGs is a promising tool to constrain the 
chemical evolution of galaxies and to determine their distances, which can be applied to galaxies clearly beyond 
the Local Group. Using the relationship between flux-weighted gravity and luminosity we were able to determine a 
new distance to M81, which compares well with TRGB distances. While there is also agreement with HST Cepheid 
distances within the error margins, our results with regard to metallicity and metallicity gradient confirmed 
previous studies that the systematic differences between distance moduli obtained from inner and outer field 
Cepheids (found in M33, M81, M101, NGC 4258) are very likely not caused by a metallicity dependence of the 
period-luminosity relationship of Cepheids. There must be another reason for these systematic differences.

An independent check of distances obtained with either the TRGB or Cepheids is important for future work.
We note that besides the importance 
for characterizing the physics of galaxies in the Local Volume accurate distances and a careful discussion 
of the systematics of stellar distance determination methods are crucial for constraining the dark energy 
equation-of-state parameter w = p/($\rho$c$^{2}$). As is well known \citep{macri06}, the determination 
of cosmological parameters from the cosmic microwave background is affected by degeneracies in parameter 
space and cannot provide strong constraints on the value of H$_{0}$ \citep{spergel06, tegmark04}.
Only if additional assumptions are made, for instance that the universe is flat, H$_{0}$ can be predicted with high 
precision (i.e. 2\%) from the observations of the cosmic microwave background, baryonic acoustic 
oscillations and type I high redshift supernovae. If these assumptions are relaxed, then much larger 
uncertainties are introduced \citep{spergel07, komatsu09}. The uncertainty of the 
determination of w is related to the uncertainty of H$_{0}$ through $\Delta$w/w $\approx$ 2$\Delta$H$_{0}$/H$_{0}$.
Thus, an independent determination of H$_{0}$ with an accuracy of 5\% will allow the uncertainty of w to 
be reduced to ±0.1. While extremely promising steps towards this goal have been made by \citet{macri06} and  
\citet{riess09a,riess09b,riess11} using the maser galaxy NGC 4258 as a new anchor point and HST IR Cepheid
photometry of recent SNIa galaxies out to 30 Mpc, it is clear that the complexity of this approach
requires additional and independent tests. Crucial contributions which can be made using BSGs besides 
independent distance determinations are to investigate the role of metallicity and interstellar extinction.

We have also shown that the determination of metallicities for individual supergiant stars beyond the Local Group is possible. 
In this way, we can determine galaxy metallicities and metallicity gradients avoiding the systematic 
uncertainties of \hii~region strong line methods. This can be used as an independent way to directly 
measure the mass-metallicity relationship of galaxies and to correlate metallicity gradients with 
galactic properties such as mass, angular momentum and morphological type. But it can also be used to 
find out about systematic uncertainties of \hii~region strong line method calibrations and to identify 
the more reliable ones or to develop a new one tested with BSG metallicities. Moreover, in combination 
with metallicity information of an older population of stars obtained through the analysis of CMDs or 
the spectroscopy of PNe the chemical evolution history of galaxies can be investigated. In the case of 
the disk of M81 we have found an indication of a late enrichment of heavy elements, which is 
significantly different from the Milky Way. We have also provided the first mass-metallicity 
relationship for star forming galaxies solely based on stellar spectroscopy.


\acknowledgments
This work was supported by the National Science Foundation under grant AST-1008798 to RPK and FB.
Moreover, RPK acknowledges support by the Alexander-von-Humboldt Foundation and the hospitality 
of the Max-Planck-Institute for Astrophysics in Garching and the University Observatory Munich, 
where part of this work was carried out.
WG and GP gratefully acknowledge financial support for this work from the
Chilean Center for Astrophysics FONDAP 15010003, and from the BASAL
Centro de Astrofisica y Tecnologias Afines (CATA) PFB-06/2007.
All members of our team want to thank the Keck staff astronomers for their 
dedicated first-class professional support, when the observations were planned and carried out. 
Last but not least we also acknowledge the usage of the Hyperleda database (http://leda.univ-lyon1.fr).

We acknowledge the discussion of this work with  Drs. Lucas Macri, Richard Henry and Ortwin Gerhard.
Most importantly, we wish thank our anonymous referee for the extremely careful review of the 
manuscript and for many very helpful suggestions to improve this paper.

The data presented in this work were obtained at the W.M. Keck Observatory, which is operated as a scientific 
partnership among the California Institute of Technology, the University of California and the National Aeronautics and 
Space Administration. The Observatory was made possible by the generous financial support of the W.M. Keck Foundation.

The authors wish to recognize and acknowledge the very significant cultural role and reverence that the summit of 
Mauna Kea has always had within the indigenous Hawaiian community.  We are most fortunate to have the opportunity to 
conduct observations from this mountain.




{\it Facilities:} \facility{Keck (LRIS)}, \facility{HST (ACS)}.

\clearpage




\clearpage


\begin{deluxetable}{cccccccccc}
\tabletypesize{\scriptsize}
\tablecolumns{10}
\tablewidth{0pt}
\tablecaption{M81 - Spectroscopic targets}

\tablehead{
\colhead{No.}     &
\colhead{name}            &
\colhead{$\alpha_{2000}$}           &
\colhead{$\delta_{2000}$}     &
\colhead{R/R$_{25}$$^{\tablenotemark{a}}$}     &
\colhead{sp.t.}             &
\colhead{m$_{V}$}     &
\colhead{B-V} &
\colhead{D$_{B}$}    & 
\colhead{}\\
\colhead{}   &
\colhead{}         &
\colhead{ h min  sec}         &
\colhead{ \arcdeg~ \arcmin~ \arcsec }     &
\colhead{}   &
\colhead{}         &
\colhead{mag}         &
\colhead{mag}		&
\colhead{dex}			&
\colhead{}\\[1mm]
\colhead{(1)}	&
\colhead{(2)}	&
\colhead{(3)}	&
\colhead{(4)}	&
\colhead{(5)}	&
\colhead{(6)}	&
\colhead{(7)}	&
\colhead{(8)}	&
\colhead{(9)}	&
\colhead{(10)}}
\startdata
\\[-1mm]
0  &  Z1 & 9 55 30.580 & 69 12 22.716 & 0.83 & B7   & 20.946 & 0.052 & 0.126 &                   \\[2pt]
1  &  Z2 & 9 55 34.965 & 69 11 58.524 & 0.81 & B4   & 21.493 & 0.011 & 0.084 &                   \\[2pt]
2  &  Z3 & 9 55 25.022 & 69 11 16.908 & 0.69 & B7   & 20.847 & 0.093 & 0.122 &                   \\[2pt]
3  &  Z4 & 9 55 12.686 & 69 10 50.880 & 0.61 & B9   & 20.305 & 0.198 &       & \tablenotemark{b} \\[2pt]
4  &  Z5 & 9 55 26.224 & 69 10 16.896 & 0.60 & B3   & 21.452 & 0.059 & 0.056 &                   \\[2pt]
5  &  Z6 & 9 55 18.204 & 69 09 51.984 & 0.53 & B7   & 21.206 & 0.185 & 0.156 &                   \\[2pt]
6  &  Z7 & 9 55 30.948 & 69 09 33.696 & 0.55 & A3   & 21.454 & 0.244 & 0.539 &                   \\[2pt]
7  &  Z9 & 9 55 10.488 & 69 08 27.132 & 0.41 & A4   & 21.296 & 0.358 &       & \tablenotemark{b} \\[2pt]
8  & Z10 & 9 55 34.353 & 69 08 39.552 & 0.48 & B9   & 21.236 & 0.185 & 0.230 &                   \\[2pt]
9  & Z11 & 9 55 21.518 & 69 08 15.864 & 0.39 & B1   & 21.327 & 0.048 &       &                   \\[2pt]
10 & Z12 & 9 55 35.100 & 69 08 16.908 & 0.45 & B4   & 21.020 & 0.185 & 0.049 &                   \\[2pt]
11 & Z13 & 9 55 29.344 & 69 07 48.432 & 0.37 & B1.5 & 21.113 & 0.113 &       &                   \\[2pt]
12 & Z14 & 9 55 31.761 & 69 07 39.036 & 0.36 & B7   & 21.371 & 0.207 & 0.127 &                   \\[2pt]
13 & Z15 & 9 55 34.783 & 69 07 31.440 & 0.37 & B0.5 & 20.495 & 0.136 &       &                   \\[2pt]
14 & Z16 & 9 55 43.579 & 69 07 18.768 & 0.42 & B4   & 19.979 & 0.135 & 0.002 &                   \\[2pt]
15 & Z17 & 9 55 39.237 & 69 06 35.172 & 0.69 & A4   & 21.165 & 0.372 & 0.586 &                   \\[2pt]
16 & Z18 & 9 55 43.442 & 69 06 18.972 & 0.33 & B9   & 20.402 & 0.211 & 0.171 &                   \\[2pt]
17 & Z20 & 9 55 46.972 & 69 05 48.516 & 0.32 & A1   & 20.330 & 0.305 & 0.110 &                   \\[2pt]
18 &  C6 & 9 54 35.976 & 69 05 00.168 & 0.74 & A1   & 20.784 & 0.306 & 0.288 &                   \\[2pt]
19 &  C9 & 9 54 51.542 & 69 05 33.288 & 0.51 & B6   & 21.217 & 0.071 & 0.129 &                   \\[2pt]
20 & C11 & 9 54 49.214 & 69 06 17.640 & 0.53 & B9   & 20.412 & 0.296 & 0.178 &                   \\[2pt]
21 & C13 & 9 54 36.451 & 69 07 15.708 & 0.69 & B2   & 21.152 & 0.038 & 0.015 &                   \\[2pt]
22 & C14 & 9 54 35.196 & 69 07 43.752 & 0.70 & B3   & 21.414 & 0.039 & 0.055 &                   \\[2pt]
23 & C16 & 9 54 54.530 & 69 08 14.892 & 0.51 & B9   & 21.245 & 0.337 & 0.186 &                   \\[2pt]
24 & C20 & 9 54 51.079 & 69 09 43.992 & 0.60 & B9   & 20.411 & 0.259 & 0.249 &                   \\[2pt]
25 & C21 & 9 55 18.777 & 69 09 52.668 & 0.53 & B8   & 20.395 & 0.105 &       & \tablenotemark{b} \\[2pt]
\enddata
\tablenotetext{a}{Galactocentric distance, in units of R$_{25}$ = 11.99 arcmin $\simeq$ 12.09 kpc (distance modulus 27.70 mag). A position angle PA = 157\arcdeg, an inclination i =57\arcdeg and central coordinates $\alpha_{2000}$ = 9h55min33.2sec, $\delta_{2000}$ = 69\arcdeg 3\arcmin 55\arcsec were assumed (Hyperleda data base, Paturel et al., 2003)}
\tablenotetext{b}{no near UV spectral coverage; no Balmer jump measured}
\end{deluxetable}

\clearpage

\begin{deluxetable}{cccccccccc}
\tabletypesize{\scriptsize}
\tablecolumns{10}
\tablewidth{0pt}
\tablecaption{Stellar Parameters}

\tablehead{
\colhead{No.}     &
\colhead{name}            &
\colhead{\teff}           &
\colhead{log g}     &
\colhead{log g$_{F}$}     &
\colhead{[Z]}             &
\colhead{E(B-V)}     &
\colhead{BC} &
\colhead{m$_{bol}$}    & 
\colhead{}\\
\colhead{}   &
\colhead{}         &
\colhead{K}         &
\colhead{cgs}     &
\colhead{cgs}   &
\colhead{dex}         &
\colhead{mag}         &
\colhead{mag}		&
\colhead{mag}			&
\colhead{}\\[1mm]
\colhead{(1)}	&
\colhead{(2)}	&
\colhead{(3)}	&
\colhead{(4)}	&
\colhead{(5)}	&
\colhead{(6)}	&
\colhead{(7)}	&
\colhead{(8)}	&
\colhead{(9)}	&
\colhead{(10)}}
\startdata
\\[-1mm]
 0 &  Z1 & 12500$\rm^{910}_{1020}$ & 1.75$\rm^{0.17}_{0.20}$ & 1.36$\rm^{0.05}_{0.06}$ & -0.07$\pm{0.15}$ & 0.13 & -0.75 & 19.78$\pm{0.18}$ &                   \\[2pt]  
 1 &  Z2 & 15000$\rm^{730}_{820}$  & 2.20$\rm^{0.13}_{0.15}$ & 1.50$\rm^{0.05}_{0.05}$ & -0.01$\pm{0.20}$ & 0.15 & -1.15 & 19.88$\pm{0.16}$ &                   \\[2pt]  
 2 &  Z3 & 12500$\rm^{960}_{1090}$ & 1.73$\rm^{0.18}_{0.21}$ & 1.34$\rm^{0.05}_{0.06}$ & -0.09$\pm{0.10}$ & 0.17 & -0.75 & 19.56$\pm{0.19}$ &                   \\[2pt]  
 3 &  Z4 & 10000$\rm^{500}_{500}$  & 1.45$\rm^{0.17}_{0.18}$ & 1.45$\rm^{0.08}_{0.09}$ &                  & 0.22 & -0.28 & 19.36$\pm{0.13}$ & \tablenotemark{a} \\[2pt] 
 4 &  Z5 & 16000$\rm^{890}_{1020}$  & 2.15$\rm^{0.14}_{0.17}$ & 1.33$\rm^{0.05}_{0.05}$ &  0.03$\pm{0.20}$ & 0.21 & -1.29 & 19.54$\pm{0.17}$ & \tablenotemark{b} \\[2pt]   
 5 &  Z6 & 12500$\rm^{670}_{680}$  & 1.95$\rm^{0.15}_{0.15}$ & 1.56$\rm^{0.05}_{0.06}$ &  0.08$\pm{0.10}$ & 0.28 & -0.73 & 19.61$\pm{0.15}$ &                   \\[2pt]
 6 &  Z7 &  8500$\rm^{180}_{130}$  & 1.40$\rm^{0.17}_{0.14}$ & 1.68$\rm^{0.13}_{0.11}$ & -0.03$\pm{0.20}$ & 0.23 &  0.02 & 20.76$\pm{0.10}$ &                   \\[2pt]  
 7 &  Z9 &  8300$\rm^{200}_{200}$  & 1.20$\rm^{0.19}_{0.24}$ & 1.52$\rm^{0.15}_{0.20}$ &                  & 0.31 &  0.05 & 20.35$\pm{0.10}$ & \tablenotemark{a} \\[2pt]  
 8 & Z10 & 11000$\rm^{440}_{460}$  & 1.75$\rm^{0.13}_{0.14}$ & 1.58$\rm^{0.06}_{0.07}$ &  0.11$\pm{0.10}$ & 0.25 & -0.45 & 20.03$\pm{0.13}$ &                   \\[2pt]  
 9 & Z11 & 22000$\rm^{1000}_{1000}$& 2.62$\rm^{0.18}_{0.18}$ & 1.25$\rm^{0.10}_{0.10}$ &                  & 0.24 & -2.12 & 18.47$\pm{0.15}$ & \tablenotemark{c} \\[2pt]  
10 & Z12 & 15000$\rm^{1100}_{1420}$& 1.95$\rm^{0.17}_{0.22}$ & 1.25$\rm^{0.05}_{0.05}$ &  0.07$\pm{0.10}$ & 0.30 & -1.17 & 18.92$\pm{0.20}$ &                   \\[2pt]  
11 & Z13 & 21000$\rm^{1000}_{1000}$& 2.54$\rm^{0.18}_{0.18}$ & 1.25$\rm^{0.10}_{0.10}$ &                  & 0.29 & -2.01 & 18.19$\pm{0.16}$ & \tablenotemark{c} \\[2pt]  
12 & Z14 & 13000$\rm^{780}_{870}$  & 1.90$\rm^{0.15}_{0.17}$ & 1.44$\rm^{0.05}_{0.05}$ &  0.21$\pm{0.15}$ & 0.31 & -0.83 & 19.59$\pm{0.17}$ &                   \\[2pt]  
13 & Z15 & 25000$\rm^{1000}_{1000}$& 2.64                    & 1.05                    &                  & 0.32 & -2.45 & 17.04$\pm{0.17}$ &\tablenotemark{c,d}\\[2pt] 
14 & Z16 & 15000$\rm^{1000}_{1000}$& 1.80$\rm^{0.16}_{0.17}$ & 1.10$\rm^{0.05}_{0.05}$ &  0.09$\pm{0.10}$ & 0.27 & -1.29 & 18.08$\pm{0.17}$ & \tablenotemark{b} \\[2pt]
15 & Z17 &  8300$\rm^{120}_{140}$  & 1.35$\rm^{0.14}_{0.18}$ & 1.67$\rm^{0.12}_{0.15}$ &  0.14$\pm{0.15}$ & 0.34 &  0.05 & 20.13$\pm{0.10}$ &                   \\[2pt]
16 & Z18 & 11500$\rm^{770}_{680}$  & 1.65$\rm^{0.17}_{0.17}$ & 1.41$\rm^{0.06}_{0.06}$ &  0.15$\pm{0.10}$ & 0.27 & -0.56 & 18.99$\pm{0.17}$ &                   \\[2pt]
17 & Z20 & 12000$\rm^{1160}_{1340}$& 1.55$\rm^{0.23}_{0.27}$ & 1.23$\rm^{0.05}_{0.05}$ &                  & 0.36 & -0.69 & 18.51$\pm{0.23}$ &                   \\[2pt] 
18 &  C6 &  9250$\rm^{560}_{300}$  & 1.20$\rm^{0.21}_{0.14}$ & 1.34$\rm^{0.11}_{0.09}$ &  0.00$\pm{0.10}$ & 0.28 & -0.16 & 19.75$\pm{0.23}$ &                   \\[2pt] 
19 &  C9 & 13000$\rm^{780}_{870}$  & 1.90$\rm^{0.15}_{0.17}$ & 1.44$\rm^{0.05}_{0.05}$ &                  & 0.17 & -0.83 & 19.86$\pm{0.17}$ &                   \\[2pt]
20 & C11 & 11000$\rm^{670}_{720}$  & 1.55$\rm^{0.17}_{0.19}$ & 1.38$\rm^{0.06}_{0.07}$ &  0.04$\pm{0.15}$ & 0.34 & -0.47 & 18.88$\pm{0.15}$ &                   \\[2pt]   
21 & C13 & 17000$\rm^{830}_{960}$  & 2.27$\rm^{0.13}_{0.15}$ & 1.35$\rm^{0.05}_{0.05}$ &                  & 0.19 & -1.49 & 19.09$\pm{0.16}$ & \tablenotemark{b} \\[2pt]	
22 & C14 & 17000$\rm^{700}_{790}$  & 2.37$\rm^{0.12}_{0.13}$ & 1.45$\rm^{0.05}_{0.05}$ &                  & 0.19 & -1.49 & 19.33$\pm{0.15}$ & \tablenotemark{b} \\[2pt]  
23 & C16 & 11000$\rm^{670}_{720}$  & 1.55$\rm^{0.17}_{0.19}$ & 1.38$\rm^{0.06}_{0.07}$ &                  & 0.38 & -0.47 & 19.58$\pm{0.15}$ &                   \\[2pt]  
24 & C20 & 10500$\rm^{470}_{520}$  & 1.60$\rm^{0.15}_{0.17}$ & 1.52$\rm^{0.07}_{0.08}$ & 0.08$\pm{0.10}$  & 0.30 & -0.36 & 19.12$\pm{0.15}$ &                   \\[2pt]  
25 & C21 & 12500$\rm^{500}_{500}$  & 1.75$\rm^{0.12}_{0.12}$ & 1.36$\rm^{0.05}_{0.05}$ &                  & 0.19 & -0.75 & 19.07$\pm{0.13}$ & \tablenotemark{a} \\[2pt]
\enddata
\tablenotetext{a}{no D$_{B}$, \teff ~from spectral type}
\tablenotetext{b}{\teff ~from D$_{B}$ and \siii, \siiii, \siiv}
\tablenotetext{c}{\teff ~from \siii, \siiii, \siiv}
\tablenotetext{d}{extreme HII contamination of Balmer lines}
\end{deluxetable}

\clearpage

\begin{deluxetable}{cccccccc}
\tabletypesize{\scriptsize}
\tablecolumns{8}
\tablewidth{0pt}
\tablecaption{Absolute magnitudes, luminosities, radii and masses}

\tablehead{
\colhead{No.}     &
\colhead{name}            &
\colhead{M$_{V}$}           &
\colhead{M$_{bol}$}     &
\colhead{log L/L$_\odot$}     &
\colhead{R}             &
\colhead{M$_{spec}$}     &
\colhead{M$_{evol}$}\\
\colhead{}   &
\colhead{}         &
\colhead{mag}         &
\colhead{mag}     &
\colhead{dex}   &
\colhead{R$_{\odot}$}         &
\colhead{M$_{\odot}$}         &
\colhead{M$_{\odot}$}\\[1mm]	
\colhead{(1)}	&
\colhead{(2)}	&
\colhead{(3)}	&
\colhead{(4)}	&
\colhead{(5)}	&
\colhead{(6)}	&
\colhead{(7)}	&
\colhead{(8)}}	
\startdata
\\[-1mm]
 0 &  Z1 & -7.16 & -7.92 & 5.07$\pm{0.07}$ &  73.1$\pm{6.2}$ & 10.9 & 17.0\\[2pt]
 1 &  Z2 & -6.67 & -7.82 & 5.03$\pm{0.06}$ &  48.5$\pm{3.5}$ & 13.5 & 16.4\\[2pt]
 2 &  Z3 & -7.38 & -8.14 & 5.16$\pm{0.08}$ &  80.9$\pm{7.1}$ & 12.6 & 18.2\\[2pt]
 3 &  Z4 & -8.08 & -8.34 & 5.24$\pm{0.05}$ & 138.6$\pm{8.4}$ & 19.7 & 18.2\\[2pt]
 4 &  Z5 & -6.90 & -8.16 & 5.16$\pm{0.07}$ &  49.8$\pm{3.9}$ & 12.7 & 18.3\\[2pt]
 5 &  Z6 & -7.36 & -8.09 & 5.14$\pm{0.06}$ &  79.1$\pm{5.6}$ & 20.2 & 17.9\\[2pt]
 6 &  Z7 & -6.96 & -6.94 & 4.68$\pm{0.04}$ & 100.7$\pm{4.9}$ &  9.2 & 12.8\\[2pt]
 7 &  Z9 & -7.37 & -7.35 & 4.84$\pm{0.04}$ & 127.6$\pm{6.1}$ &  9.4 & 14.3\\[2pt]
 8 & Z10 & -7.24 & -7.67 & 4.97$\pm{0.05}$ &  84.2$\pm{4.9}$ & 14.5 & 15.7\\[2pt]
 9 & Z11 & -7.12 & -9.23 & 5.59$\pm{0.06}$ &  43.2$\pm{3.0}$ & 28.2 & 27.0\\[2pt]
10 & Z11 & -7.61 & -8.78 & 5.41$\pm{0.08}$ &  75.5$\pm{6.9}$ & 18.4 & 22.4\\[2pt]
11 & Z13 & -7.49 & -9.51 & 5.70$\pm{0.06}$ &  53.9$\pm{3.9}$ & 36.5 & 30.1\\[2pt]
12 & Z14 & -7.29 & -8.11 & 5.14$\pm{0.07}$ &  73.8$\pm{3.9}$ & 15.7 & 18.0\\[2pt]
13 & Z15 & -8.20 &-10.66 & 6.16$\pm{0.04}$ &  64.6$\pm{3.3}$ & 66.1 & 52.6\\[2pt]
14 & Z16 & -8.56 & -9.62 & 5.75$\pm{0.07}$ & 111.1$\pm{8.6}$ & 28.3 & 30.7\\[2pt]
15 & Z17 & -7.59 & -7.57 & 4.93$\pm{0.04}$ & 141.2$\pm{6.6}$ & 16.2 & 15.3\\[2pt]
16 & Z18 & -8.14 & -8.71 & 5.38$\pm{0.07}$ & 124.3$\pm{9.6}$ & 25.1 & 21.9\\[2pt]
17 & Z20 & -8.49 & -9.19 & 5.58$\pm{0.09}$ & 142.4$\pm{15.3}$& 26.1 & 26.0\\[2pt]
18 &  C6 & -7.78 & -7.95 & 5.08$\pm{0.05}$ & 135.4$\pm{8.4}$ & 10.5 & 17.1\\[2pt]
19 &  C9 & -7.01 & -7.84 & 5.04$\pm{0.07}$ &  65.2$\pm{5.0}$ & 12.2 & 16.5\\[2pt]
20 & C11 & -8.34 & -8.82 & 5.43$\pm{0.06}$ & 142.9$\pm{10.1}$& 26.3 & 22.7\\[2pt]
21 & C13 & -7.14 & -8.61 & 5.34$\pm{0.06}$ &  54.3$\pm{4.0}$ & 19.9 & 21.2\\[2pt]
22 & C14 & -6.88 & -8.37 & 5.25$\pm{0.06}$ &  48.6$\pm{3.3}$ & 20.1 & 19.5\\[2pt]
23 & C16 & -7.63 & -8.12 & 5.15$\pm{0.06}$ & 103.5$\pm{7.3}$ & 13.8 & 18.0\\[2pt]
24 & C20 & -8.22 & -8.58 & 5.33$\pm{0.05}$ & 140.4$\pm{8.4}$ & 28.5 & 21.0\\[2pt]
25 & C21 & -7.89 & -8.63 & 5.35$\pm{0.05}$ & 101.4$\pm{6.0}$ & 21.0 & 21.3\\[2pt]
\enddata
\end{deluxetable}

\clearpage

\begin{deluxetable}{cccc}
\tabletypesize{\scriptsize}
\tablecolumns{4}
\tablewidth{0pt}
\tablecaption{Mass-metallicity Relationship of Galaxies}

\tablehead{
\colhead{Galaxy}            &
\colhead{log M$_{stars}/M_{\odot}$}           &
\colhead{[Z]}     &
\colhead{source}\\[1mm]     
\colhead{(1)}	&
\colhead{(2)}	&
\colhead{(3)}	&
\colhead{(4)}}	
\startdata
\\[-1mm]
   M81  & 10.93 &  0.08 &  \tablenotemark{a,b}     \\[2pt]
   M31  & 10.98 &  0.04 &  \tablenotemark{c,d,e,f} \\[2pt]
    MW  & 10.81 &  0.00 &  \tablenotemark{g,h}     \\[2pt]
   M33  &  9.55 & -0.15 &  \tablenotemark{i,j}     \\[2pt]
 NGC300 &  9.00 & -0.36 &  \tablenotemark{k,l}     \\[2pt]
   LMC  &  9.19 & -0.36 &  \tablenotemark{i,m}     \\[2pt] 
   SMC  &  8.67 & -0.65 &  \tablenotemark{i,n,o}   \\[2pt]
NGC6822 &  8.23 & -0.50 &  \tablenotemark{i,p}     \\[2pt]
NGC3109 &  8.13 & -0.93 &  \tablenotemark{i,q}     \\[2pt]
   WLM  &  7.67 & -0.87 &  \tablenotemark{i,r}     \\[2pt]
 Sex A  &  7.43 & -1.00 &  \tablenotemark{i,s}     \\[2pt] 
\enddata
\tablenotetext{a}{\citet{deblok08}}
\tablenotetext{b}{this work}
\tablenotetext{c}{\citet{chemin09}}
\tablenotetext{d}{\citet{przybilla08b}}
\tablenotetext{e}{\citet{trundle02}}
\tablenotetext{f}{\citet{smartt01}}
\tablenotetext{g}{\citet{sofue09}}
\tablenotetext{h}{\citet{przybilla08}}
\tablenotetext{i}{\citet{woo08}}
\tablenotetext{j}{\citet{u09}}
\tablenotetext{k}{\citet{kent87}}
\tablenotetext{l}{\citet{kud08}}
\tablenotetext{m}{\citet{hunter07}}
\tablenotetext{n}{\citet{schiller10}}
\tablenotetext{o}{\citet{trundle05}}
\tablenotetext{p}{\citet{venn01}}
\tablenotetext{q}{\citet{evans07}}
\tablenotetext{r}{\citet{urbaneja08}}
\tablenotetext{s}{\citet{kaufer04}}
\end{deluxetable}

\clearpage




\begin{thebibliography}{}

\bibitem[Allende Prieto et al.(2001)]{allende01} Allende Prieto, C., Lambert, D.~L., \& Asplund, M.\ 2001, \apjl, 556, L63 
\bibitem[Appleton et al.(1981)]{appleton81} Appleton, P.~N., Davies, R.~D., \& Stephenson, R.~J.\ 1981, \mnras, 195, 327 
\bibitem[Barker et al.(2009)]{barker09} Barker, M.~K., Ferguson, A.~M.~N., Irwin, M., Arimoto, N., \& Jablonka, P.\ 2009, \aj, 138, 1469 
\bibitem[B{\`e}land et al.(1988)]{beland88} B{\`e}land, S., Boulade, O., \& Davidge, T.\ 1988, Bulletin d'information du telescope Canada-France-Hawaii, 19, 16 
\bibitem[Bono et al.(2008)]{bono08} Bono, G., Caputo, F., Fiorentino, G., Marconi, M., \& Musella, I.\ 2008, \apj, 684, 102 
\bibitem[Bresolin(2003)]{bresolin03} Bresolin, F.\ 2003, in ``Stellar Candles for the Extragalactic Distance Scale'', Lecture Notes in Physics, 635, 
eds. D. Alloin \& W. Gieren, p. 149-174
\bibitem[Bresolin et al.(2005)]{bresolin05} Bresolin, F., Pietrzy{\'n}ski, G., Gieren, W., \& Kudritzki, R.~P.\ 2005, \apj, 634, 1020
\bibitem[Bresolin et al.(2006)]{bresolin06} Bresolin, F., Pietrzy{\'n}ski, G., Urbaneja, M.~A., et al.\ 2006, \apj, 648, 1007 
\bibitem[Bresolin et al.(2007)]{bresolin07} Bresolin, F., Urbaneja, M.~A., Gieren, W., Pietrzy{\'n}ski, G., \& Kudritzki, R.-P.\ 2007, \apj, 671, 2028 
\bibitem[Bresolin et al.(2009)]{bresolin09} Bresolin, F., Gieren, W., Kudritzki, R.-P., et al.\ 2009, \apj, 700, 309 
\bibitem[Bresolin et al.(2010)]{bresolin10} Bresolin, F., Stasi{\'n}ska, G., V{\'{\i}}lchez, J.~M., Simon, J.~D., 
\& Rosolowsky, E.\ 2010, \mnras, 404, 1679 
\bibitem[Bresolin(2011)]{bresolin11} Bresolin, F.\ 2011, \apj, 729, 56 
\bibitem[Brooks et al.(2007)]{brooks07} Brooks, A.~M., Governato, F., Booth, C.~M., et al.\ 2007, \apjl, 655, L17 
\bibitem[Cardelli et al.(1989)]{cardelli89} Cardelli, J.~A., Clayton, G.~C., \& Mathis, J.~S.\ 1989, \apj, 345, 245 
\bibitem[Chandar et al.(2001)]{chandar01} Chandar, R., Ford, H.~C., \& Tsvetanov, Z.\ 2001, \aj, 122, 1330 
\bibitem[Chemin et al.(2009)]{chemin09} Chemin, L., Carignan, 
C., \& Foster, T.\ 2009, \apj, 705, 1395 
\bibitem[Chiappini et al.(2001)]{chiappini01} Chiappini, C., Matteucci, F., \& Romano, D.\ 2001, \apj, 554, 1044 
\bibitem[Colavitti et al.(2008)]{colavitti08} Colavitti, E., Matteucci, F., \& Murante, G.\ 2008, \aap, 483, 401 
\bibitem[Daflon \& Cunha(2004)]{daflon04} Daflon, S., \& Cunha, K.\ 2004, \apj, 617, 1115 
\bibitem[Dalcanton et al.(2009)]{dalcanton09} Dalcanton, J.~J., Williams, B.~F., Seth, A.~C., et al.\ 2009, \apjs, 183, 67 
\bibitem[Davidge(2009)]{davidge09} Davidge, T.~J.\ 2009, \apj, 697, 1439 
\bibitem[Dav{\'e} et al.(2011a)]{dave11a} Dav{\'e}, R., Oppenheimer, B.~D., \& Finlator, K.\ 2011, \mnras, 415, 11 (a) 
\bibitem[Dav{\'e} et al.(2011b)]{dave11b} Dav{\'e}, R., Finlator, K., \& Oppenheimer, B.~D.\ 2011, \mnras, 416, 1354 (b)
\bibitem[de Blok et al.(2008)]{deblok08} de Blok, W.~J.~G., Walter, F., Brinks, E., et al.\ 2008, \aj, 136, 2648 
\bibitem[Deharveng et al.(2000)]{deharveng00} Deharveng, L., Pe{\~n}a, M., Caplan, J., \& Costero, R.\ 2000, \mnras, 311, 329 
\bibitem[De Lucia et al.(2004)]{delucia04} De Lucia, G., Kauffmann, G., \& White, S.~D.~M.\ 2004, \mnras, 349, 1101 
\bibitem[de Rossi et al.(2007)]{derossi07} de Rossi, M.~E., Tissera, P.~B., \& Scannapieco, C.\ 2007, \mnras, 374, 323 
\bibitem[Denicol{\'o} et al.(2002)]{denicolo02} Denicol{\'o}, G., Terlevich, R., \& Terlevich, E.\ 2002, \mnras, 330, 69 
\bibitem[Durrell et al.(2010)]{durell10} Durrell, P.~R., Sarajedini, A., \& Chandar, R.\ 2010, \apj, 718, 1118 
\bibitem[Evans et al.(2007)]{evans07} Evans, C.~J., Bresolin, F., Urbaneja, M.~A., et al.\ 2007, \apj, 659, 1198 
\bibitem[Fiorentino et al.(2002)]{fiorentino02} Fiorentino, G., Caputo, F., Marconi, M., \& Musella, I.\ 2002, \apj, 576, 402 
\bibitem[Fiorentino et al.(2007)]{fiorentino07} Fiorentino, G., Marconi, M., Musella, I., \& Caputo, F.\ 2007, \aap, 476, 863 
\bibitem[Finlator \& Dav{\'e}(2008)]{finlator08} Finlator, K., \& Dav{\'e}, R.\ 2008, \mnras, 385, 2181 
\bibitem[Freedman et al.(1994)]{freedman94} Freedman, W.~L., Hughes, S.~M., Madore, B.~F., et al.\ 1994, \apj, 427, 628 
\bibitem[Freedman et al.(2001)]{freedman01} Freedman, W.~L., Madore, B.~F., Gibson, B.~K., et al.\ 2001, \apj, 553, 47 
\bibitem[Fuhrmann(2011)]{fuhrmann11} Fuhrmann, K.\ 2011, \mnras, 414, 2893 
\bibitem[Garnett \& Shields(1987)]{garnett87} Garnett, D.~R., \& Shields, G.~A.\ 1987, \apj, 317, 82 
\bibitem[Garnett et al.(1997)]{garnett97} Garnett, D.~R., Shields, G.~A., Skillman, E.~D., Sagan, S.~P., \& Dufour, R.~J.\ 1997, \apj, 489, 63 
\bibitem[Garnett(2004)]{garnett04} Garnett, D.~R.\ 2004, Cosmochemistry.~The melting pot of the elements, 171 
\bibitem[Gerke et al.(2011)]{gerke11} Gerke, J.~R., Kochanek, C.~S., Prieto, J.~L., Stanek, K.~Z., \& Macri, L.~M.\ 2011, arXiv:1103.0549 
\bibitem[Grevesse \& Sauval(1998)]{grevesse98} Grevesse, N., \& Sauval, A.~J.\ 1998, \ssr, 85, 161 
\bibitem[Heckman et al.(1990)]{heckman90} Heckman, T.~M., Armus, L., \& Miley, G.~K.\ 1990, \apjs, 74, 833 
\bibitem[Henry \& Howard(1995)]{henry95} Henry, R.~B.~C., Howard, J.~W.\ 1995, \apj, 438, 170 
\bibitem[Henry et al.(2010)]{henry10} Henry, R.~B.~C., Kwitter, K.~B., Jaskot, A.~E., et al.\ 2010, \apj, 724, 748 
\bibitem[Herrero et al.(1992)]{herrero92} Herrero, A., Kudritzki, R.~P., Vilchez, J.~M., et al.\ 1992, \aap, 261, 209 
\bibitem[Horne(1986)]{horne86} Horne, K.\ 1986, \pasp, 98, 609 
\bibitem[Hou et al.(2000)]{hou00} Hou, J.~L., Prantzos, N., \& Boissier, S.\ 2000, \aap, 362, 921 
\bibitem[Humphreys et al.(2008)]{humphreys08} Humphreys, E.~M.~L., Reid, M.~J., Greenhill, L.~J., Moran, J.~M., 
\& Argon, A.~L.\ 2008, \apj, 672, 800 
\bibitem[Hunter et al.(2007)]{hunter07} Hunter, I., Dufton, P.~L., Smartt, S.~J., et al.\ 2007, \aap, 466, 277 
\bibitem[Kaufer et al.(2004)]{kaufer04} Kaufer, A., Venn, K.~A., Tolstoy, E., Pinte, C., \& Kudritzki, R.-P.\ 2004, \aj, 127, 2723
\bibitem[Kennicutt et al.(1998)]{kennicutt98} Kennicutt, R.~C., Jr., Stetson, P.~B., Saha, A., et al.\ 1998, \apj, 498, 181 
\bibitem[Kent(1987)]{kent87} Kent, S.~M.\ 1987, \aj, 93, 816 
\bibitem[Kewley \& Dopita(2002)]{kewley02} Kewley, L.~J., \& Dopita, M.~A.\ 2002, \apjs, 142, 35 
\bibitem[Kewley \& Ellison(2008)]{kewley08} Kewley, L.~J., \& Ellison, S.~L.\ 2008, \apj, 681, 1183 
\bibitem[Kobulnicky \& Kewley(2004)]{kobulnicky04} Kobulnicky, H.~A., \& Kewley, L.~J.\ 2004, \apj, 617, 240 
\bibitem[K{\"o}ppen et al.(2007)]{koeppen07} K{\"o}ppen, J., Weidner, C., \& Kroupa, P.\ 2007, \mnras, 375, 673 
\bibitem[Komatsu et al.(2009)]{komatsu09} Komatsu, E., Dunkley, J., Nolta, M.~R., et al.\ 2009, \apjs, 180, 330 
\bibitem[Kudritzki \& Urbaneja(2009)]{kud09} Kudritzki, R.~P., \& Urbaneja, M.~A.\ 2009, Massive Stars: From Pop III and GRBs to the Milky Way.~Space Telescope Science Institute Symposium Series No.~20.~Edited by Mario Livio and Eva Villaver.~ Cambridge University Press, 2009, ISSN 9780521762632, p.126-151
\bibitem[Kudritzki et al.(2008)]{kud08} Kudritzki, R.-P., Urbaneja, M.~A., Bresolin, F., et al.\ 2008, \apj, 681, 269 
\bibitem[Kudritzki et al.(2003)]{kud03} Kudritzki, R.~P., Bresolin, F., \& Przybilla, N.\ 2003, \apjl, 582, L83 
\bibitem[Lequeux et al.(1979)]{lequeux79} Lequeux, J., Peimbert, M., Rayo, J.~F., Serrano, A., \& Torres-Peimbert, S.\ 1979, \aap, 80, 155 
\bibitem[Lee et al. (2006)]{lee06} Lee, H., Skillman, E.D., Cannon, J.M., et al.\ 2006, \apj, 647, 970
\bibitem[Luck et al.(1998)]{luck98} Luck, R.~E., Moffett, T.~J., Barnes, T.~G., \& Gieren, W. \ 1998, \aj, 115, 605 
\bibitem[Luck et al.(2006)]{luck06} Luck, R.~E., Kovtyukh, V.~V., \& Andrievsky, S.~M.\ 2006, \aj, 132, 902 
\bibitem[Luck et al.(2011)]{luck11} Luck, R.~E., Andrievsky, S.~M., Kovtyukh, V.~V., Gieren, W., \& Graczyk, D.\ 2011, \aj, 142, 51
\bibitem[Maciel \& Costa(2009)]{maciel08} Maciel, W.~J., \& Costa, R.~D.~D.\ 2009, IAU Symposium, 254, 38P 
\bibitem[Maciel et al.(2010)]{maciel09} Maciel, W.~J., Costa, R.~D.~D., \& Idiart, T.~E.~P.\ 2010, \aap, 512, A19 
\bibitem[Macri et al.(2006)]{macri06} Macri, L.~M., Stanek, K.~Z., Bersier, D., Greenhill, L.~J., \& Reid, M.~J.\ 2006, \apj, 652, 1133 
\bibitem[Madore(1982)]{madore82} Madore, B.~F.\ 1982, \apj, 253, 575
\bibitem[Maiolino et al.(2008)]{maiolino08} Maiolino, R., Nagao, T., Grazian, A., et al.\ 2008, \aap, 488, 463 
\bibitem[Majaess et al. (2011)]{majaess11} Majaess, D., Turner, D., \& Gieren, W. \ 2011, \apj, 741, L36
\bibitem[Marconi et al.(2005)]{marconi05} Marconi, M., Musella, I., \& Fiorentino, G.\ 2005, \apj, 632, 590 
\bibitem[Martin(1997)]{martin97} Martin, C.~L.\ 1997, \apj, 491, 561 
\bibitem[McCommas et al.(2009)]{mccommas09} McCommas, L.~P., Yoachim, P., Williams, B.~F., et al.\ 2009, \aj, 137, 4707 
\bibitem[McGaugh(1991)]{mcgaugh91} McGaugh, S.~S.\ 1991, \apj, 380, 140 
\bibitem[Meynet \& Maeder(2003)]{meynet03} Meynet, G., \& Maeder, A.\ 2003, \aap, 404, 975 
\bibitem[Mould \& Sakai(2008)]{mould08} Mould, J., \& Sakai, S.\ 2008, \apjl, 686, L75 
\bibitem[Mould \& Sakai(2009)]{mould09} Mould, J., \& Sakai, S.\ 2009, \apj, 697, 996 
\bibitem[Naab \& Ostriker(2006)]{naab06} Naab, T., \& Ostriker, J.~P.\ 2006, \mnras, 366, 899 
\bibitem[Oke(1990)]{oke90} Oke, J.~B.\ 1990, \aj, 99, 1621 
\bibitem[Oke et al.(1995)]{oke95} Oke, J.~B., Cohen, J.~G., Carr, M., et al.\ 1995, \pasp, 107, 375 
\bibitem[Paturel et al.(2003)]{paturel03} Paturel, G., Petit, C., Prugniel, P., et al.\ 2003, \aap, 412, 45 
\bibitem[Pagel et al.(1979)]{pagel79} Pagel, B.~E.~J., Edmunds, M.~G., Blackwell, D.~E., Chun, M.~S., \& Smith, G.\ 1979, \mnras, 189, 95 
\bibitem[Pettini \& Pagel(2004)]{pettini04} Pettini, M., \& Pagel, B.~E.~J.\ 2004, \mnras, 348, L59 
\bibitem[Pilyugin(2001)]{pilyugin01} Pilyugin, L.~S.\ 2001, \aap, 374, 412 
\bibitem[Pilyugin \& Thuan(2005)]{pilyugin05} Pilyugin, L.~S., \& Thuan, T.~X.\ 2005, \apj, 631, 231 
\bibitem[Prantzos \& Boissier(2000)]{prantzos00} Prantzos, N., \& Boissier, S.\ 2000, \mnras, 313, 338 
\bibitem[Press et al.(1992)]{press92} Press, W.~H., Teukolsky, S.~A., Vetterling, W.~T., 
\& Flannery, B.~P.\ 1992, Cambridge: University Press, |c1992, 2nd ed.,  
\bibitem[Przybilla et al.(2006)]{przybilla06} Przybilla, N., Butler, K., Becker, S.~R., \& Kudritzki, R.~P.\ 2006, \aap, 445, 1099 
\bibitem[Przybilla et al.(2008a)]{przybilla08} Przybilla, N., Nieva, M.~F., Heber, U., \& Butler, K.\ 2008, \apjl, 684, L103 (a) 
\bibitem[Przybilla et al.(2008b)]{przybilla08b} Przybilla, N., Butler, K., \& Kudritzki, R.-P.\ 2008, The Metal-Rich Universe, ed. 
G. Israelian, \& G. Meynet (Cambridge: Cambridge University Press), 332  (b)
\bibitem[Riess et al.(2009a)]{riess09a} Riess, A.~G., Macri, L., Li, W., et al.\ 2009, \apjs, 183, 109 (a)
\bibitem[Riess et al.(2009b)]{riess09b} Riess, A.~G., Macri, L., Casertano, S., et al.\ 2009, \apj, 699, 539 (b) 
\bibitem[Riess et al.(2011)]{riess11} Riess, A.~G., Macri, L., Casertano, S., et al.\ 2011, \apj, 730, 119 
\bibitem[Rizzi et al.(2007)]{rizzi07} Rizzi, L., Tully, R.~B., Makarov, D., et al.\ 2007, \apj, 661, 815 
\bibitem[Rolleston et al.(2000)]{rolleston00} Rolleston, W.~R.~J., Smartt, S.~J., Dufton, P.~L., \& Ryans, R.~S.~I.\ 2000, \aap, 363, 537 
\bibitem[Romaniello et al.(2008)]{romaniello08} Romaniello, M., Primas, F., Mottini, M., et al.\ 2008, \aap, 488, 731 
\bibitem[Rood et al.(2007)]{rood07} Rood, R.~T., Quireza, C., Bania, T.~M., Balser, D.~S., 
\& Maciel, W.~J.\ 2007, From Stars to Galaxies: Building the Pieces to Build Up the Universe, 374, 169 
\bibitem[Ro{\v s}kar et al.(2008)]{roskar08} Ro{\v s}kar, R., Debattista, V.~P., Quinn, T.~R., Stinson, G.~S., \& Wadsley, J.\ 2008, \apjl, 684, L79 
\bibitem[Sakai et al.(2004)]{sakai04} Sakai, S., Ferrarese, L., Kennicutt, R.~C., Jr., \& Saha, A.\ 2004, \apj, 608, 42 
\bibitem[S{\'a}nchez-Bl{\'a}zquez et al.(2009)]{sanchez09} S{\'a}nchez-Bl{\'a}zquez, P., Courty, S., Gibson, B.~K., 
\& Brook, C.~B.\ 2009, \mnras, 398, 591 
\bibitem[Santiago-Cort{\'e}s et al.(2010)]{santiago10} Santiago-Cort{\'e}s, M., Mayya, Y.~D., \& Rosa-Gonz{\'a}lez, D.\ 2010, \mnras, 405, 1293 
\bibitem[Schiller \& Przybilla(2008)]{schiller08} Schiller, F., \& Przybilla, N.\ 2008, \aap, 479, 849 
\bibitem[Schiller (2010)]{schiller10} Schiller, F., 2010, {\em Quantitative Spectroscopy of BA-type Supergiants in the Small Magellanic Cloud}, thesis, Friedrich-Alexander-University Erlangen-Nuernberg, Germany
\bibitem[Schlegel et al.(1998)]{schlegel98} Schlegel, D.~J., Finkbeiner, D.~P., \& Davis, M.\ 1998, \apj, 500, 525 
\bibitem[Scowcroft et al.(2009)]{scowcroft09} Scowcroft, V., Bersier, D., Mould, J.~R., \& Wood, P.~R.\ 2009, \mnras, 396, 1287 
\bibitem[Shappee \& Stanek (2011)]{shappee11} Shappee, B.~J, \& Stanek, K.~Z., 2011, \apj, 733, 124
\bibitem[Skillman(1998)]{skillman98} Skillman, E.~D.\ 1998, Stellar astrophysics for the local group: VIII Canary Islands Winter School 
of Astrophysics, 457 
\bibitem[Smartt et al.(2001)]{smartt01} Smartt, S.~J., Crowther, P.~A., Dufton, P.~L., et al.\ 2001, \mnras, 325, 257 
\bibitem[Sofue et al.(2009)]{sofue09} Sofue, Y., Honma, M., \& Omodaka, T.\ 2009, \pasj, 61, 227 
\bibitem[Spergel(2006)]{spergel06} Spergel, D.\ 2006, APS April Meeting Abstracts, 5002 
\bibitem[Spergel et al.(2007)]{spergel07} Spergel, D.~N., Bean, R., Dor{\'e}, O., et al.\ 2007, \apjs, 170, 377 
\bibitem[Stanghellini et al.(2010)]{stanghellini10} Stanghellini, L., Magrini, L., Villaver, E., \& Galli, D.\ 2010, \aap, 521, A3 (a)
\bibitem[Stanghellini \& Haywood(2010)]{stanghellini10b} Stanghellini, L., \& Haywood, M.\ 2010, \apj, 714, 1096 (b)
\bibitem[Stauffer \& Bothun(1984)]{stauffer84} Stauffer, J.~R., \& Bothun, G.~D.\ 1984, \aj, 89, 1702 
\bibitem[Storm et al.(2011)]{storm11} Storm, J., Gieren, W., Fouque, P., et al.\ 2011, arXiv:1109.2016 
\bibitem[Tegmark et al.(2004)]{tegmark04} Tegmark, M., Strauss, M.~A., Blanton, M.~R., et al.\ 2004, \prd, 69, 103501 
\bibitem[Tikhonov et al.(2005)]{tikhonov05} Tikhonov, N.~A., Galazutdinova, O.~A., \& Drozdovsky, I.~O.\ 2005, \aap, 431, 127 
\bibitem[Tremonti et al.(2004)]{tremonti04} Tremonti, C.~A., Heckman, T.~M., Kauffmann, G., et al.\ 2004, \apj, 613, 898 
\bibitem[Trundle et al.(2002)]{trundle02} Trundle, C., Dufton, P.~L., Lennon, D.~J., Smartt, S.~J., \& Urbaneja, M.~A.\ 2002, \aap, 395, 519 
\bibitem[Trundle \& Lennon(2005)]{trundle05} Trundle, C., \& Lennon, D.~J.\ 2005, \aap, 434, 677 
\bibitem[Tully et al.(2009)]{tully09} Tully, R.~B., Rizzi, L., Shaya, E.~J., et al.\ 2009, \aj, 138, 323 
\bibitem[U et al.(2009)]{u09} U, V., Urbaneja, M.~A., Kudritzki, R.-P., et al.\ 2009, \apj, 704, 1120 
\bibitem[Urbaneja et al.(2008)]{urbaneja08} Urbaneja, M.~A., Kudritzki, R.-P., Bresolin, F., et al.\ 2008, \apj, 684, 118 
\bibitem[Urbaneja et al.(2005a)]{urbaneja05} Urbaneja, M.~A., Herrero, A., Bresolin, F., et al.\ 2005, \apj, 622, 862 
\bibitem[Urbaneja et al.(2005b)]{urbaneja05b} Urbaneja, M.~A., Herrero, A., Kudritzki, R.P., et al.\ 2005, \apj, 635, 311 
\bibitem[Venn et al.(2001)]{venn01} Venn, K.~A., Lennon, D.~J., Kaufer, A., et al.\ 2001, \apj, 547, 765 
\bibitem[Wiersma et al.(2009)]{wiersma09} Wiersma, R.~P.~C., Schaye, J., \& Smith, B.~D.\ 2009, \mnras, 393, 99 
\bibitem[Williams et al.(2009)]{williams09} Williams, B.~F., Dalcanton, J.~J., Seth, A.~C., et al.\ 2009, \aj, 137, 419 
\bibitem[Woo et al.(2008)]{woo08} Woo, J., Courteau, S., \& Dekel, A.\ 2008, \mnras, 390, 1453 
\bibitem[Yin et al.(2009)]{yin09} Yin, J., Hou, J.~L., Prantzos, N., et al.\ 2009, \aap, 505, 497 
\bibitem[Zaritsky et al.(1994)]{zaritsky94} Zaritsky, D., Kennicutt, R.~C., Jr., \& Huchra, J.~P.\ 1994, \apj, 420, 87 

\end{thebibliography}
\end{document}